\documentclass[preprint,12pt]{elsarticle}

\usepackage{lscape}
\usepackage{longtable}
\usepackage{amsmath,amssymb}
\usepackage{graphicx}

\usepackage{amssymb}
\usepackage{mathtools}

\usepackage[mathlines]{lineno}

\begin{document}

\begin{frontmatter}

\title{A particle-based hybrid code for planet formation}

\author[ucla,jpl]{Ryuji Morishima\corref{cor1}}
\ead{Ryuji.Morishima@jpl.nasa.gov}

\cortext[cor1]{Corresponding author}

\address[ucla]{University of California, Los Angeles, Institute of Geophysics and Planetary Physics, Los Angeles, CA  90095, USA}
\address[jpl]{Jet Propulsion Laboratory/California Institute of Technology,
Pasadena, CA 91109, USA}

\begin{abstract}
We introduce a new particle-based hybrid code for planetary accretion.
The code uses an $N$-body routine for interactions with planetary embryos 
while it can handle a large number of planetesimals using 
a super-particle approximation, 
in which a large number of small planetesimals are represented by a small number of tracers. 
Tracer-tracer interactions are handled by a statistical routine which uses the phase-averaged stirring and collision rates.
We compare hybrid simulations with analytic predictions and pure $N$-body simulations for various problems 
in detail and find good agreements for all cases.
The computational load on the portion of the statistical routine is comparable to or less than that for the $N$-body routine.
The present code includes an option of hit-and-run bouncing but not fragmentation, which remains for future work.

\end{abstract}


\end{frontmatter}


\section{Introduction}
Terrestrial planets and cores of giant planets are generally considered to have formed through accretion of many small 
bodies called planetesimals.
To simulate accretion processes of planets, two methods, which are complementary to each other, have been applied.

The first one is $N$-body simulations in which orbits of all bodies are numerically integrated
and gravitational accelerations due to other bodies are calculated in every time step
(e.g., Kokubo and Ida, 1996; Chambers and Wetherill, 1998; Richardson et al., 2000; Morishima et al., 2010).
$N$-body simulations are accurate and can automatically handle any complicated phenomena, such as
resonant interactions and spatially non-uniform distributions of planetesimals.
Gravity calculations are accelerated by such as tree-methods (Richardson et al., 2000) or special hardwares (Kokubo and Ida, 1996; Grimm and Stadel, 2014),
and a large time step can be used with sophisticated integrators, such as Mixed Variable Symplectic (MVS) or Wisdom-Holman integrators
(Duncan et al., 1998; Chambers, 1999).
Even with these novel techniques, $N$-body simulations are computationally intense and 
the number of particles or the number of time steps in a simulation is severely limited.

The second approach is statistical calculations in which
 planetesimals are placed in two dimensional (distance and mass) Eulerian grids, 
 and the time evolutions of the number and the mean velocity of an ensemble of planetesimals in each grid are calculated 
 using the phase-averaged collision and stirring rates
(e.g., Greenberg et al., 1978; Wetherill and Stewart, 1989, 1993; Inaba et al., 2001; Morbidelli et al., 2009; Kobayashi et al., 2010). 
While this approach does not directly follow orbits of individual particles,  it can handle many particles, 
even numerous collisional fragments.
Largest bodies, called planetary embryos, are handled in a different manner than small bodies, 
taking into account their orbital isolation. The mutual orbital separation between neighboring embryos
is usually assumed to be 10 mutual Hill radii.

The last assumption is not always guaranteed, particularly in the late stage of planetary accretion
(e.g., Chambers and Wetherill, 1998).
To handle orbital evolution of embryos more accurately,  Eulerian hybrid codes \footnote{The term "hybrid" means 
a combination of two different approaches/schemes. While this term might be used as a combination of 
Eulerian grids and Lagrangian embryos by the authors, we use the term for a combination 
of statistical calculations and $N$-body calculations.} have been developed 
(Spaute et al., 1991; Weidenschilling et al., 1997; Bromley and Kenyon, 2006; Glaschke et al., 2014),
in which small planetesimals are still handled by the Eulerian approach whereas
orbital evolutions of largest embryos are individually followed by such as $N$-body integrations.
Gravitational and collisional interactions between embryos and small planetesimals are handled using analytic prescriptions. 
Glaschke et al. (2014) took into account radial diffusion of planetesimals due to gravitational scattering
of embryos and their code can approximately handle gap opening around embryos' orbits. 
  
A Lagrangian hybrid method has also been introduced by Levison et al. (2012) (LDT12 hereafter).  In their LIPAD code, 
a large number of planetesimals are represented by a small number of Lagrangian tracers. This type of approach is  
called a super-particle approximation and is also employed in modeling of debris disks (Kral et al., 2013; Nesvold et al., 2013)
and planetesimal formation (Johansen et al., 2007; Rein et al., 2010). 
Orbits of individual tracers are directly followed by numerical integrations,  
and interactions between planetesimals (stirring and collisions) in tracers are handled by a statistical routine.
Embryos are represented by single particles and the accelerations of any bodies due to gravity of embryos are handled in the $N$-body routine.
Lagrangian hybrid methods have several advantages than Eulerian hybrid methods.
For example, Lagrangian methods can accurately handle spatial inhomogeneity in a planetesimal disk (e.g., 
gap opening by embryos), planetesimal-driven migration, resonant interactions between embryos and small planetesimals, 
and eccentric ringlets.  
A drawback of Lagrangian methods would be computational cost as 
orbits of all tracers need to be directly integrated. Therefore, it is desirable that 
Lagrangian hybrid methods can handle various problems accurately 
even with a moderately small number of tracers.

In this paper, we develop a new Lagrangian hybrid code for planet formation.
While we follow the basic concept of particle classes introduced by LDT12,
recipes used in our code are different from those used in the LIPAD code in many places.
The largest difference appears in the methods to handle viscous stirring and dynamical friction.
The LIPAD code solves pair-wise interactions while our method gives the accelerations of tracers using 
the phase-averaged stirring and dynamical friction rates.  
While the LIPAD code conducts a series of three body integrations in the stirring routine (in the shear dominant regime)
and in the routine of sub-embryo migration during a simulation,
our code avoids them by using unique routines.
The complete list of comparison with the LIPAD code turns out to be rather long and is given in Appenedix~G. 

In Section~2,  we explain our method.
In Section~3,  we show various tests of the new hybrid code and  
compare them with analytic estimates and 
pure $N$-body simulations. 
The computational speed and limitations of our code are discussed in Section~4. 
The summary is given in Section~5.
For the sake of clarity, specific derivations are deferred to appendices.

\section{Method}

\subsection{Particle classes}

\begin{figure}
\begin{center}
\includegraphics[width=.8\textwidth]{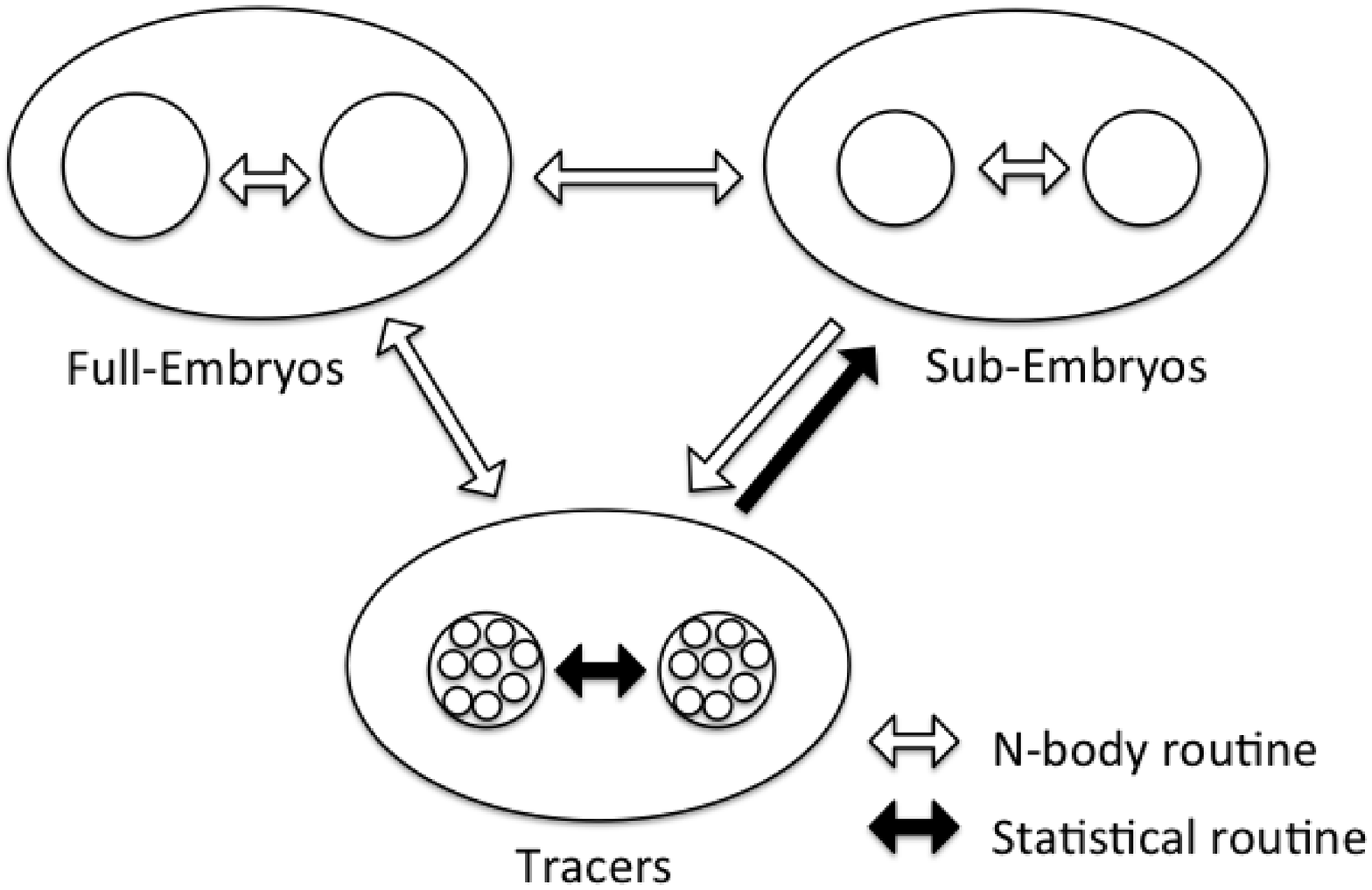}
\end{center}

Fig.~1. Schematic illustration of our hybrid code.  Accelerations due to gravity of embryos are handled by the $N$-body routine. 
Tracer-tracer interactions and back reaction of tracers on sub-embryos including collision are handled by 
the statistical routine.

\end{figure}
The particle classes in our hybrid code are the same as those introduced by LDT12.
The code has three classes of particles: tracers, sub-embryos, and full-embryos (Fig.~1). 
In the present paper, we do not consider fragmentation of planetesimals and dust-tracers are not introduced.
A tracer represents a large number of equal-mass planetesimals on roughly the same orbits.
The mass of a planetesimal and the number of planetesimals in the tracer $i$ ($i$ for the index of the tracer) 
are defined to be $m_i$ and $k_i$. Therefore, the tracer mass is given by $k_i m_i$.
Through collisional growth, $m_i$ increases and $k_i$ decreases
while $k_i m_i$ remains close to its original value. 
We allow mass exchanges between tracers through collisions so the tracer mass, $k_i m_i$, 
is not necessarily fixed but has the upper and lower limits, $m_{t,{\rm max}}$ and $m_{t,{\rm min}}$ ($m_{t,{\rm min}} < k_i m_i < m_{t,{\rm max}}$). 
We employ $m_{t,{\rm max}} = 2.0 m_{t0}$ and $m_{t,{\rm min}} = 0.1m_{t0}$ in this paper,  
where $m_{t0}$ is the minimum mass of a sub-embryo.
The mass $m_{t0}$ is usually the initial mass of a tracer, which is the same for all tracers at the beginning of a simulation.
If $k_i = 1$ and $m_i \ge m_{t0}$,  the tracer is promoted to a sub-embryo. 
If $m_i \ge f_{\rm f} m_{t0}$, the sub-embryo is further promoted to a full-embryo, 
where we employ the numerical factor $f_{\rm f}$ of 100, as recommended by LDT12.
The number $k_i$ is an integer in our model.


Orbits of any types of particles are directly integrated.
We use a Mixed Variable Symplectic integrator known as SyMBA (Duncan et al., 1998), which 
can handle close encounters between particles. This integrator is also used by LDT12. 
The collisional and gravitational interactions between full-embryos with tracers 
are directly handled in every time step of orbital integrations, as is the case of pure $N$-body simulations. 
On the other hand, tracer-tracer interactions are handled in a statistical routine which is described in subsequent sections in great detail.
While the time step for the orbital integration $\delta t$ is $\sim 10^{-2}$ of the orbital period, 
the time step $\Delta t$ for the statistical routine can be taken to be much larger
as long as $\Delta t $ is sufficiently smaller than the stirring and collisional timescales. 
In this paper, we employ $\Delta t = 30 \delta t $ for all test simulations. 
We confirmed that almost the same outcome is obtained  even with
a smaller $\Delta t $ for all the test simulations.

In interactions between a tracer and a full-embryo in the $N$-body routine, 
the tracer is assumed to be a single particle with the mass equivalent to the total mass of planetesimals in the tracer.
This is not a problem as long as the embryo is sufficiently massive compared with tracers.
However,  if the embryo mass is similar to tracer masses as is the case immediately after its promotion, and if
embryo-tracer interactions are handled by the direct $N$-body routine,  the embryo suffers artificially strong kicks from tracers. 
To avoid this issue, sub-embryos are introduced. Accelerations of sub-embryos due to gravitational interactions with tracers are handled by  
the statistical routine whereas accelerations of tracers due to gravitational interactions with sub-embryos are 
handled by the direct $N$-body routine (Fig.~1). 
Collisions of planetesimals in tracers with sub-embryos 
are also handled in the statistical routine to avoid artificially large mass jumps in sub-embryos, contrary to LDT12 (see Appendix~G for more discussion).

\subsection{Neighbor search for calculating surface number densities}

\begin{figure}
\begin{center}
\includegraphics[width=.63\textwidth]{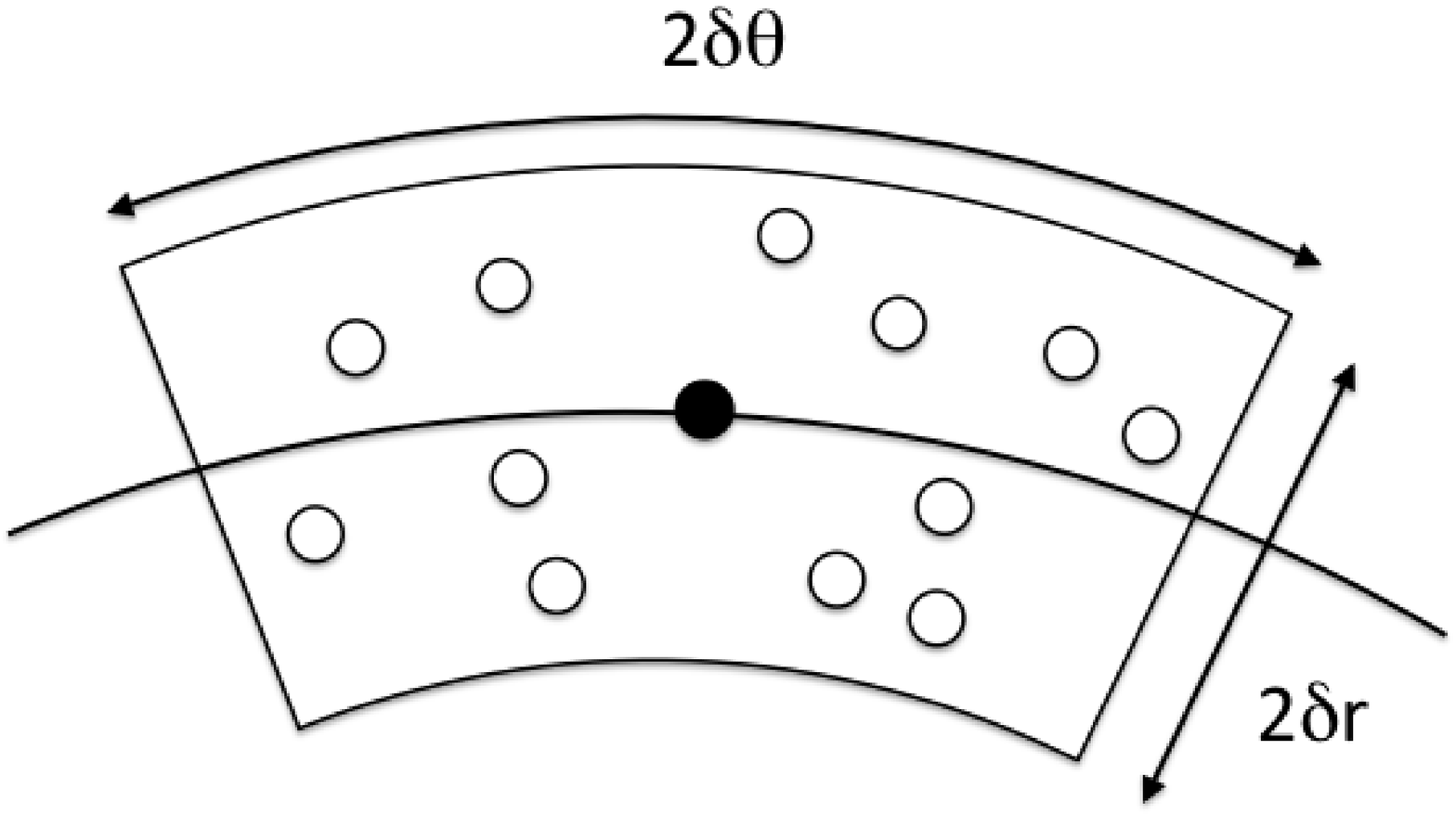}
\end{center}

Fig.~2. Schematic illustration of neighboring tracer search. The filled circle is the target tracer and open
circles are interloping tracers (interlopers). The surface number density of planetesimals 
in an interloper is simply given by the number of planetesimal in the interloper divided by the area of the region between 
two arcs.

\end{figure}

Our statistical routine uses the surface number density 
of nearby tracers. We first describe how we determine the surface number density
for tracer-tracer interactions. The surface number density of tracers for 
interactions between a sub-embryo and tracers 
is derived in a similar way, but slightly modified so that close and distant encounters
are treated differently (Section~2.5).

Consider a target tracer $i$, surrounded by nearby interloping tracers.
The cylindrical co-ordinate system $(r,\theta, z)$ is introduced and 
we consider a curved region, called the region $i$, located at $z=0$ and centered at the position of the tracer $i$ (Fig.~2). 
The radial half width and the angular half width in the $\theta$ direction of the region $i$ 
are given by $\delta r$ and $\delta \theta$, respectively.
The area of this region is given by  
\begin{equation}
S_i = 4 r_i \delta r \delta \theta.\label{eq:ss}
\end{equation}
For neighboring tracer search,  tracers are sorted into 2D $(r,\theta)$ grids and 
only nearby cells of the target tracer are checked whether other tracers are in the designated region.
If the interloper $j$ is in the region $i$, 
the surface number density of planetesimals in the interloper $j$ around the target $i$ is simply expressed as 
\begin{equation}
n_j  = \frac{k_j}{S_i}. \label{eq:nj}
\end{equation}
The density $n_j$ is an instantaneous density.
At other times, the interloper $j$ may not be in the region $i$ but
other interlopers having planetesimal masses similar to $m_j$ may be in the region $i$.
After averaging over time, the encounter rate with planetesimals in a certain mass range 
is expected to converge regardless of the size of $S_i$.

A small $\delta r $ is favorable to capture a small radial structure accurately.
However, it is not reasonable to use $\delta r $ smaller than 
a typical radial length of gravitational influence of a planetesimal,
which is usually about $\sim 10 R_{\rm H}$.
Here,  $R_{\rm H}$ is the Hill radius given by 
\begin{equation}
R_{\rm H} = a_i \left(\frac{m_i}{3M_{\odot}}\right)^{1/3},
\end{equation}
where $a_i$ is the semi-major axis of the tracer $i$ and $M_{\odot}$ is the 
mass of the central star.
Since the mass of a possible largest planetesimal is $m_{t0}$, we employ 
\begin{equation}
\delta r =  10 a_i \left(\frac{m_{t0}}{3M_{\odot}}\right)^{1/3}, \label{eq:drt0}
\end{equation}
for tracer-tracer interactions.
For interactions between a sub-embryo and tracers, 
a larger $\delta r$ needs to be used and we employ 10 Hill radii of a possible largest sub-embryo with the mass of $f_{\rm f}m_{t0}$. 

An advantage of the choice of $\delta r$ given by Eq.~(\ref{eq:drt0}) is that we can omit the procedure for orbital isolation of planetesimals.
Orbital isolation of the largest bodies occurs if the sum of orbital spaces of these bodies is less than the width of the region of interest (Wetherill and Stewart, 1993).
If a planetesimal is assumed to occupy an orbital space of 10 Hill radii (Kokubo and Ida, 1998) and if the tracer masses are $m_{t0}$, 
the sum of the orbital spaces of planetesimals 
in a pair of tracers are always larger than $2\delta r$. 
Since our model allows a tracer mass lower than $m_{t0}$, isolation can potentially occur if the tracer masses 
and the numbers of planetesimals in the tracers are both small. Such cases are rather rare and we can ignore them.
Therefore, orbital isolation can occur only for embryos and that is handled by the $N$-body routine. 
An additional discussion is given in Appendix~G. 

A small $\delta \theta$ is favorable to handle non-axisymmetric structures, in which 
the surface density is not azimuthally uniform.
Such structures usually appear if  external massive bodies exist
(Kortenkamp and Wetherill, 2000; Nagasawa et al., 2005; Levison and Morbidelli, 2007; Queck et al., 2007).
A small $\delta \theta$ can also be chosen, if we want to suppress the number of interlopers to save computational costs (Section~4.1).
In this paper, we usually adopt $\delta \theta = 0.5 \pi$, but dependency on $\delta \theta$ is also examined.

Even if the tracer $j$ is in the region $i$,  the tracer $i$ is occasionally outside the region $j$
used in neighboring search around the tracer $j$. 
To make sure that the tracer $i$ and the tracer $j$ are mutually counted (or excluded) 
as interlopers, the number density $n_j$ is evaluated using Eq.~(\ref{eq:nj}) 
only if $m_i \ge m_j $ or $i < j$ for $m_i = m_j$.
Otherwise and only if the tracer $i$ is in the region $j$, 
$n_j$ is given by $k_j/S_j$, where $S_j$ is the area of the region $j$. 

Although we only considered encounters with interlopers, planetesimals in the target 
$i$ may encounter with other planetesimals in the same target if $k_i > 1$. 
Appendix~A describes how we correct this effect.
This effect is found to be too small to be identified.

\subsection{Hill's approximation and quasi ergodic hypothesis}
Let the semimajor axis, the orbital eccentricity, the orbital inclination, the longitudes of pericenter 
and  ascending node, and the mean longitude of the target be $a_i$, $e_i$, $i_i$, $\varpi_i$, $\Omega_i$, and $\lambda_i$,
and those for the interloper be $a_j$, $e_j$, $i_j$, $\varpi_j$, $\Omega_j$, and $\lambda_j$.
The difference between the semimajor axes is defined as $d_{ij} = a_j-a_i$.
When the target encounters with the interloper, they may collide with each other at a certain probability.
Their orbital elements are also modified by mutual gravitational scattering. 
To handle these processes in the statistical routine in a simple manner, we make following assumptions.
\begin{enumerate}
\item We use the collision probability and the mean change rates of orbital elements 
derived by previous studies which usually solved Hill's equations.
The equations of motion are reduced to Hill's equations if Hill's approximation 
is applied (Hill, 1878; Nakazawa et al., 1989).
The conditions to apply Hill's approximation are 
\begin{enumerate}
\item $e_i$, $i_i$, $e_j$, $i_j$ $\ll 1$, 
\item $m_i$, $m_j$ $\ll M_{\odot}$, and 
\item $|d_{ij}| \ll a_i$.
\end{enumerate}

\item 
The semimajor axis differences $d_{ij}$ and the phases of the interlopers, 
$\varpi_j$, $\Omega_j$, $\lambda_j$ are uniformly distributed.
Namely, if a certain interloper in the region $i$ is found, we assume that 
other interlopers with the same $e_j$, $i_j$, but different 
$d_{ij}$, $\varpi_j$, $\Omega_j$, and $\lambda_j$ are 
distributed uniformly over the region $i$.

\item The longitudes,  $\varpi_i$ and $\Omega_i$, of the target change randomly and large 
enough on the evolution timescales of  $e_i$ and $i_i$. 
This approximation is unnecessary if Approximation 2 is valid and either $e_i \ll e_j$ or $i_i \ll i_j$, 
because the collision and stirring rates do not depend on $\varpi_i$ and $\Omega_i$ in such a situation.

\end{enumerate}

Assumption 1 (Hill's approximation) seems to work well in most of cases of planetary accretion.
Greenzweig and Lissauer (1990) showed that the collision probability of a planet with 
planetesimals on circular and co-planar orbits agrees with that based on Hill's approximation  within $1\%$ 
as long as $m_i/M_{\odot} < 10^{-4}$. Our test simulations for late stage planetary accretion (Section~3.5) 
indicate that Hill's approximation also works well even at moderately high orbital eccentricities ($e_j \sim 0.3$).

Assumption 2 is acceptable if  many planetesimals mutually interact 
via gravitational scattering and collisions without orbital isolation. These interactions lead 
to randomization of spatial positions and phases.
We extensively adopt these uniformities even if the target is a
sub-embryo isolated from other embryos, but with some approximated cares (Section~2.6). 

Assumption 3 is likely to be reasonable since 
the timescale of phase change is comparable 
to the stirring timescale in general (e.g, Tanaka and Ida, 1996). 
With Assumptions 2 and 3, we can assume that
the time averaged stirring and collision rates of
an individual target are equivalent to the stirring and collision rates averaged over phase space.
While this equivalence resembles the so-called ergodic hypothesis,  the difference is that we assume   
the equivalence on a short timescale.
Although Assumption 3 may not be strictly correct,  it greatly simplifies the problem and 
it is worthwhile to examine its validity or usefulness.

\begin{figure}
\begin{center}
\includegraphics[width=0.7\textwidth]{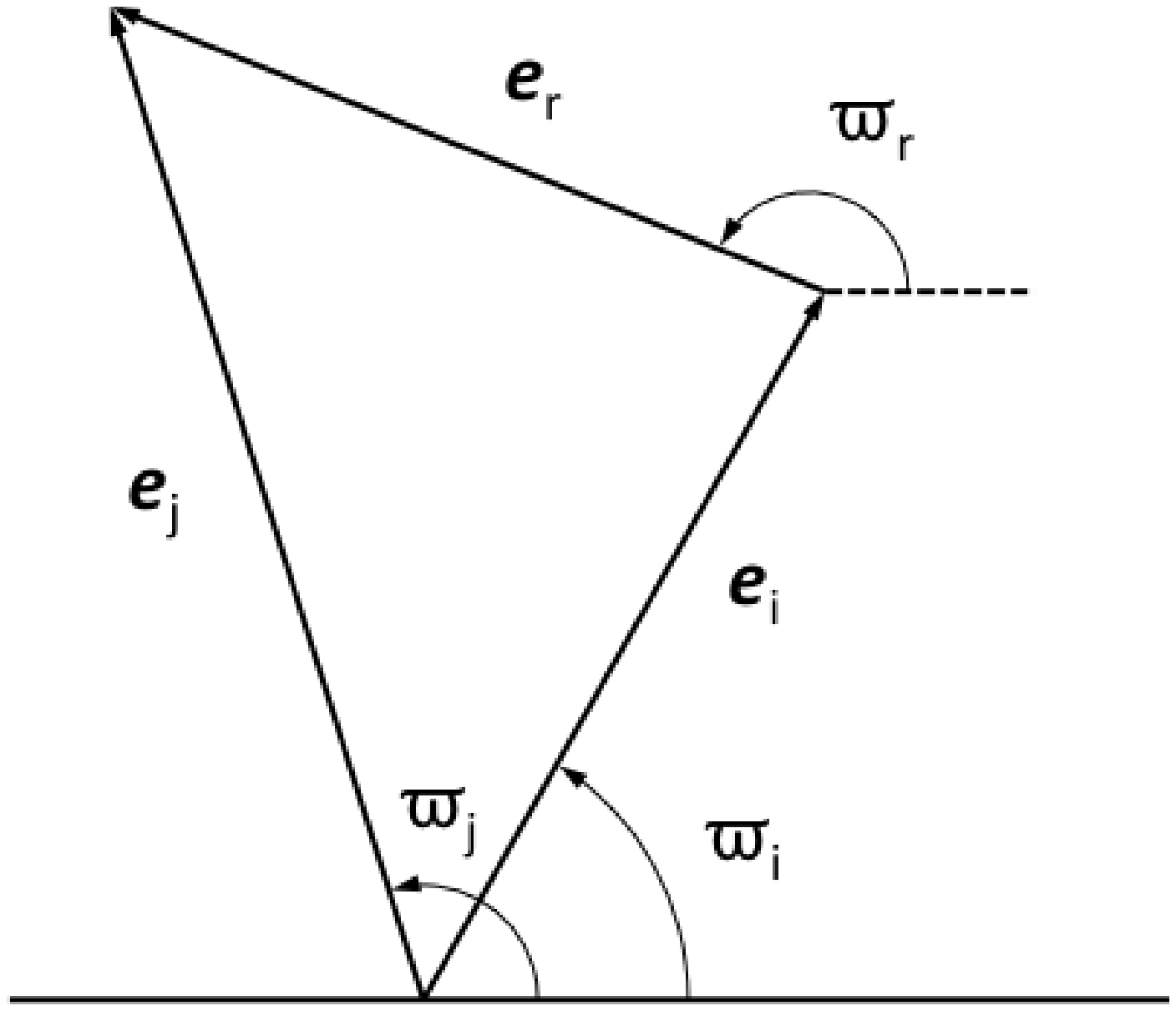}
\end{center}

Fig.~3. Definitions of the eccentricity vectors. The inclination vectors are defined in a similar manner.

\end{figure}

Integrations over distributions of orbital elements of interlopers 
can be transformed to integrations over relative orbital elements of interlopers with respect to 
the target (Nakazawa et al., 1989). 
The eccentricity and inclination vectors of the target $i$ are given as 
\begin{equation}
\begin{aligned}
\mbox{\boldmath $e$}_i  &= (e_i \cos{\varpi_i}, e_i \sin{\varpi_i}),  \\
\mbox{\boldmath $i$}_i  &= (i_i \cos{\Omega_i}, i_i \sin{\Omega_i}).
\end{aligned}
\end{equation}
The same characters but with the subscript $j$ represent the eccentricity and inclination vectors for the interloper $j$. 
The relative eccentricity and inclination vectors are given as  (Fig.~3)
\begin{equation}
\begin{aligned}
\mbox{\boldmath $e$}_{\rm r}   & = (e_{\rm r} \cos{\varpi_{\rm r}}, e_{\rm r} \sin{\varpi_{\rm r}})  & &\hspace{-0.7em } =   \mbox{\boldmath $e$}_j -  \mbox{\boldmath $e$}_i,  \\
\mbox{\boldmath $i$}_{\rm r}    &  = (i_{\rm r} \cos{\Omega_{\rm r}}, i_{\rm r} \sin{\Omega_{\rm r}}) & &\hspace{-0.7em } =    \mbox{\boldmath $i$}_j -  \mbox{\boldmath $i$}_i,  \label{eq:relei}
\end{aligned}
\end{equation}
where $\varpi_{\rm r}$ and $\Omega_{\rm r}$ are the longitudes for the relative vectors.

Nakazawa et al. (1989) adopted Assumptions 1-3 and 
derived the number of planetesimals $\Delta N_j$ colliding with the target $i$ during a time interval $\Delta t$ as [their Eq.~(49)]
\begin{equation}
\begin{multlined}
\frac{\Delta N_j}{\Delta t} (e_i, i_i, \langle e_j^2 \rangle, \langle i_j^2 \rangle) =  \\
 n_j a_{ij}^2 h_{ij}^2 \omega_{\rm K}  \int f(e_{\rm r},i_{\rm r},e_i,i_i,\langle e_j^2 \rangle, \langle i_j^2 \rangle)P_{\rm col}(e_{\rm r},i_{\rm r}) de_{\rm r}di_{\rm r},
\end{multlined}
\end{equation}
where  $\langle e_j^2 \rangle$ and $\langle i_j^2 \rangle$ are the mean squares of  $e_j$ and $i_j$ of interlopers, 
$n_j$ is the surface number density of interlopers, $a_{ij} = (a_i+a_j)/2$ is the mean semi-major axis, 
$\omega_{\rm K}$ is the orbital frequency at $a_{ij}$,
$f (e_{\rm r},i_{\rm r},e_i,i_i,\langle e_j^2 \rangle, \langle i_j^2 \rangle)$ is the distribution function of $e_{\rm r}$ and  $i_{\rm r}$ normalized as
$\int f de_{\rm r} di_{\rm r} = 1$,
and $P_{\rm col}(e_{\rm r},i_{\rm r})$ is the non-dimensional collision rate averaged over $d_{ij}$, $\varpi_{\rm r}$, and $\Omega_{\rm r}$ (Appendix~B.1). 
In the above, the reduced mutual Hill radius, $h_{ij}$, is given as
\begin{equation}
h_{ij} = \left(\frac{m_i + m_j}{3M_{\odot}}\right)^{1/3}. \label{eq:rhill}
\end{equation}

The expression of $f$ for the case of the Rayleigh distributions of $e_j$ and $i_j$ is given by Eq.~(35) of Nakazawa et al. (1989).
In our method, on the other hand, $f$ for a certain interloper is given by the Dirac delta function. Thus,  the expected number of planetesimals 
$\Delta N_j$ in the interloper $j$ colliding with the target $i$ during $\Delta t$ is given as
\begin{equation}
\frac{\Delta N_j}{\Delta t} (e_{\rm r}, i_{\rm r}) =  
 n_j a_{ij}^2 h_{ij}^2 \omega_{\rm K}  P_{\rm col}(e_{\rm r},i_{\rm r}), \label{eq:dndt}
\end{equation}
where $n_j$ is given by Eq.~(\ref{eq:nj}).
The stirring rates can be derived in a similar manner using the non-dimensional stirring rates (Section~2.5). 
In our method, the averaging over distributions of eccentricities and inclinations of interlopers will be made by 
averaging over multiple interlopers.
Although it is not necessary to assume the distribution functions of eccentricities and inclinations of interlopers in calculations of the collision and stirring rates, 
our stirring routine naturally reproduces Rayleigh distributions (Section~3.2). 

If the system is perturbed by a massive external body and the forced eccentricity is much larger than free eccentricities, 
the phases,  $\varpi_j$ and $\Omega_j$, may not be uniformly distributed. Even in such a case, our approach may still be applied
by replacing the eccentricity and inclination vectors by the free eccentricity and inclination vectors whose directions are uniformly distributed, 
provided that the forced eccentricity vectors are roughly the same between the target and interlopers.
Thus, while uniformities of $\varpi_{\rm r}$ and  $\Omega_{\rm r}$ are assumed, we do not necessarily assume
uniformities of the relative longitudes $\varpi_{ij} = \varpi_{j}-\varpi_{i}$ and $\Omega_{ij} = \Omega_{j} - \Omega_i$.
The averaging over $\varpi_{ij}$ and $\Omega_{ij}$ is done by averaging over interlopers [$n_j = n_j(e_{\rm r},i_{\rm r}, \varpi_{ij},\Omega_{ij})$]. 
While this relaxing of the assumption does not alter Eq.~(\ref{eq:dndt}), it is important for dynamical friction (Section~2.5).
The forced eccentricity vectors are not necessarily the same between the target and interlopers, for example, 
if the mass-dependent dissipative force works. In such a situation, our approach is not valid, although it is still worthwhile 
to examine how inaccurate it is.

\subsection{Collision}
\subsubsection{Judgement of collision}
The expected change rate of the mass $m_i$ of a planetesimal in the tracer $i$ 
due to collisions with planetesimals in the interloper $j$ is simply 
given by multiplying the planetesimal mass $m_j$ to Eq.~(\ref{eq:dndt})
(Ida and Nakazawa, 1989; Greenzweig and Lissauer, 1990; Inaba et al., 2001)
\begin{equation}
\left(\frac{dm_i}{dt}\right)_j= n_j m_j a_{ij}^2 h_{ij}^2  \omega_{\rm K}  P_{\rm col}, \label{eq:dm}
\end{equation}

To avoid double counting, we only consider a case for $m_i \ge m_j $ or $i < j$ for $m_i = m_j$.
Using the mass change rate, the probability that the mass of a planetesimal in the target $i$ increases by $\Delta m_i$ 
due to a collision with a planetesimal in the interloper $j$ during the time step $\Delta t$ 
is given as $(dm_i/dt)_j(\Delta t/\Delta m_i)$. If this probability is larger than 
a random number which takes between 0 and 1, we assume that all planetesimals in the target $i$ 
collide with planetesimals in the interloper.  The same procedure is applied to all other interlopers in the region $i$ to check whether 
collisions occur between planetesimals in the target and those in interlopers.

If a collision occurs and mutually colliding bodies merge, 
the changes in  the masses and the numbers of planetesimals in the tracers are given as follows:
\begin{eqnarray}
m_{ib}= m_i + \Delta m_i \hspace{0.5em} {\rm and} \hspace{0.5em} k_{ib} = k_i  \hspace{0.5em} ({\rm for \hspace {0.3em} target}), \\
m_{jb} = m_j \hspace{0.5em} {\rm and} \hspace{0.5em} k_{jb} = k_j - \frac{\Delta m_i}{m_j} k_i  \hspace{0.5em} ({\rm for \hspace {0.3em} interloper}), \label{eq:inte}
\end{eqnarray}
where the characters with the additional subscript $b$ mean those after the merging.
Note that the total mass is conserved in the collisions: $k_{ib}m_{ib} + k_{jb}m_{bj} =  k_im_i + k_jm_j$.

Before determining whether a collision occurs or not, we  need to choose an appropriate $\Delta m_i$.
The mass $\Delta m_i$ is usually the mass of the interloping planetesimal $m_j$, 
but can take different values for convenience as follows: 
\begin{equation}
\Delta m_i  = 
\left\{ \begin{array}{ll} 
{\rm Max} [m_j, {\rm int}(0.01m_{i}/m_j)m_j]
& \mbox{Case (a); $k_{jb} > 0$}, \\ 
k_j m_{j}/k_i
& \mbox{Case (b); $k_{jb} = 0$ },
\end{array}\right.
\label{eq:delm}
\end{equation}
where int() means the decimals are dropped off for the number in the parentheses.
The form in the upper row of Eq.~(\ref{eq:delm}) means that if $m_j$ is smaller than the threshold mass, 
we do not consider individual collisions but 
the accumulated mass change of the target planetesimal. We adopt the threshold mass of $0.01 m_{i}$.
We first estimate the post-collision mass of the interloper $k_{jb}m_j$ using Eq.~(\ref{eq:inte}) and $\Delta m_i$
for Case (a) of Eq.~(\ref{eq:delm}) for any interlopers.
If the resulting mass, $k_{jb}m_j$, is found to be smaller than the lower limit of the tracer mass, $m_{t,{\rm min}}$, 
we use $\Delta m_i$ of Case (b) given 
by the lower row in Eq.~(\ref{eq:delm}) instead of that of Case (a).
If a collision occurs for Case (b), all the planetesimals in the interloper $j$ merge with those in the target $i$ 
and the interloper is deleted.  
Therefore, the mass of any tracer is kept to be lager than $m_{t,{\rm min}}$. 
While we adopt $m_{t,{\rm min}} = 0.1 m_{t0}$,  this value can be arbitrarily chosen: a small lower limit 
may be able to give accurate mass evolution while increase 
of the number of tracers can be suppressed with a large lower limit.

\subsubsection{Mass and orbit changes due to merging} 
The procedure to change masses and orbital elements of planetesimals 
due to collisional merging is schematically summarized in Fig.~4. 
For Case (a) in  Eq.~(\ref{eq:delm}),  
in which $k_{jb}> 0$ and $k_{jb}m_{j} > m_{t,{\rm min}}$, only some of planetesimals in the interloper $j$ collide with planetesimals in the target $i$.
In this case, we split the interloper $j$ into two tracers: the interloper $j$ and the impactor. 
The number of planetesimals in the interloper after the split is given by $k_{jb}$ in Eq.~(\ref{eq:inte}) and 
the position and velocity of the post-split interloper are unchanged during the collision between the target and the impactor.
For Case (b), all the planetesimal in the interloper are involved in the collision and we call the interloper the impactor. 
In both cases,  the planetesimal mass and the number of planetesimals in the impactor are given by 
$\Delta m_i$ and $k_i$, respectively, and the collision process is almost the same except some small differences.

\begin{figure}
\begin{center}
\includegraphics[width=1.\textwidth]{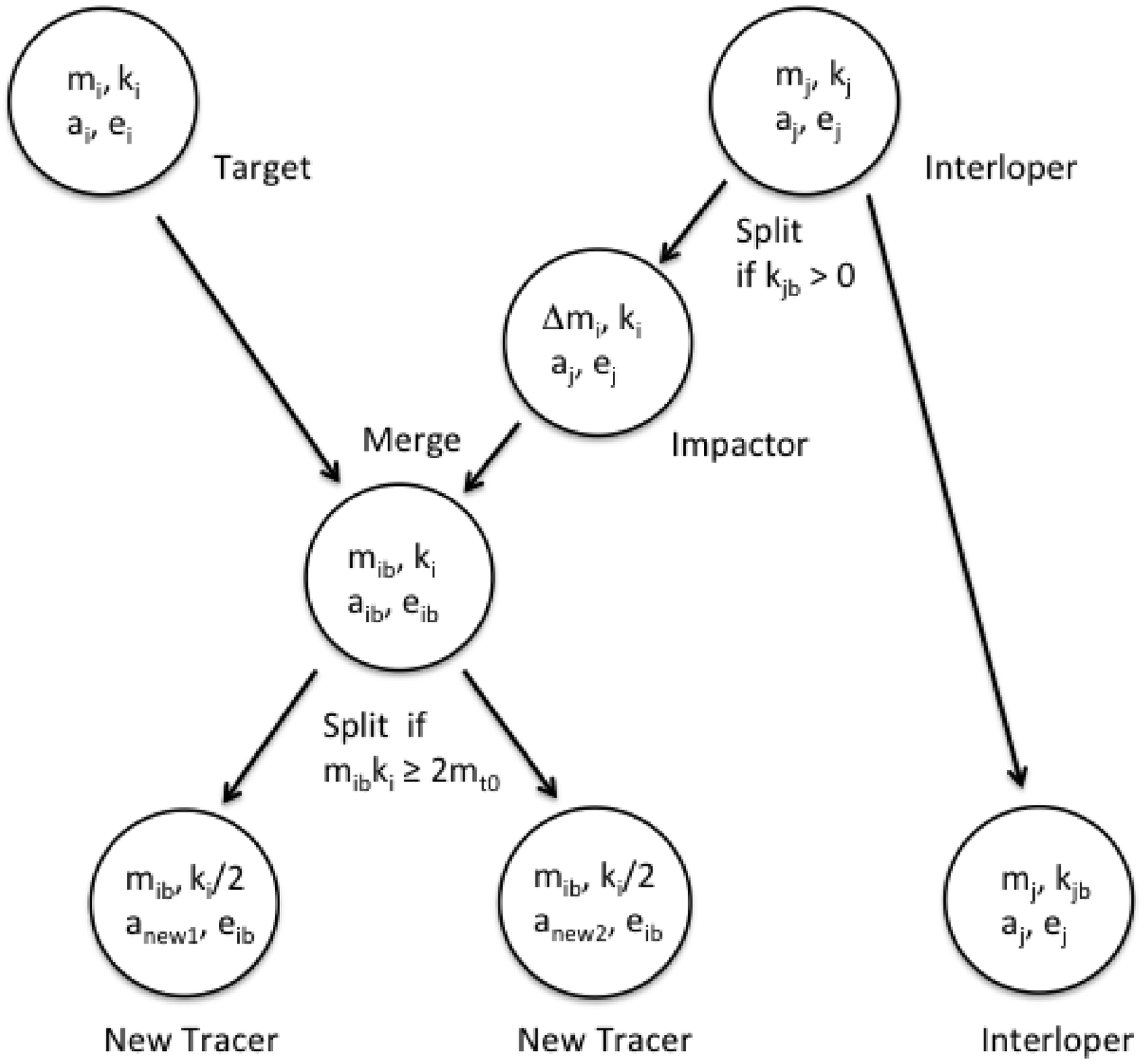}
\end{center}

Fig.~4. Schematic illustration of the procedure to handle collisional merging between planetesimals in
two tracers (the target and the interloper). The figure shows how the masses of planetesimal ($m_i$ and $m_j$), 
the numbers of planetesimals in the tracers ($k_i$ and $k_j$), the semimajor axes ($a_i$ and  $a_j$), and 
the orbital eccentricities ($e_i$ and  $e_j$) change during the procedure.

\end{figure} 

When the target and the impactor are judged to collide, they do not physically contact each other.  Even their stream lines 
do not usually cross each other.  Since we assume uniformity in semimajor axis of an interloper, 
it is possible to make two stream lines cross each other by changing the semi-major axis of the impactor without changing other orbital elements.
The crossing point of the stream lines is the location where the impact takes place.  
This impact point can be found as follows. 
First, $\theta$ and $z/r$ of the target and the impactor are aligned by changing the true anomaly of the target $f_i$ by 
$\Delta f_i$ and that for the impactor $f_j$ by $\Delta f_j$.
The angular changes, $\Delta f_i$ and $\Delta f_j$, are derived by solving the following set of equations:
 \begin{eqnarray}
\varpi_i + f_i + \Delta f_i = \varpi_j + f_j + \Delta f_j = \theta_{\rm col}, \label{eq:xx0} \\
\sin{i_i}\sin{(\theta_{\rm col}-\Omega_i)} = \sin{i_j}\sin{(\theta_{\rm col}-\Omega_j)}, \label{eq:xx}
\end{eqnarray}
where $\theta_{\rm col}$ is the longitude of the impacting position 
measured from the reference direction $\theta = 0$.
Eqs.~(\ref{eq:xx0}) and (\ref{eq:xx}) give two solutions for $\Delta f_i$ mutually shifted by $\pi$, and we choose one with a smaller absolute value. 
Then, the true anomalies are changed by moving the colliding pair along their Keplerian orbits.
Second,  the semimajor axis of the impactor is scaled by $r_i/r_j$.
This is done by scaling the impactor position by $r_i/r_j$ and the velocity by $\sqrt{r_j/r_i}$.
The positions of the target and the impactor become almost identical through these procedures without changing their eccentricity and 
inclination vectors. 

Then, the merging between the target and the impactor takes place.
The position and velocity of the target after the merging with the impactor are given by 
the mass weighted mean values (i.e., the mass center for the position).
After that, the semi-major axis of the target is adjusted to $a_{ib}$ by scaling the position and the velocity 
so that the $z$-component of the total orbital angular momentum of the target and the interloper 
is conserved during the merging:
\begin{equation}
\begin{multlined}
k_{i}m_{ib}\sqrt{a_{ib}(1-e_{ib}^2)}\cos{i_{ib}} = \\
k_{i}m_{i}\sqrt{a_{i}(1-e_{i}^2)}\cos{i_{i}} + (k_{j}-k_{jb})m_{j}\sqrt{a_{j}(1-e_{j}^2)}\cos{i_{j}}, 
\end{multlined}
\end{equation}
where $e_{ib}$ and $i_{ib}$ are the orbital eccentricity and inclination of the target after the merging.

If the target tracer mass is equal to or exceeds the upper limit,  $m_{t,{\rm max}} (= 2 m_{t0})$, we split it into two tracers.
If $k_{ib} = 2$ and $m_{ib} > 2 m_{t,{\rm min}}$, we also split the tracer to avoid simultaneous emerging of two sub-embryos.
The numbers of planetesimals in the split  tracers are given by $k_{ib}/2$
if $k_{ib}$ is even, and $k_{ib}/2 \pm 0.5$ if $k_{ib}$ is odd. 
We give weak velocity changes, symmetrically in momenta, to the two new tracers 
so that their relative eccentricity and inclination are about $h_{ij}$. 
This will avoid an encounter velocity artificially too low between them in the next encounter.
The semi-major axes of the two tracers are adjusted so that 
the orbital separation between two tracers is randomly chosen between 0 to $\delta r$
but the $z$-component of the total orbital angular momentum is conserved. For Case (b), 
we set the semi-major axis of one of the new tracers be the one for the deleted interloper 
instead of choosing the orbital separation randomly.
Finally, the true anomalies of the two tracers (or one if no splitting is made) are changed 
by $-\Delta f_i$ and $-\Delta f_j$. 

Through the processes described above, tracer masses remain between $m_{t,{\rm min}}$ and $m_{t,{\rm max}}$.
The number of tracers remains unchanged or increases for Case (a) of Eq.~(\ref{eq:delm}) whereas it remains unchanged or decreases for Case (b).
As a result,  the total number of tracers remains similar to the initial value even after many collisions between planetesimals.
Our several experiments show that increase of the tracer number from the initial value is suppressed less than 30$\%$ in most cases. 
Our current code also has an option of hit-and-run collisions and the procedure
is described in detail in Appendix~E. In this case, neither the number of planetesimals nor that of tracers changes.

\subsection{Stirring and radial diffusion}
\subsubsection{Rates of stirring and radial diffusion}

The orbital elements of planetesimals  evolve due to gravitational interactions between them.
The interactions are separated into viscous stirring (VS) and dynamical friction (DF).
Viscous stirring increases eccentricities and inclinations in expense of the tidal potential (i.e., radial diffusion) 
and dynamical friction leads to energy equipartition between particles (Ida, 1990, Stewart and Ida, 2000; Ohtsuki et al., 2002).
Dynamical friction can be separated into stirring and damping parts (DFS and DFD) and sometimes 
the latter part only is called dynamical friction (Binney and Tremaine, 1987; Rafikov, 2003; LDT12).
We handle the stirring and damping parts of dynamical friction separately as shown below.

The evolution of eccentricity $e_i$ and inclination $i_i$ of the target $i$ 
due to viscous stirring and dynamical friction caused by interactions with the interloper $j$ is given as (Ida, 1990; Tanaka and Ida, 1996)
\begin{eqnarray}
\left(\frac{de_i^2}{dt}\right)_{{\rm VS},j} &=& C_1  P_{\rm VS},  \label{eq:devs} \\
\left(\frac{de_i^2}{dt}\right)_{{\rm DFS},j} &=& \frac{C_1}{h_{ij}^2} {e}_{\rm r}^2 P_{\rm DF}, \label{eq:dedfs} \\
\left(\frac{de_i^2}{dt}\right)_{{\rm DFD},j} &=& - \frac{C_1}{\nu_jh_{ij}^2} \left(e_i^2 - e_i e_j\cos{\varpi_{ij}} \right)  P_{\rm DF}, \label{eq:dedfd} \\
\left(\frac{di_i^2}{dt}\right)_{{\rm VS},j}  &=& C_1 Q_{\rm VS} ,  \label{eq:divs} \\ 
\left(\frac{di_i^2}{dt}\right)_{{\rm DFS},j} &=& \frac{C_1}{h_{ij}^2} i_{\rm r}^2 P_{\rm DF}, \label{eq:didfs} \\
\left(\frac{di_i^2}{dt}\right)_{{\rm DFD},j} &=& - \frac{C_1}{\nu_jh_{ij}^2} \left( i_i^2 - i_i i_j\cos{\Omega_{ij}}\right) P_{\rm DF}, \label{eq:didfd} 
\end{eqnarray}
where $C_1 = n_j \nu_j^2 h_{ij}^4 a_{ij}^2 \omega_{\rm K}$, $\nu_j = m_j/(m_i+m_j)$, $P_{\rm VS}$, $Q_{\rm VS}$, and $P_{\rm DF}$ are the non-dimensional rates 
of viscous stirring and dynamical friction. The expressions of these rates are given in Appendix B.2. 
Since we do not assume uniformities of $\varpi_{ij} = \varpi_{j}-\varpi_{i}$ and $\Omega_{ij} = \Omega_{j} - \Omega_i$ 
to handle eccentric and/or inclined ringlets, as discussed in Section~2.3, 
the terms with these relative angles remain in the expressions for dynamical friction. 
The way to split dynamical friction into the two terms is explained in detail in Appendix~C.

The  evolutions of $e_i$ due to viscous stirring and dynamical friction are then given as 
\begin{eqnarray}
\left(\frac{de_i^2}{dt}\right)_{\rm VS} &=& \sum_j \left(\frac{de_i^2}{dt}\right)_{{\rm VS},j},   \\
\left(\frac{de_i^2}{dt}\right)_{\rm DFS} &=& \sum_j \left(\frac{de_i^2}{dt}\right)_{{\rm DFS},j},  
\hspace{1em} \left(\frac{de_i^2}{dt}\right)_{\rm DFD} = \sum_j \left(\frac{de_i^2}{dt}\right)_{{\rm DFD},j}. 
\end{eqnarray}
Similar expressions can be obtained for $i_i$. 

Mutual encounters between planetesimals causes random walk in their semimajor axes.
The radial diffusion coefficient $D_i$ for the target $i$ is approximately given by (Appendix~D)
\begin{equation}
D_i = \sum_j \left[C_1a_{ij}^2(P_{\rm VS} + Q_{\rm VS})\right]. \label{eq:di0}
\end{equation}  
Ignoring the effect of curvature, the probability function of the change in $a_i$ after the time step $\Delta t$ is given by
\begin{equation}
P(\Delta a_i) = \frac{1}{\sqrt{4\pi D_i \Delta t}} \exp{\left(-\frac{(\Delta a_i)^2}{4 D_i \Delta t}\right)}, \label{eq:pda}
\end{equation}  
where the function is normalized as $\int^{\infty}_{-\infty} P(\Delta a_i) d\Delta a_i = 1$.
We choose $\Delta a_i$ at random with a weight given by $P(\Delta a_i)$. 
If $D_i < 0$, we set $\Delta a_i = 0$. This is rare but can occur if $e_{\rm r} \gg i_{\rm r}$.

\subsubsection{Conversion of stirring rates to accelerations}
The stirring and damping rates of orbital elements need to be converted into the acceleration of the tracer.
Since the time step for the statistical routine $\Delta t$ is usually much smaller than the stirring time, $e_{i}^2(de_i^2/dt)^{-1}$,
the absolute values of the changes in the orbital elements during $\Delta t$  are much smaller than the magnitudes of the orbital elements themselves. 
Consider a situation that encounters between planetesimals occur randomly in directions of velocities.
A certain acceleration occurs to a planetesimal in an encounter, and in another encounter
an acceleration with the same magnitude to the first encounter but in the opposite direction can occur.
In first order, the changes of orbital elements cancel out in these two encounters. 
However, in second order,  the orbital eccentricity and inclination increase a little on average, because
the orbital elements are slightly changed after the first encounter. 
This is the way how viscous stirring works, and 
to handle it, we introduce second order Gauss planetary equations.
The stirring part of dynamical friction is also handled by these equations.
On the other hand, the damping part of dynamical friction is
handled by the standard Gauss planetary equations of first order (e.g., Murray and Dermott, 1999).
If we use the standard Gauss planetary equations for the stirring part of dynamical friction,
the standard deviations of orbital eccentricity and inclination become unphysically small for equal-mass planetesimals or 
mass-dependent segregation of these orbital elements occurs for a case of a mass distribution. 
We assume that $e_i \ll 1$ and $i_i \ll 1$ in the following formulation. Note that these assumptions are already used in Hill's approximation.

In this subsection only, we drop the subscript $i$ for the tracer $i$ to avoid long, confusing subscripts. 
The expected changes of orbital elements during the time step $\Delta t$ are 
$\Delta a$, $\Delta e^2_{\rm VS} + \Delta e^2_{\rm DFS}$, $\Delta e_{\rm DFD}$, $\Delta i^2_{\rm VS} + \Delta i^2_{\rm DFS}$, and  $\Delta i_{\rm DFD}$,
where $\Delta e^2_{\rm VS} = (de^2/dt)_{{\rm VS}} \Delta t $,  $\Delta e^2_{\rm DFS} = (de^2/dt)_{{\rm DFS}} \Delta t $,  
$\Delta e_{\rm DFD} = (e^2 + (de^2/dt)_{{\rm DFD}} \Delta t)^{1/2} - e$,
and similar expressions are obtained for the inclination.

The change of the velocity vector is split into three components as 
\begin{equation}
\Delta \mbox{\boldmath $v$} = 
\Delta v_{\rm R} \mbox{\boldmath $x$}_{\rm R} + \Delta v_{\rm T} \mbox{\boldmath $x$}_{\rm T} + \Delta v_{\rm N} \mbox{\boldmath $x$}_{\rm N},  
 \end{equation}    
where $\Delta v_{\rm R}$ and $\Delta v_{\rm T}$ are the components of the velocity change in the radial and tangential directions in the orbital plane,  
$\Delta v_{\rm N}$ is the velocity change normal to the orbital plane,
$\mbox{\boldmath $x$}_{\rm R}$,  $\mbox{\boldmath $x$}_{\rm T}$, and $\mbox{\boldmath $x$}_{\rm N}$ are unit vectors.
The velocity vector of the tracer itself is given as 
$\mbox{\boldmath $v$} = v_{\rm R} \mbox{\boldmath $x$}_{\rm R} +  v_{\rm T} \mbox{\boldmath $x$}_{\rm T}$, where 
$v_{\rm R}$ and $v_{\rm T}$ are the radial and tangential components.
All the components of the velocity change are further split into the stirring and damping terms with subscripts S and D, respectively:
\begin{equation}
\begin{aligned}
\Delta v_{\rm R} = \Delta v_{\rm R,S} + \Delta v_{\rm R,D},\\
\Delta v_{\rm T} = \Delta v_{\rm T,S} + \Delta v_{\rm T,D},\\
\Delta v_{\rm N} = \Delta v_{\rm N,S} + \Delta v_{\rm N,D}.
\end{aligned}
\end{equation}
In our stirring routine, all the components of the stirring part are chosen using random numbers, but their averaged 
functional forms follow Gaussian distributions with the zero mean values:
$\langle \Delta v_{\rm R,S} \rangle =  \langle \Delta v_{\rm T,S} \rangle =  \langle \Delta v_{\rm N,S} \rangle = 0$,
where the angle brackets in this subsection mean averaging over multiple velocity changes.
The acceleration due to an individual encounter may not necessarily follow Gaussian. On the other hand,
since we handle the time averaged accelerations of tracers due to multiple encounters that occur in a random fashion,
their distributions are limited to Gaussian.

First we consider the velocity changes in the orbital plane.
The energy conservation gives the following equation, if we take up to the second order terms
\begin{eqnarray}
&& \frac{\mu}{a} \left(\frac{\Delta a}{a} -  \left(\frac{\Delta a}{a}\right)^2  \right) =    \nonumber \\
&& 2 v_{\rm R} \Delta v_{\rm R}  + 2 v_{\rm T} \Delta v_{\rm T} + (\Delta v_{\rm R})^2  + (\Delta v_{\rm T})^2 + (\Delta v_{\rm N})^2,  \label{eq:enec}
\end{eqnarray}
where $\mu = \omega_{\rm K}^2 a^3$.
If we ignore the second order terms and the first term in the r.h.s. of  Eq.~(\ref{eq:enec}) as $v_{\rm R}  \ll v_{\rm T} $, 
this gives  $\Delta v_{\rm T,S}$ as 
\begin{equation}
 \Delta v_{\rm T,S} = \frac{\mu \Delta a}{2a^2v_{\rm T}}.
 \end{equation}
 Note again that $\Delta a$ is given by Eq.~(\ref{eq:pda}) and thus $\Delta v_{\rm T,S}$ is Gaussian.
 The angular momentum conservation gives 
\begin{eqnarray}
&& -a \mu (2e\Delta e +(\Delta e)^2) + \Delta a \mu (1-e^2-2e\Delta e) = \nonumber \\
&& R^2\left(2v_{\rm T} \Delta v_{\rm T} + (\Delta v_{\rm T})^2 + (\Delta v_{\rm N})^2\right),  \label{eq:angc}
\end{eqnarray} 
where $\Delta e$ is the change of $e$ and $R$ is the distance from the central star.  
Inserting Eq.~(\ref{eq:angc}) into  Eq.~(\ref{eq:enec}), we have 
\begin{eqnarray}
&& \frac{\mu \Delta a}{a^2}\left(1-\frac{\Delta a}{a} - \frac{a^2}{R^2}(1-e^2-2e\Delta e)\right) +\frac{\mu a (2e \Delta e + (\Delta e)^2)}{R^2} = \nonumber \\
&& 2 v_{\rm R} \Delta v_{\rm R} + (\Delta v_{\rm R})^2. 
\label{eq:angc2}
\end{eqnarray}   

Averaging Eq.~(\ref{eq:angc2}) over time, the first order terms vanish and we have only the second order terms.
Using $\Delta e^2 = 2e \Delta e + (\Delta e)^2$, we have 
\begin{eqnarray}
\frac{\mu}{a}  \langle \Delta e^2 \rangle 
&=& \frac{\mu \langle{(\Delta a)^2}\rangle}{a^3} + \langle {(\Delta v_{\rm R})^2 \rangle} \nonumber \\
&=&  \frac{2\mu}{a} (\Delta e^2_{\rm VS} + \Delta i^2_{\rm VS}) + \langle {(\Delta v_{\rm R})^2 \rangle}, \label{eq:de2}
\end{eqnarray}   
where we used $a^2 \simeq R^2$ and $\langle{(\Delta a)^2}\rangle/2 \simeq a^2(\Delta e_{\rm VS}^2 + \Delta i_{\rm VS}^2)$ [$\simeq D_i \Delta t$;  Eq.~(\ref{eq:di0}) or Eq.~(\ref{eq:db2})], 
and ignored the term proportional to $\Delta a  \Delta e $ as it is small enough as compared with other terms for $e \ll 1$. 
Inserting $\langle \Delta e^2 \rangle = \Delta e^2_{\rm VS} + \Delta e^2_{\rm DFS}$ into Eq.~(\ref{eq:de2}), we have 
\begin{equation}
 \langle (\Delta v_{\rm R,S})^2 \rangle = \frac{\mu}{a}\left(\Delta e^2_{\rm DFS} - \Delta e^2_{\rm VS} - 2\Delta i^2_{\rm VS}\right). \label{eq:vrs}
\end{equation}
We give $\Delta v_{\rm R,S}$ assuming it is Gaussian.
If  the r.h.s side of Eq.~(\ref{eq:vrs}) is negative, we set  $\Delta v_{\rm R,S} = 0$, 
add the r.h.s divided by $\mu/a$ to  $(de^2/dt)_{{\rm DFD}} \Delta t$, 
and re-calculate $\Delta e_{\rm DFD}$, which will be handled as shown below. 
This occurs in the shear dominant regime [$(e_{\rm r}^2 + i_{\rm r}^2)^{1/2} < 2h_{ij}$].

To give the velocity changes  $\Delta v_{\rm R, D}$ and $\Delta v_{\rm T, D}$ for the damping parts of dynamical friction,
we use the standard first order Gauss planetary equation for $e$ (Murray and Dermott, 1999) that can also be derived from  Eq.~(\ref{eq:angc2}):
\begin{equation}
\Delta e = \sqrt{\frac{a(1-e^2)}{\mu}} \left[\Delta v_{\rm R} \sin{f} + \Delta v_{\rm T} (\cos{f} + \cos{E})\right], \label{eq:dedf}
\end{equation} 
where $f$ and $E$ are the true and eccentric anomalies.
We employ the following forms 
\begin{equation}
\Delta v_{\rm R,D} = \Delta v_{\rm R0} \sin{f}, \hspace{0.3em}   \Delta v_{\rm T,D} = \frac{1}{2}  \Delta v_{\rm R0} \cos{f}.  \label{eq:dedfc}
\end{equation} 
Setting $ \Delta e =  \Delta e_{\rm DFD} $ in Eq.~(\ref{eq:dedf}),  we have 
\begin{equation}
\Delta v_{\rm R0} = \sqrt{\frac{\mu}{a(1-e^2)}} \Delta e_{\rm DFD},  \label{eq:dedf0}
\end{equation} 
where we assumed $\cos{f} \simeq \cos{E}$.

Next, we consider the velocity change perpendicular to the orbital plane.   
The specific angular momentum vector $\mbox{\boldmath $h$} = (h_x,h_y,h_z)$ after  
the velocity change $\Delta v_{\rm N}$  in the inertial Cartesian coordinate system 
are given as    
\begin{equation}
\begin{pmatrix}
h_x \\
h_y \\
h_z
\end{pmatrix}
=  \mbox{\boldmath $P$}_3 \mbox{\boldmath $P$}_2 \mbox{\boldmath $P$}_1
\begin{pmatrix}
R\sin{f} \Delta v_{\rm N} \\
-R\cos{f}  \Delta v_{\rm N} \\
(h^2 - (R \Delta v_{\rm N})^2)^{1/2}
\end{pmatrix},
\label{eq:ang4} 
\end{equation}
where $ h = |\mbox{\boldmath $h$}|$, 
$\mbox{\boldmath $P$}_3$, $\mbox{\boldmath $P$}_2$, and $\mbox{\boldmath $P$}_1$ are the rotation matrices 
given by Eqs.~(2.119) and (2.120) of Murray and Dermott (1999).
Using $h_z = h \cos{(i + \Delta i)}$, the $z$-component of Eq.~(\ref{eq:ang4}) is given as 
\begin{equation}
\Delta i \sin{i}  + \frac{(\Delta i)^2}{2}\cos{i} = \frac{R}{h} \Delta v_{\rm N} \cos{(w + f)} \sin{i} + \frac{1}{2}\frac{R^2}{h^2} (\Delta v_{\rm N})^2 \cos{i},\label{eq:ang5} 
\end{equation}
where $\Delta i$ is the change of $i$ and $w = \varpi - \Omega$ is the argument of pericenter.
This equation leads to the standard Gauss planetary equation for $i$ (Murray and Dermott, 1999),
if we ignore the second order terms.
Averaging Eq.~(\ref{eq:ang5}) over time, the first term in the r.h.s. vanishes. 
Setting the mean change of $i^2$ as $\langle \Delta i^2 \rangle = \Delta i^2_{\rm VS} + \Delta i^2_{\rm DFS}$ and using $i \ll 1$, we have 
\begin{equation}
\langle (\Delta v_{\rm N,S})^2 \rangle = \frac{h^2}{a^2} (\Delta i^2_{\rm VS} +  \Delta i^2_{\rm DFS}).
\end{equation}
We give $\Delta v_{\rm N,S}$, assuming it is Gaussian.
For the damping part of dynamical friction, we adopt the form 
\begin{equation}
\Delta v_{\rm N,D} = \Delta v_{\rm N0} \cos{(w + f)}. 
\end{equation}
Ignoring the second order terms in Eq.~(\ref{eq:ang5}), and setting $ \Delta i  =  \Delta i_{\rm DFD}  $, we have 
\begin{equation}
\Delta v_{\rm N0} = \frac{2h}{a}\Delta i_{\rm DFD},
\end{equation}
where we took averaging over the orbital period and used $\int \cos^2{(w + f)} df/(2\pi) = 1/2$.

\subsection{Treatments of sub-embryos}
If the planetesimal mass $m_i$ is equal to or exceeds $m_{t0}$, the tracer is promoted to a sub-embryo.
The acceleration of a sub-embryo due to gravitational interactions with a single planetesimal in 
a tracer is taken into account in the $N$-body routine to hold the correct mutual Hill radius.
On the other hand, the acceleration of the sub-embryo due to interactions with other $k_j-1$ planetesimals in the tracer is 
handled by the statistical routine. 
Collisions of planetesimals with sub-embryos are also handled in the statistical routine.
A large embryo tends to form a gap around in its feeding zone (Tanaka and Ida, 1997), 
although other nearby embryos, if they exit, tend to fill the gap. 
The surface density in the feeding zone usually decreases through this process.
We take into account a possible difference of surface densities inside and outside of the feeding zone whereas
spatial variation of the surface density inside of the feeding zone is not considered. 
This assumption allows us to still use the stirring and collision rates averaged over phases and semi-major axis for interactions between 
a sub-embryo and tracers.   

\subsubsection{Collisions and gravitational stirring}
A collision between the sub-embryo $i$ and the tracer $j$ occurs only if $E_{\rm J} > 0$, where $E_{\rm J}$ 
is the Jacobi integral of Hill's equations given by 
\begin{eqnarray}
E_{\rm J} &=& \left(\frac{1}{2}(e_{\rm r}^2 + i_{\rm r}^2)a_{i}^2-\frac{3}{8}(d_{ij})^2 + \frac{9}{2}a_{i}^2h_{ij}^2\right)\omega_{\rm K}^2 \nonumber \\
&=& \frac{3}{8}\left( (d_{E0})^2-(d_{ij})^2\right)\omega_{\rm K}^2, \label{eq:jaco}
\end{eqnarray}  
where $d_{ij} = a_j-a_i$ and $d_{E0}$ is the half width of the feeding zone given by
\begin{equation}
d_{E0}= \left(\frac{4}{3}\frac{(e_{\rm r}^2 + i_{\rm r}^2)}{h_{ij}^2} + 12 \right)^{1/2}a_{i}h_{ij}. \label{eq:de0}
\end{equation}  
Even if $E_{\rm J} > 0$, collisions do not occur with trojans or if $|d_{ij}|$ is too small  (see Fig.~2 of Ohtsuki and Ida, 1998).
Ida and Nakazawa (1989) showed that collisions occur only if  
\begin{equation}
|d_{ij}| > d_{\rm min} = [8/(2.5+2.0e_{\rm r}/h_{ij})]^{1/2}a_ih_{ij},
\end{equation} 
where $d_{\rm min}$ is the minimum orbital separation for occurrence of collision.
We do not consider interactions with trojans in the statistical routine.
Planetesimals in the feeding zone ($E_{\rm J} > 0$ and $|d_{ij}| > d_{\rm min}$) 
can potentially enter the Hill sphere of the embryo 
and orbits of planetesimals are strongly deflected by close encounters with the embryo.

For tracer-tracer interactions, the number density is 
calculated by Eqs.~ (\ref{eq:ss}) and (\ref{eq:nj}). 
If a sub-embryo is the target, however, we search for planetesimals in the feeding zone.  
Their surface number density is given by  
\begin{equation}
n_{j} = \frac{k_{j}}{4r_i (d_{E0}-d_{\rm min}) \delta \theta}\frac{T_j}{T_{j,{\rm in}}}. \label{eq:nj2}
\end{equation}
We miss some of tracers in the feeding zone if they are radially outside of the region $i$ of neighboring search. 
The factor $T_j/T_{j,{\rm in}}$ in Eq.~(\ref{eq:nj2}) is introduced to correct this effect, 
where $T_j$ is the orbital period of the tracer $j$, and $T_{j,{\rm in}}$ is the period 
in which the tracer $j$ is radially in the region $i$ during $T_j$ provided that the tracer $j$ is 
azimuthally in the range of the region $i$.
The expression of $T_{j,{\rm in}}$ is given in Appendix~F.
In the stirring routine, we use $k_j-1$ in Eq.~(\ref{eq:nj2}) instead of $k_j$ as the back reaction from a single planetesimal is taken into account 
in the $N$-body routine. 
Using  Eq.~(\ref{eq:nj2}), collisional growth and velocity changes of sub-embryos are handled in the statistical routine 
in the same manner as tracer-tracer interactions. 
If there is no spatial density variation between 
the inside and outside of the feeding zone, 
the same collision and stirring rates are derived whether  
the target is a tracer or a sub-embryo.

Even if the mutual distance between the centers of a tracer and a sub-embryo is less than the sum of the radii of the constituent planetesimal and the sub-embryo,
orbital integrations of both bodies are continued in the $N$-body routine using softened forces that avoid singularity.  
We employ the K1 kernel of Dehnen (2001) for the softening.

Gravitational interactions occur even if $E_{\rm J} < 0$
due to distant encounters (Weidenschilling 1989, Hasegawa and Nakazawa, 1990).
We evaluate the change rates of the orbital elements of the sub-embryo due to individual 
distant encounters without assuming uniformity in $d_{ij}$.
Let the changes of the squared relative orbital eccentricity and inclination 
due to a single distant encounter
be $\delta e_{\rm r}^2$ and $ \delta i_{\rm r}^2$.
Hasegawa and Nakazawa (1990) derived the changes 
$\langle \delta e_{\rm r}^2 \rangle$ and $\langle \delta i_{\rm r}^2 \rangle$ 
averaged over $\varpi_{\rm r}$ and $\Omega_{\rm r}$ [their Eqs.~(38) and (39)].
The averaged changes of $e_i^2$ and $i_i^2$ of the sub-embryo $i$ during a single encounter with a planetesimal in the interloper $j$
are given by $ \nu_j^2 \langle \delta e_{\rm r}^2 \rangle$ and $ \nu_j^2 \langle \delta i_{\rm r}^2 \rangle$ (Ida, 1990).
Using the Keplerian shear velocity given by (Bromley and Kenyon, 2011) 
\begin{equation}
v_{\rm sh} = \frac{3}{2}\omega_{\rm K}|d_{ij}| \left(1-\frac{5 d_{ij}}{4 a_{i}}\right), \label{eq:vsh}
\end{equation} 
the frequency of encounters is given as $v_{\rm sh}n_{j} \delta r = v_{\rm sh}(k_j-1)/(4a_{i}\delta \theta)$, where we used Eq.~(\ref{eq:nj}) and 
$\delta r$ in this case is the radial extent of planetesimals in the tracer $j$ 
and is assumed to be infinitesimally small. 
The change rates of $e_i^2$ and $i_i^2$ due to distant encounters with the interloper $j$ are then given as 
\begin{eqnarray}
\left(\frac{de_i^2}{dt}\right)_{{\rm VS}, j} &=&  \nu_j^2 \langle \delta e_{\rm r}^2 \rangle v_{\rm sh} n_{j} \delta r \nonumber \\
&=& \frac{243}{8} R_1^2 \nu_j^2 h_{ij}^6  \left(\frac{a_{i}}{d_{ij}}\right)^3 \omega_{\rm K} \frac{k_j-1}{\delta \theta} \label{eq:dedis}, \\
\left(\frac{di_i^2}{dt}\right)_{{\rm VS}, j} &=&  \frac{243}{8}\left(\frac{4}{81} + \frac{R_3^2}{4}\right) i^2 \nu_j^2 h_{ij}^6  \left(\frac{a_{i}}{d_{ij}}\right)^5 \omega_{\rm K} \frac{k_j-1}{\delta \theta},  \label{eq:didis}  
\end{eqnarray}
where $R_1 = 0.747$ and $R_3 = 0.210$. 
The rates given by Eqs.~(\ref{eq:dedis}) and (\ref{eq:didis}) are added to Eqs.~(\ref{eq:devs}) and (\ref{eq:divs}) if $E_{\rm J} < 0$.
Dynamical friction due to distant encounters is negligible (Ida, 1990). 

\subsubsection{Planetesimal-driven migration}
An embryo gravitationally scatters surrounding planetesimals.
Back reaction on the embryo causes its radial migration. 
Individual gravitational encounters usually cause a random walk of the embryo.
However,  if the accumulated torque is not cancelled out,  the embryo migrates over a long distance.
This is called planetesimal-driven migration (Ida et al., 2000; Kirsh et al., 2009).
Planetesimal-driven migration of full-embryos is automatically taken into account in the $N$-body routine.
We developed a unique method to handle planetesimal-driven migration of sub-embryos explained in the following.

In our code, the angular momentum change $\Delta L_{N, i}$ of the sub-embryo $i$ due to gravitational interactions with tracers during $\Delta t$
is stored in the $N$-body routine.
If $\Delta L_{N, i}$ is released instantaneously by giving the tangential acceleration, 
the sub-embryo may migrate unrealistically too fast 
due to strong kicks from tracers that have masses similar to the sub-embryo. 
The direction of migration may change rapidly in such a situation.
This problem can be avoided by limiting the angular momentum released 
during $\Delta t$ to the theoretical prediction,  $\Delta L_{S, i}$, derived from the statistical routine. 
If $|\Delta L_{N, i}| \le |\Delta L_{S, i}|$, $\Delta L_{N, i}$ is fully released.
If $|\Delta L_{N, i}| > |\Delta L_{S, i}|$, on the other hand, the remaining portion $\Delta L_{N, i} - \Delta L_{S,i}$ 
is added to $\Delta L_{N, i}$ stored during the next time step interval. 
What is done by the scheme is basically time averaging of the torque on the sub-embryo over a timescale much longer than $\Delta t$
so that strong spikes of the torque due to close encounters are smoothed out.
More explanations are given in Section~3.3, together with test results.


The predicted angular momentum change,  $\Delta L_{S, i}$, is determined from the surface density and 
velocities of planetesimals in neighboring tracers.
Bromley and Kenyon (2011, 2013) and Ormel et al. (2012) showed 
three different modes of planetesimal-driven migration, 
referred to as Type I -- III migration analogous to gas-driven migration. 
Type I and III migration occurs when an embryo interacts with planetesimals in close encounters.
Type I  migration is considered to occur when the velocity dispersion of planetesimals is large, and 
its migration rate is very slow as is determined by the difference between torques from outer and inner disks.
On the other hand, Type III migration is fast self-sustained migration that occurs when the velocity dispersion of planetesimals is relatively small, 
and the embryo migrates due to the torque from the one side (inner or outer side) of the disk, leaving 
a relatively enhanced disk in the other side (Ida et al., 2000; Kirsh et al., 2009; Levison et al., 2010; Capobianco et al., 2011).
We ignore Type I migration in the evaluation below, 
as the migration rate is well represented by the formula for Type III migration
even at large velocity dispersions (Kirsh et al., 2009; Bromley and Kenyon, 2011).
Type II migration occurs when an embryo opens up a gap in the planetesimal disk so 
the torque is primarily from planetesimals in distant encounters.
This torque is particularly important if the asymmetries
of the surface density and the velocity dispersion
between the inner and outer disks are large.
Such a situation occurs if an embryo in a gaseous disk shepherds 
outer small planetesimals that rapidly spiral toward the central star without the embryo.

The radial migration rate of  Type III planetesimal-driven migration due to an interaction with the tracer $j$ is (Ida et al., 2000)
\begin{eqnarray}
\left(\frac{da_i}{dt}\right)_{III,j} &=& \pm n_j \nu_j h_{ij}^3 a_{i}^3 \omega_{\rm K} R_{\rm MG}, \label{eq:dat3j}
\end{eqnarray}  
where the sign is positive if $a_j > a_i$ and vice versa, and 
$R_{\rm MG}$ is the non-dimensional migration rate given in Appendix~B.3. 
If $\Delta L_{N,i}$ stored in the $N$-body routine is positive/negative, we count only  tracers
with semimajor axes larger/smaller than $a_i$ because planetesimals in close encounters cause attractive forces. 
In the evaluation of $n_j$ in Eq.~(\ref{eq:dat3j}), we use $0.5S_i(1 \pm 0.5\delta r/r_i)$ instead of $S_i$ used in Eq.~(\ref{eq:nj}).
The Type III migration rate is then given as 
\begin{equation}
\left(\frac{da_i}{dt}\right)_{III} = f_{\rm slow}\sum_j  \left(\frac{da_i}{dt}\right)_{III, j}, \label{eq:dat3} 
\end{equation}
where $f_{\rm slow}$ is the factor that represents slowing down for massive embryos. 
The slowing down occurs when the embryo cannot migrate over the feeding zone
during the synodic period for that width, as cancellation of angular momentum exchange 
occurs in the second encounter (Ida et al., 2000, Kirsh et al., 2009).
Bromley and Kenyon (2011) showed that the slowing down factor 
is given by a migration length during the synodic period for the half-width of the 
feeding zone $d_{\rm f}$
relative to $d_{\rm f}$, if this ratio is less than unity: 
\begin{equation}
f_{\rm slow} = {\rm Min}\left[\frac{4\pi a_i}{3d_{\rm f}^2\omega_{\rm K}}\sum_j  \left(\frac{da_i}{dt}\right)_{III, j}, 1\right]. \label{eq:slow} 
\end{equation}
We employ $d_{\rm f} = 1.8 a_i h_{ij}$ after Bromley and Kenyon (2011). This length is somewhat shorter than the actual half-length of 
the feeding zone, $d_{E0}$ [Eq.~(\ref{eq:de0})],  
and the reason for this value being applied is because the strong attractive force comes from a relatively narrow region. 
If the embryo mass is small enough, $f_{\rm slow} = 1$; in this regime, the migration rate is independent of the embryo mass.

One may think that the migration rate can be simply estimated by 
summing up the torques from both sides in Eqs.~(\ref{eq:dat3j}) and (\ref{eq:dat3}) rather than summing up for one side
and that $\Delta L_{N,i}$ calculated in the $N$-body routine is unnecessary.
However, we find that such an approach can work only if the embryo is massive enough so that $f_{\rm slow} < 1$.
In this case, the all tracers in the feeding zone strongly interact with 
the embryo and this produces large velocity asymmetry between the unperturbed side and the other side that the embryo migrated through.
On the other hand, if the embryo is small, it migrates over the feeding zone 
through interactions with a small fraction of tracers in the feeding zone,
and only small velocity asymmetry is produced.
In such a situation, the evaluated torques 
from both sides roughly cancel out each other 
and the fast migration cannot be reproduced. 
If we evaluate the torques from tracers in an azimuthally limited region, where 
velocity asymmetry exists, summation of both sides may work. 
However, an appropriate choice of the azimuthal width is unclear to us 
and the number of tracers in a narrow region is too few to give a reliable migration rate.
This is why we evaluate the migration rate due to the one-sided torque as an upper limit,
while the actual angular momentum change is evaluated in the $N$-body routine.

For Type II migration, we evaluate the torques from tracers in both sides of the disk as done for the stirring rate due to distant encounters. 
The change in the semimajor axis difference, $\delta d_{ij}$,  
during a single encounter for $e_{\rm r} = i_{\rm r} = 0$ is given as (Bromley and Kenyon, 2011)
\begin{equation}
\delta d_{ij} = \left(g(b) - \frac{9}{20}b^2 h_{ij} \frac{dg}{db}\right)a_i h_{ij}, 
\end{equation}
where $g(b) = -32b^{-5}$ and $b = d_{ij}/(a_i h_{ij})$. 
This formula is expected to be independent of $e_{\rm r}$ and $i_{\rm r}$ after averaged 
 over $\varpi_{\rm r}$ and $\Omega_{\rm r}$ (Hasegawa and Nakazawa, 1990).
In a similar manner of  Eq.~(\ref{eq:dedis}), the change rate of the semimajor axis of the embryo 
is given as 
\begin{eqnarray}
\left(\frac{da_i}{dt}\right)_{II, j} &=&  \nu_j  \delta d_{ij}  v_{\rm sh} n_{j} \delta r \nonumber \\
 &=&   \frac{3}{8}g(b) \left(1+ h_{ij}b\right) a_i h_{ij}^2 |b| \nu_j  \omega_{\rm K}  \frac{k_j-1}{\delta \theta},  \label{eq:dat20}  
\end{eqnarray}
The Type II migration rate is then given as 
\begin{equation}
\left(\frac{da_i}{dt}\right)_{II} = \sum_j  \left(\frac{da_i}{dt}\right)_{II, j}. \label{eq:dat2} 
\end{equation}
If the surface density is symmetric for the inner and outer disk, the first term in the parentheses in the r.h.s. of Eq.~(\ref{eq:dat20})
vanishes after summation. In this case, the embryo slowly migrate inward.
If density asymmetry exists, the first term remains and relatively fast migration takes place in either inward or outward direction.

The migration rate of the embryo suggested from the statistical routine is given as 
\begin{equation}
\frac{da_i}{dt} = \left(\frac{da_i}{dt}\right)_{III} + \left(\frac{da_i}{dt}\right)_{II}.
\end{equation}
The predicted angular momentum change during $\Delta t$ is given as 
\begin{equation}
\Delta L_{S,i} = \frac{L_i}{2 a_i}\left(\frac{da_i}{dt} \right) \Delta t,
\end{equation}
where $L_i$ is the angular momentum of the sub-embryo. 
To produce smooth Type III migration, at least a few, preferably $\sim 10$,  tracers need to exist constantly in the feeding zone in the region $i$.
If we choose a very small $\delta \theta$ for the region $i$, that makes sub-embryo migration noisy.

We find that a large $\Delta L_{N,i}$ sometimes remains unreleased  
even when an embryo reaches a disk edge.  
This usually causes  the embryo migrate artificially further away from the disk edge.
To avoid this problem,  
we remove $\Delta L_{N,i}$ quickly on the timescale of $1000$ orbital periods at the disk edge, although this violates conservation of the total angular momentum of  the system.
We regard the embryo being at the disk edge if the total mass of tracers 
in the feeding zone ahead of the embryo in the migrating direction is less than 
the embryo mass. For a small embryo, sometimes no tracer exists in the feeding zone due to statistical 
fluctuations even if the embryo is not at the disk edge. 
To avoid this issue, we also count tracers within $10 a_i h_{ij}$ ahead of the embryo even if they are not 
in the feeding zone. 

\subsection{Global gravitational force of tracers}
The  global gravitational forces of tracers are added to the equations of motion for tracers and sub-embryos.
To calculate the forces, we assume that the disk is axisymmetric and the vertical distribution is given by Gaussian for 
tracers.  For simplicity, at each radial location, we consider only one scale height of tracers with various planetesimal masses 
whereas in reality different mass groups have different scale heights. 
The scale height of planetesimals in a radial grid at $r$ is given by
\begin{eqnarray}
h_{\rm tr} (r) = \sqrt{\frac{1}{M_{\rm tr}}\sum_{\rm tracers} m_i k_i a_i^2 \sin^2{i_i}}, \label{eq:htr}
\end{eqnarray}
where the summation is done for each radial bin, and the total mass of the tracers $M_{\rm tr}$ in the grid is 
\begin{eqnarray}
M_{\rm tr}(r) =  \sum_{\rm tracers} m_i k_i. \label{eq:mtr}
\end{eqnarray}
The surface density of tracers are given by $\Sigma_{\rm tr}(r) = M_{\rm tr}(r)/(2\pi r\Delta r)$, where $\Delta r$ is 
the radial grid width.
The spatial density of tracers is given as 
\begin{equation}
\rho_{\rm tr}(r,z) = \frac{\Sigma_{\rm tr}}{\sqrt{2\pi}h_{\rm tr}}\exp{\left(-\frac{z^2}{h_{\rm tr}^2}\right)}.
\end{equation}

The $r$ and $z$ components of the tracers' global gravitational force are given by
(Nagasawa et al., 2000)
\begin{eqnarray}
f_{{\rm tr}, r}(r,z) &=& -2G \int^{\infty}_{-\infty} \int_{r'}\frac{r'}{r}
\frac{\rho_{\rm tr}(r',z')}{\sqrt{(r+r')^2 + (z-z')^2}} \label{eq:fglr}
\nonumber \\
&& \times \left[\frac{r^2-r'^2 -(z-z')^2}{(r-r')^2 + (z-z')^2}E(\xi) + K(\xi)\right]
dr'dz'
\nonumber \\
f_{{\rm tr}, z}(r,z) &=& -4G  \int^{\infty}_{-\infty} \int_{r'}
\frac{\rho_{\rm tr}(r',z')r'(z-z')}{\sqrt{(r+r')^2 + (z-z')^2}} \nonumber \\
&& \times \left[\frac{E(\xi)}{(r-r')^2 +(z-z')^2}\right]dr'dz',
\end{eqnarray}
where $G$ is the gravitational constant, $K(\xi)$ and $E(\xi)$ are the first and second kind of elliptic integrals, 
and $\xi$ is given by
\begin{equation}
\xi = \sqrt{\frac{4rr'}{(r+r')^2 + (z-z')^2}}.
\end{equation}
These components of the force are tabulated in ($r,z$) grids and interpolated for use in simulations. 
For the accelerations on sub-embryos, $k_i-1$ is used instead of $k_i$ in Eqs.~(\ref{eq:htr}) and (\ref{eq:mtr}).
It is not necessary to calculate the global gravitational force components frequently and we usually 
update them every $\sim 1000$ times the shortest orbital period of particles.

The global gravitational force modifies the precession rates of the longitudes of pericenter and ascending node; 
it usually accelerates the absolute precession rates.
The precession rates do not usually matter to dynamics of tracers in terms of such as encounter velocities.  
These rates are, however, likely to be important for stability of the system, when multiple sub-embryos appear in the system.
The embryos mutually interact through secular resonances which cause oscillations in eccentricities and inclinations of embryos.
The global gravitational force reduces these oscillation periods. As a result, the maximum eccentricities and inclinations of sub-embryos
during the oscillations are suppressed to relatively small values.
We identify this effect in simulations in Section~3.5 (results without the global gravitational force are not shown there, though). 
Although its effect is much smaller than dynamical friction, it does not seem to be negligible. 


\section{Tests of the hybrid code}

\begin{table}
\begin{center}
\footnotesize
\begin{tabular}{|ccccccc|} 

\hline
Section                                  & 3.1                      & 3.2            & 3.3                & 3.4                      & 3.5           & 3.6        \\    
Theme                                  & {\rm Collision}    & {\rm Scattering}      & {\rm Migration}           & {\rm Runaway}   & {\rm Terrestrial}  & {\rm Jovian}        \\    
Figure                                    & 5                         & 6-8             & 9,10              & 11,12                  & 13,14       & 15,16   \\  \hline
$M_{\rm Tot}$ ($M_{\oplus}$) & 0.1                      & 0.1,0.125  & 230            & 0.67                       & 2              & 200  \\ 
Region (AU)                        & 0.95-1.05           & 0.95-1.05 & 14.5-35.5 & 0.915-1.085        & 0.7-1.0   & 4-16     \\
$\alpha$                               &-1.5                      & -1.5           & -1.0            & -1.5                       & -1.5          & -1.5       \\
$N$                                        & $10^{12}$        & 1700-$2\times10^9$ & $2.3 \times 10^6$ & $8.33 \times 10^8$ &2000 & $10^{13}$ \\
$N_{\rm tr}$                         & $500,2000$           & $200,250$   & 230            & $400, 4000$           & $200, 400$ & 10000 \\ 
$N_{\rm em}$                      & 0                         & 0                 & 1                 & $>0$                   & $>0$        & $>0$     \\ 
$\langle e^2 \rangle^{1/2}$ &$0.005,0.0005$  & $10^{-4}, 10^{-5}$ & $0.01$-$0.05$ & $1.4\times 10^{-4}$ & 0.001& $ 10^{-4}$ \\
$\delta t$ (days)                 & 6                         & 6                & 60               & 6                           & 6               & 60         \\
 $\delta \theta$  ($\pi$)                  & $0.5, 0.1$ & $0.5$    & $0.5$     & $0.5, 0.25$ & $0.5$  & $0.25$ \\
 Collision                          & Rebound & N/A        & N/A             & Merging               & Merging  & Merging  \\ 
 Gas disk                    & No  & No        & No             & Yes              & No          & Yes \\ \hline  
 \end{tabular}

\end{center}

Table~1. Parameters used for hybrid simulations. 
Theme is a very brief expression of what experiments are performed.
$M_{\rm Tot}$ is the initial total mass of planetesimals.
Region represents the initial inner and outer edges of the planetesimal disk.  
$\alpha$ is the power-law exponent for the surface density of planetesimals ($ \propto r^{\alpha}$).
$N$ is the initial number of planetesimals.
$N_{\rm tr}$ is the initial total number of tracers.
$N_{\rm em}$ is the initial number of embryos ($>0$ means that no embryo is introduced in the initial state 
but embryos appear as a result of collisional growth). 
$\langle e^2 \rangle^{1/2}$ is the initial root-mean-square eccentricity of tracers.
$\delta t$ is the time step for orbital integrations. 
 $\delta \theta$ is the angular half-width of the region used in neighboring search (Fig.~2). 
 Collision represents collisional outcome. Gas disk represents whether damping effects due to a gas disk 
 are taken into account.
 In all the simulations,  the physical density of any body is 2.0 g cm$^{-3}$.
The initial distributions of $e$ and $i$ are given by the Rayleigh distributions with $\langle e^2 \rangle^{1/2}$ = 2$\langle i^2 \rangle^{1/2}$.

\end{table}

We show six different types of tests of the hybrid code:
collisional damping and stirring (Section~3.1), 
gravitational scattering and radial diffusion (Section~3.2),
planetesimal-driven migration of sub-embryos (Section~3.3), 
runaway to oligarchic growth (Section~3.4),
accretion of terrestrial planets from narrow annuli (Section~3.5), and
accretion of cores of jovian planets (Section~3.6).
The parameters used in the test simulations are summarized in Table~1.

\subsection{Collisional damping and stirring}
\begin{figure}
\begin{center}
\includegraphics[width=0.72\textwidth]{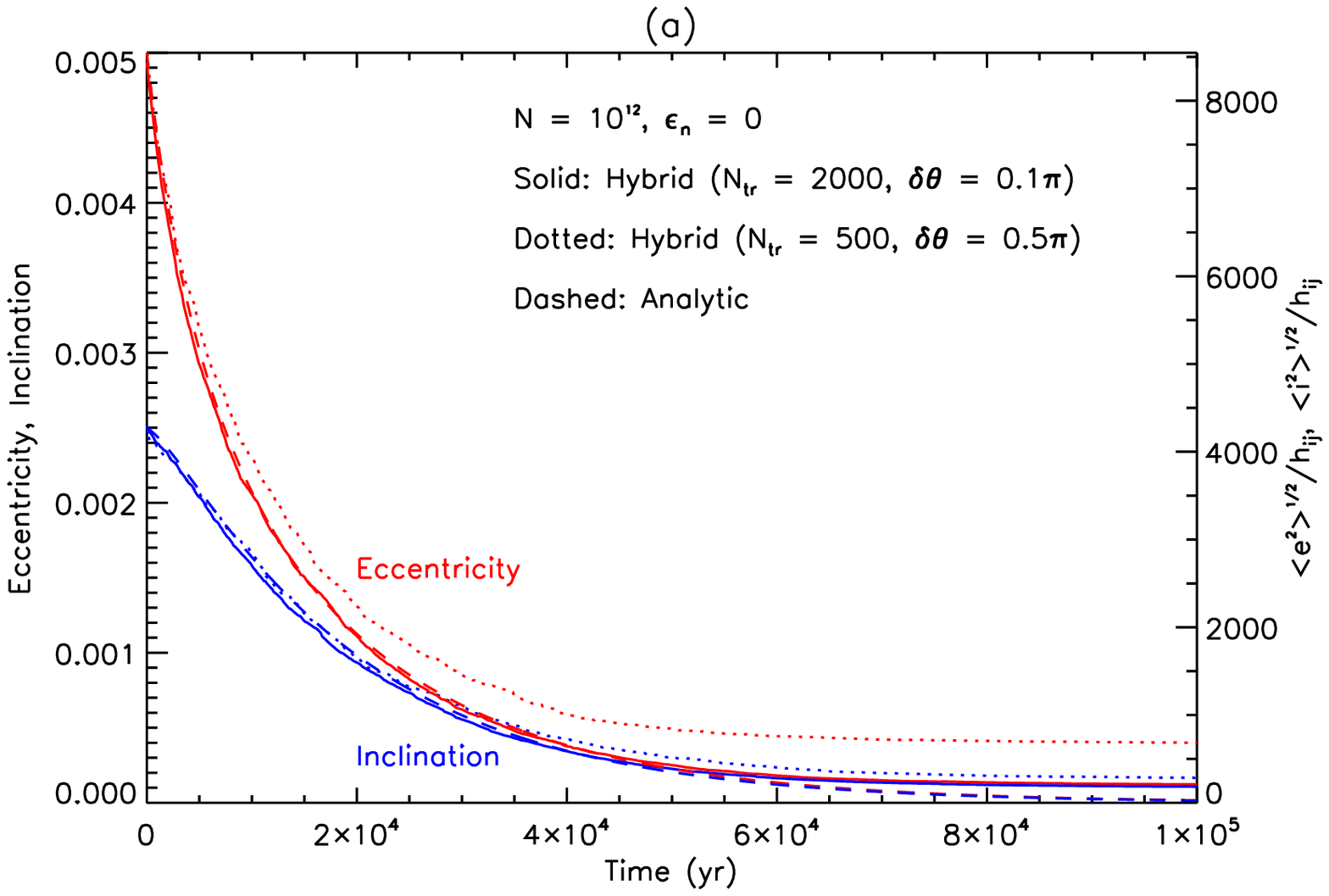}
\includegraphics[width=0.72\textwidth]{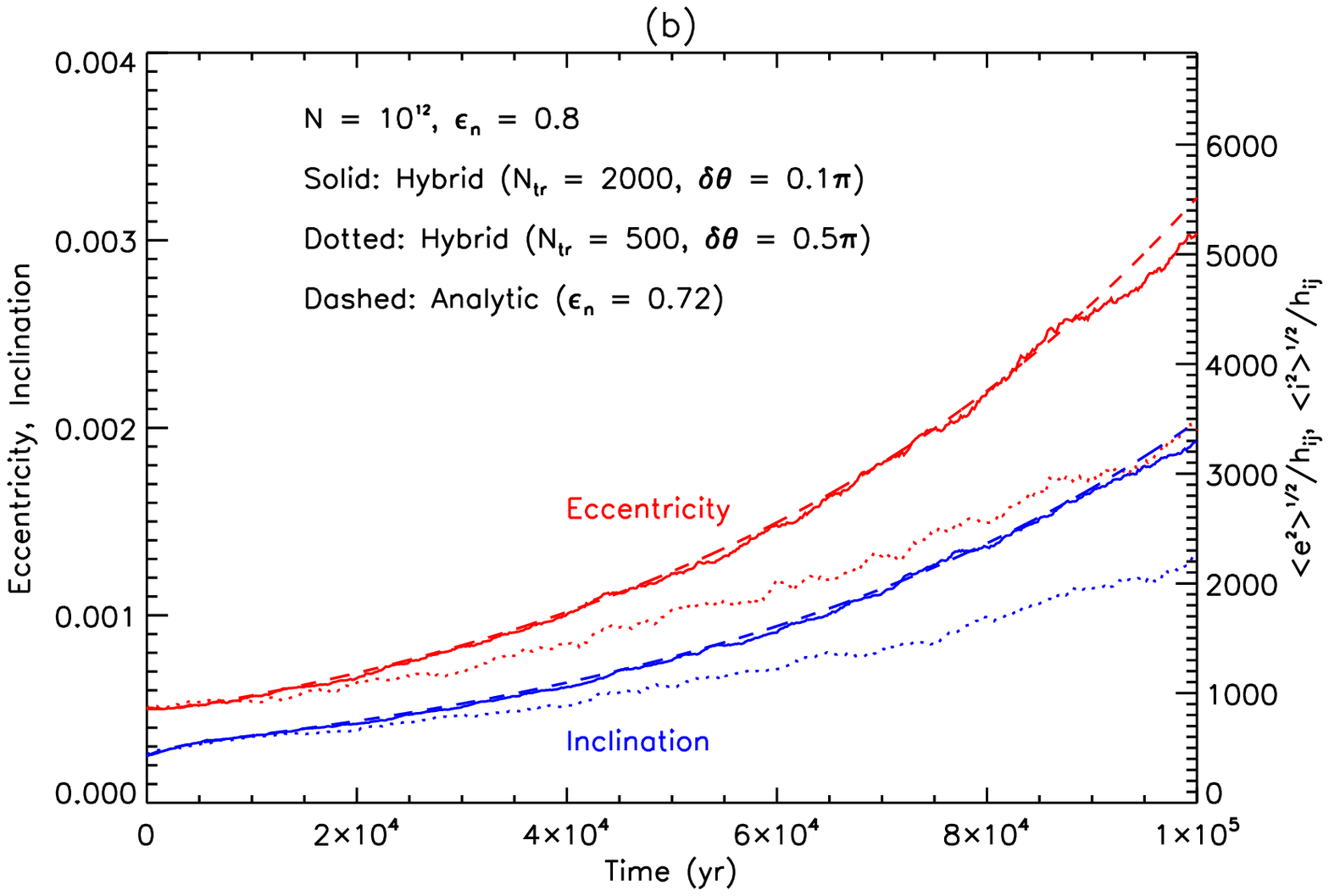}
\end{center}

Fig.~5. Time evolution of $\langle e^2 \rangle^{1/2}$ and $\langle i^2 \rangle^{1/2}$ 
of planetesimals due to mutual collisions only for (a) $\epsilon_{\rm n}$ = 0 and (b) $\epsilon_{\rm n}$ = 0.8. 
All effects related to mutual gravity are ignored.  
The total number of planetesimals is $10^{12}$ represented by 2000 (solid curves) and 500 (dotted curves) tracers.
Planetesimals are placed between 0.95 AU and 1.05 AU and the total mass  is 0.1 $M_{\oplus}$.
In the hybrid simulations, we only use tracers between 0.96 and 1.04 AU in calculations 
of $\langle e^2 \rangle^{1/2}$ and $\langle i^2 \rangle^{1/2}$  to 
avoid the effect due to radial expansion.
Evolution curves calculated using the analytic formula (Ohtsuki, 1999) are also shown by dashed curves for comparison. 
In the panel (b), $\epsilon_{\rm n}$ = 0.72 is used for the analytic curves.

\end{figure}

The first test we present is velocity evolution of tracers due to mutual collisions only
to check whether the collision probability and the velocity changes at collisions are 
calculated accurately. All mutual-gravity related effects are ignored in the test:
gravitational scattering,  gravitational focusing in the collisional probability
[the second term in the bracket of the r.h.s of Eq.~(\ref{eq:pcol})],  and enhancement of the impact velocity (Appendix~E.2). 
Collisions are assumed to result in inelastic re-bounds (hit-and-run collisions),
and post collision velocities are characterized by the normal and tangential restitution coefficients, $\epsilon_{\rm n}$ and  $\epsilon_{\rm t}$. 
Appendix~E.2 explains how to handle hit-and-run collisions between planetesimals in two tracers, with assumptions of $\epsilon_{\rm n} = 0$
and $\epsilon_{\rm t} = 1$ [see Eq.~(\ref{eq:vnd}) for more detail].
These coefficients, however, can be modified to any values.
In the tests, we assume no friction (i.e.,  $\epsilon_{\rm t} = 1$).
Collisional velocity evolution is a well-studied problem for planetary rings and  
several authors have developed analytical formulae (Goldreich and Tremaine, 1978; H\"{a}meen-Antilla, 1978;  Ohtsuki, 1992, 1999). 
We compare our simulation results with analytic estimates of Ohtsuki (1999) because of the easiness of use.


Figure~5 shows the time evolution of $\langle e^2 \rangle^{1/2}$ and $\langle i^2 \rangle^{1/2}$ of $10^{12}$ planetesimals
for (a) $\epsilon_{\rm n}$ = 0  and (b) $\epsilon_{\rm n}$ = 0.8. 
The radius of all planetesimals is 0.4 km.
The planetesimal disk extends from 0.95 to 1.05 AU and its total mass is 0.1 $M_{\oplus}$. These parameters give
the disk normal optical depth of $5 \times 10^{-5}$.
Two simulations are performed to see resolution effects for both Fig.~5a and 5b: 
one with $N_{\rm tr} = 2000$ and $\delta \theta = 0.1\pi$
and another one with $N_{\rm tr} = 500$ and $\delta \theta = 0.5\pi$, where $N_{\rm tr}$ is the total number of tracers. 
The total angular momentum, $L_{\rm tot}$, is well conserved for all test simulations shown in Fig.~5;
 $|L_{\rm tot}(t)-L_{\rm tot}(t=0)|/L_{\rm tot}(t) < 10^{-12}$.

For  $\epsilon_{\rm n} = 0$ (Fig.~5a),
we find excellent agreements between our results and  the analytic estimates of Ohtsuki (1999), in particular 
for the case of $N_{\rm tr} =  2000$. 
However, the decreases in $\langle e^2 \rangle^{1/2}$ and $\langle i^2 \rangle^{1/2}$  stall at certain values in hybrid simulations --
$\langle e^2 \rangle^{1/2} \sim 10^{-4} (4 \times 10^{-4})$ for $N_{\rm tr} =  2000$ (500) --
whereas in the analytic curves, $e$ and $i$ decrease down to $\sim s/r$, where $s$ is the particle radius.
In the hybrid simulations, eccentric ringlets form, in 
which the longitudes of pericenter of all particles are nearly aligned.
This issue is probably related to the initial condition. 
In creation of the initial positions and velocities, 
the longitudes of pericenter are randomly chosen, and the sum of the eccentricity vectors of all tracers is not exactly zero.  
The eccentricities of the ringlets in the end state are, in fact, close to the expected values of the residual eccentricity, $\langle e^2 \rangle^{1/2}(t=0)/N_{\rm tr}^{1/2}$.
Since the residual eccentricity decreases with increasing the number of tracers, 
the eccentricity of the ringlet in the end state decreases. 
Even if the residual eccentricity is fixed to be zero,
the same problem is likely to occur unless all particles effectively interact with each other.
The ringlets are not only eccentric but also inclined because a similar discussion is applied to inclination. 
The evolution curves of $e$ and $i$ also weakly depends on $\delta \theta$; a small $\delta \theta$ gives a better agreement 
with the analytic curves. It is found that relative velocities are slightly underestimated 
if $\delta \theta$ is large and radial excursions of particles are larger than $\delta r$. 
This leads a slightly lower damping rate.

Collisions cause not only damping but also stirring  as the collisional viscosity converts the tidal potential to the kinetic energy.
If  $\epsilon_{\rm n} > \epsilon_{\rm crit} = 0.627$,  where $\epsilon_{\rm crit}$ is the critical restitution coefficient, 
the stirring rate exceeds the damping rate and $\langle e^2 \rangle^{1/2}$ and $\langle i^2 \rangle^{1/2}$
 increase with collisions (Goldreich and Tremaine, 1978; Ohtsuki, 1999).
We find that our method is slightly more dissipative than the analytic estimate and
gives $\epsilon_{\rm crit} \simeq 0.7$ for reasons that we were not able to figure out.  
Figure~5b shows the same as Fig.~5a but for the case of $\epsilon_{\rm n} = 0.8$.  
It is found that our result for $\epsilon_{\rm n} = 0.8$ and $N_{\rm tr} = 2000$ (500) well agrees with the analytic estimate with $\epsilon_{\rm n} = 0.72$ (0.70).
Overall, our method can reproduce analytic results using a $\sim 10\%$ higher $\epsilon_{\rm n}$. In simulations of planetary accretion, we use 
$\epsilon_{\rm n} = 0$ for hit-and-run collisions and the offset in $\epsilon_{\rm n}$ is not a problem.
Note that orbital alignments do not occur for the case of Fig.~5b, because collisional stirring randomizes orbital phases.

\subsection{Gravitational scattering and radial diffusion}

\begin{figure}
\begin{center}
\includegraphics[width=0.75\textwidth]{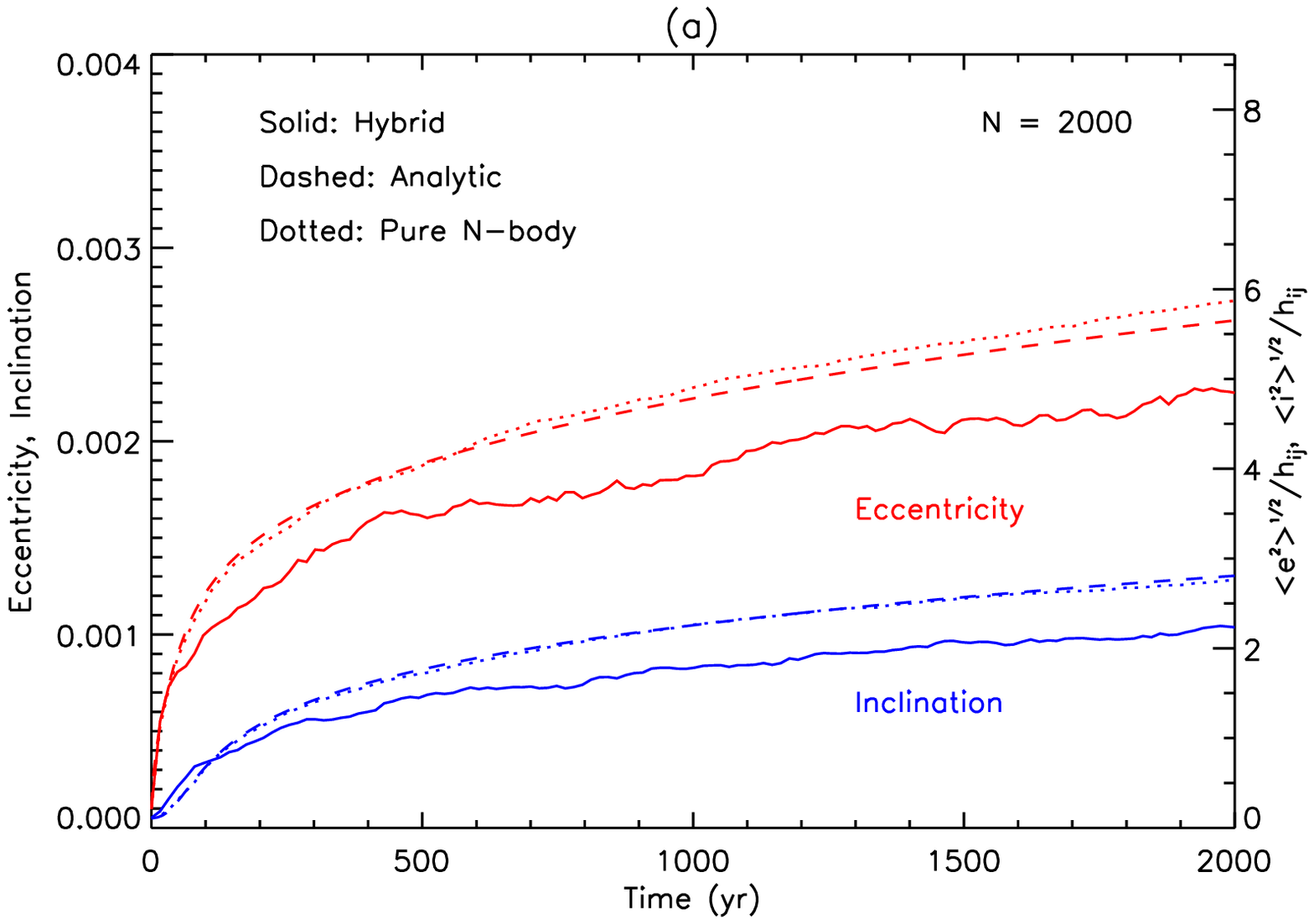}
\includegraphics[width=0.75\textwidth]{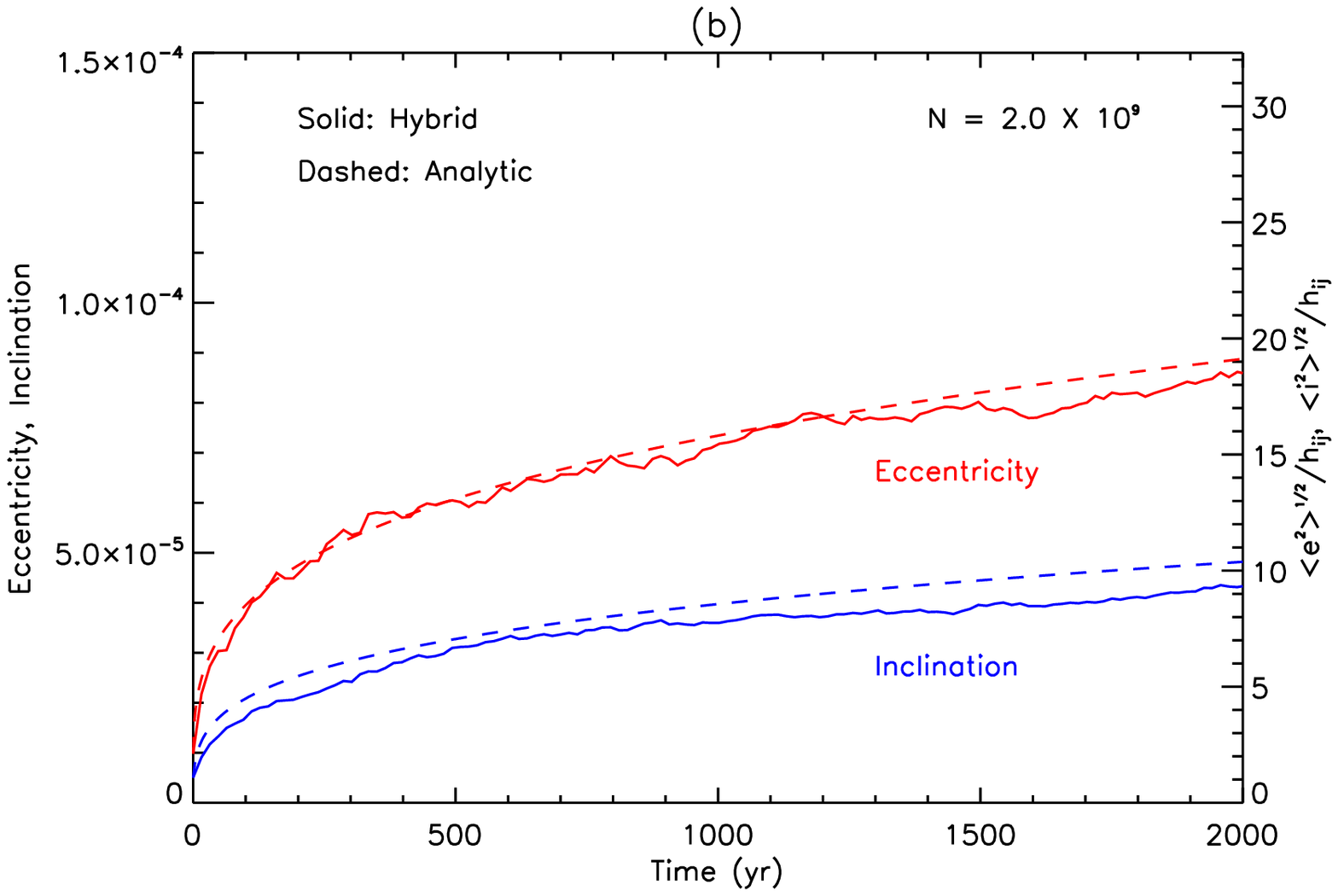}
\end{center}

Fig.~6. 

\end{figure} 

\begin{figure}
\begin{center}
\includegraphics[width=0.75\textwidth]{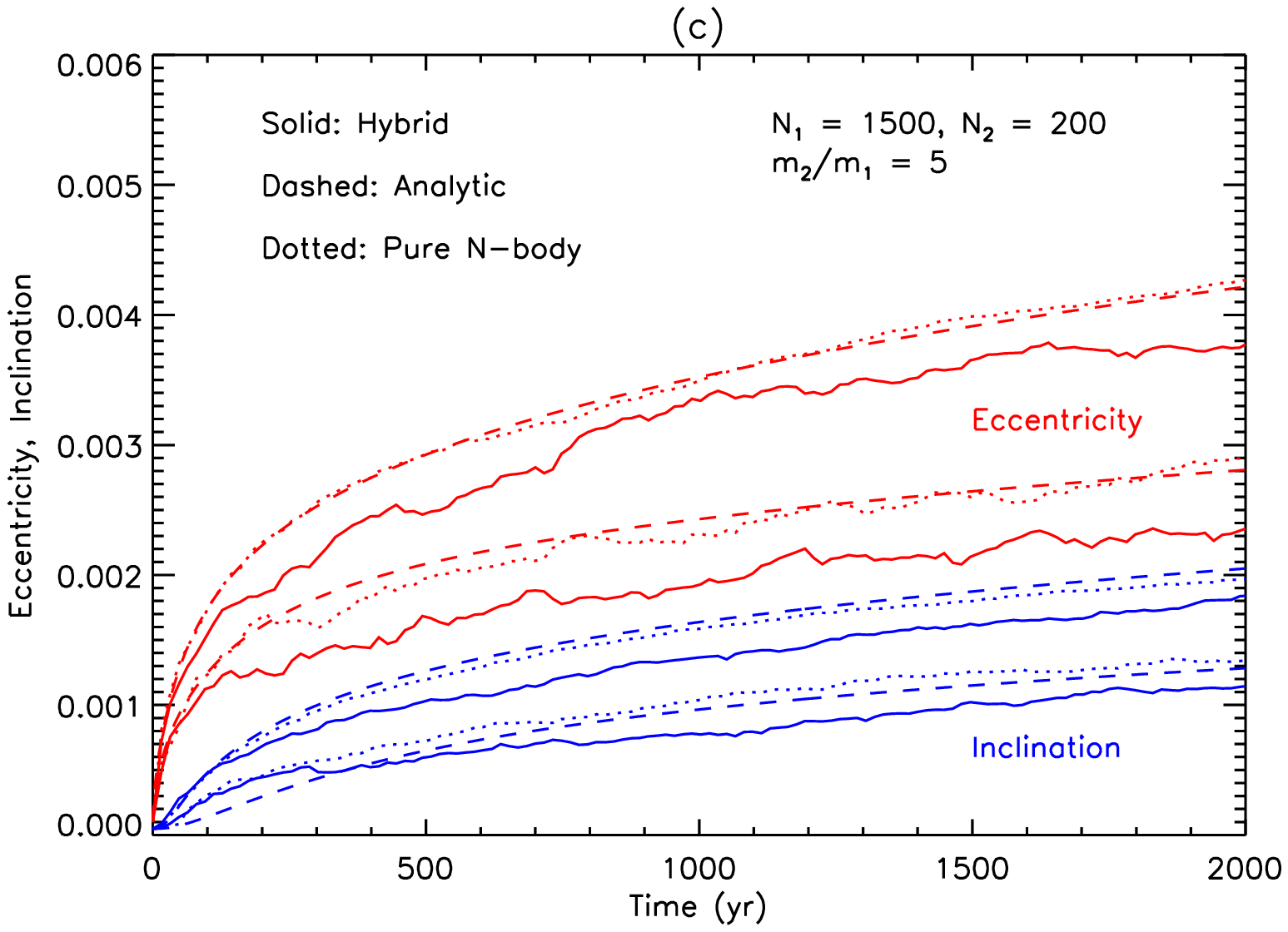}
\end{center}

Fig.~6. Time evolution of $\langle e^2 \rangle^{1/2}$ and $\langle i^2 \rangle^{1/2}$ 
of planetesimals due to gravitational interactions:
(a) $N = 2000$, $N_{\rm tr} = 200$, $M_{\rm tot} = 0.1 M_{\oplus}$,
(b) $N = 2 \times 10^9$, $N_{\rm tr} = 200$, $M_{\rm tot} = 0.1 M_{\oplus}$, 
and (c) $N = 1700$, $N_{\rm tr} = 250$, $M_{\rm tot} = 0.125 M_{\oplus}$ (see Table~1 for the characters).
The panel (c) is the case of the bimodal size distribution with 
$N_1 = 1500$, $N_2 = 200$, $N_{\rm tr,1} = 150$, and $N_{\rm tr,2} = 100$,
where the characters with the subscripts 1 and 2 are those for small and large planetesimals, and
the mass ratio of a large planetesimal to a small one is 5. 
In all cases, planetesimals are placed between 0.95 AU and 1.05 AU, and
$\langle e^2 \rangle^{1/2}$ and $\langle i^2 \rangle^{1/2}$ 
are calculated using tracers between 0.96 AU and 1.04 AU.
Evolution curves calculated using the analytic formula (Ohtsuki et al., 2002) and 
from the pure $N$-body simulations (only for the panels (a) and (c))
are also shown for comparison. 

\end{figure} 

\begin{figure}
\begin{center}
\includegraphics[width=0.49\textwidth]{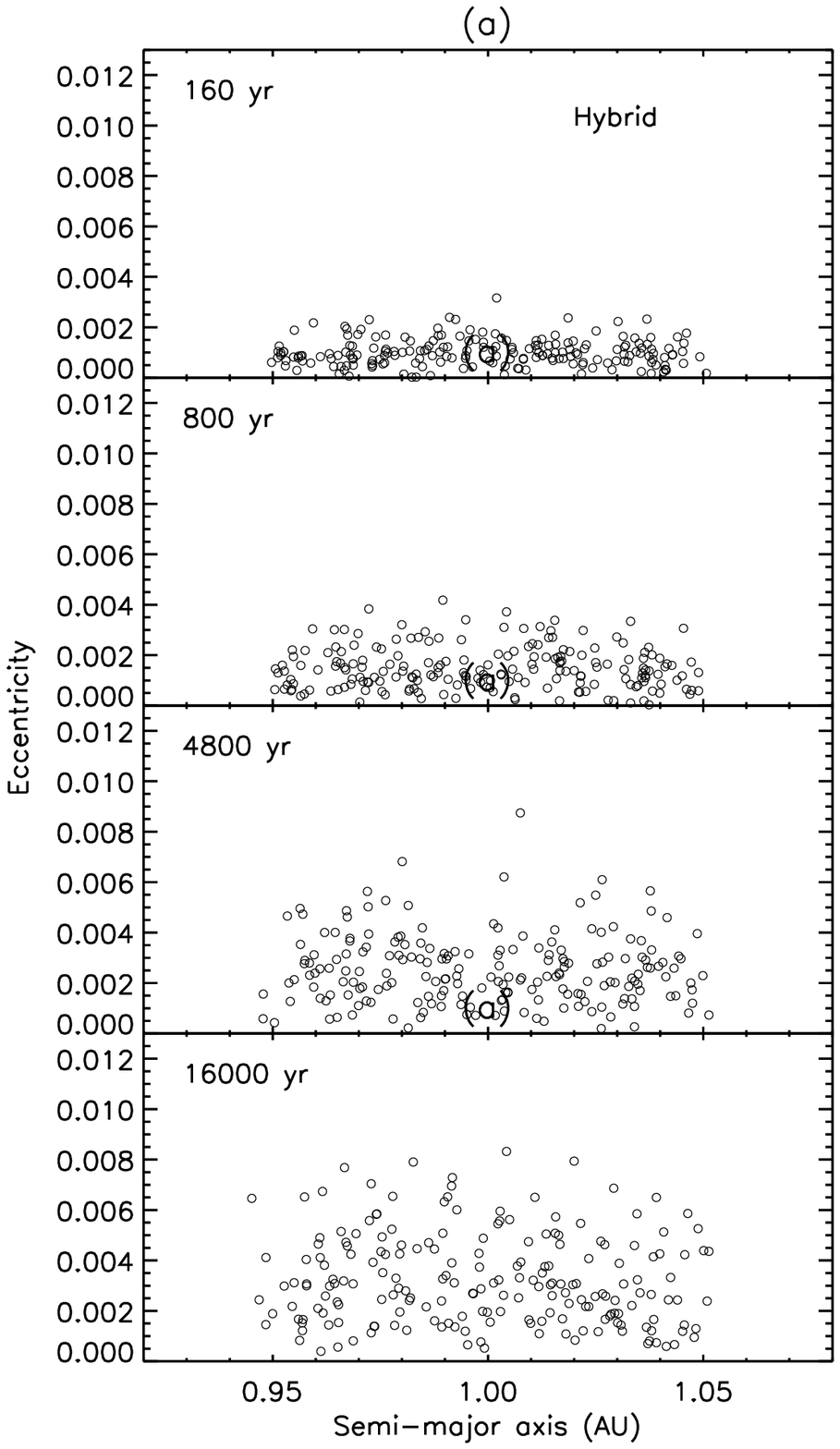}
\includegraphics[width=0.49\textwidth]{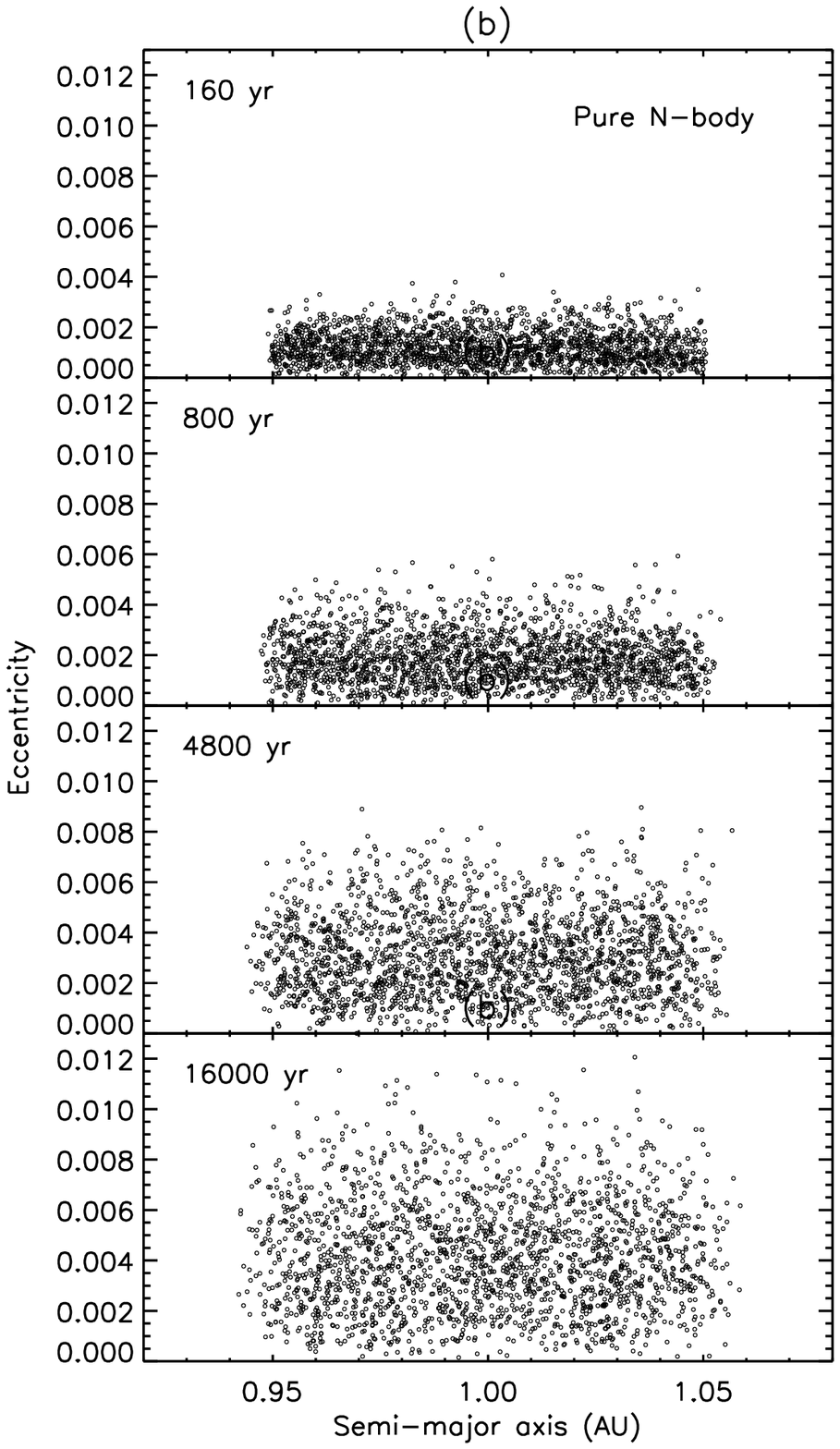}
\end{center}

Fig.~7. Snapshots on the $a-e$ plane for the simulations shown in Fig.~6a:
(a) the hybrid simulation and (b) the pure $N$-body simulation.

\end{figure}

\begin{figure}
\begin{center}
\includegraphics[width=0.73\textwidth]{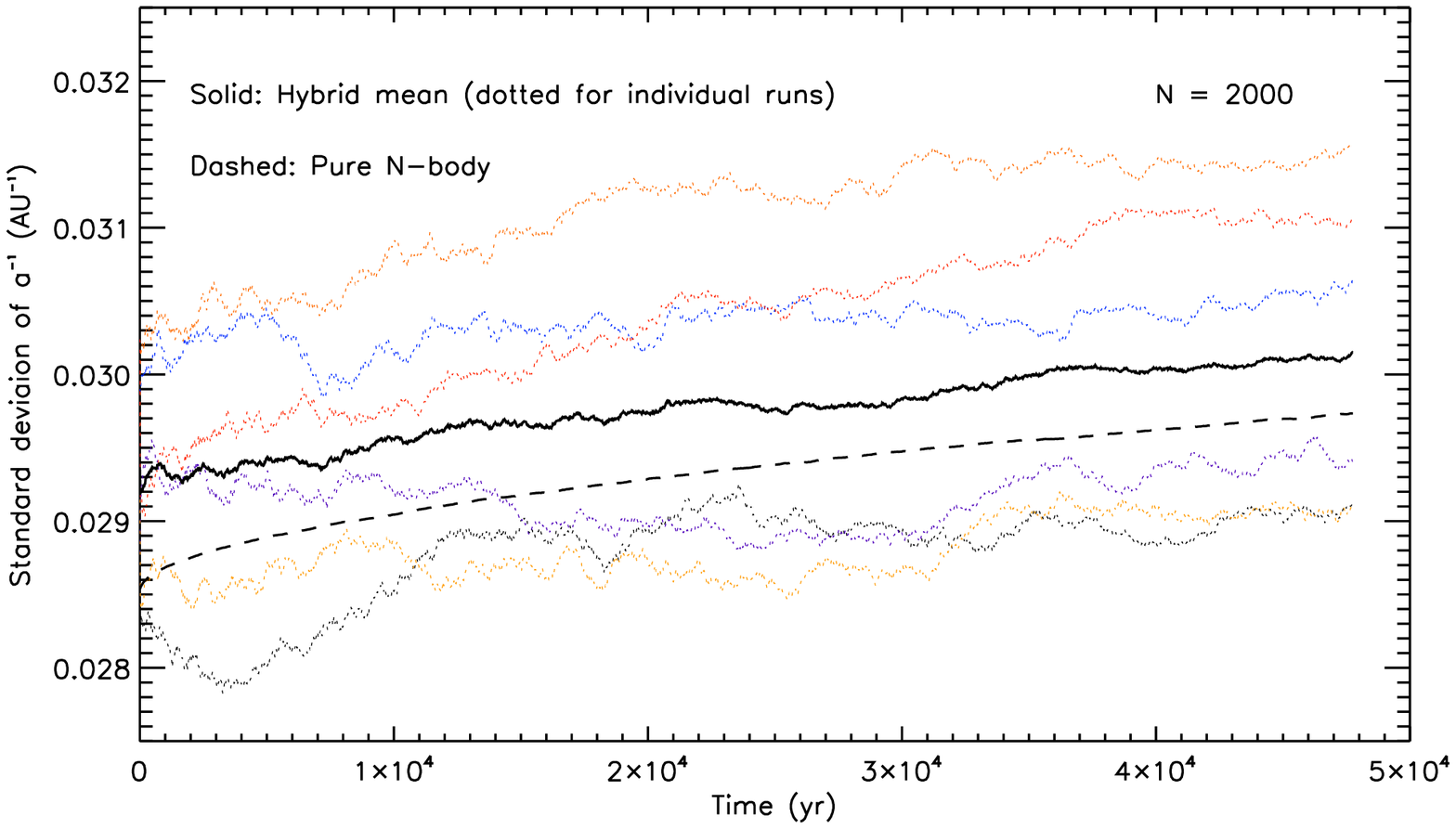}
\end{center}

Fig.~8. Time evolutions of the standard deviation of $1/a$ for the simulations shown in Fig.~7.
Results for six individual hybrid simulations are shown by colored dotted curves, 
whereas their mean value is shown by a black solid curve. 
The result for the hybrid simulation shown in Fig.~7a is a black dotted curve.
For comparison, the result from the pure N-body simulation (Fig.~7b) is shown by 
a black dashed  curve.

\end{figure} 

The next test we present is velocity evolution of tracers due to mutual gravity only in the absence of any collisional effects.
Figure~6 shows evolution of $\langle e^2 \rangle^{1/2}$ and $\langle i^2 \rangle^{1/2}$ for a planetesimal disk 
extending from  0.95 and 1.05 AU. 
The panels (a) and (b) are for mono-sized planetesimals represented by 200 tracers and 
the numbers of planetesimals $N$ are $2000$ and $2\times 10^{9}$, respectively.
The panel (c) is for a bimodal size distribution case; 
1500 small planetesimals and 200 large planetesimals are placed in the disk.
The mass ratio of a large planetesimal to a small one is 5. 
The number of tracers for small planetesimals is 150 and that for large planetesimals is 100. 
The results are compared with analytic estimates of Ohtsuki et al. (2002).
For the panels (a) and (c), we also performed pure $N$-body simulations using 
the parallel-tree code pkdgrav2 (Morishima et al., 2010) with the same initial conditions. 
The agreements between the analytic estimates and pure $N$-body simulation results 
are excellent, so results from both methods are trustable as benchmarks.

Overall, we find good agreements between the results from the hybrid code and the analytic estimates, 
although our model gives slightly lower $\langle e^2 \rangle^{1/2}$ and $\langle i^2 \rangle^{1/2}$ 
than the analytic estimates, except Fig.~6b shows a good agreement. 
We also perform the same hybrid simulations with different values of $N_{\rm tr}$ and $\delta \theta$, 
and the absolute levels of $\langle e^2 \rangle^{1/2}$ and $\langle i^2 \rangle^{1/2}$ at a certain time are found to be 
almost independent of these parameters,
unless $N_{\rm tr}$ and $\delta \theta$ are too small.
The underestimates of $\langle e^2 \rangle^{1/2}$ and $\langle i^2 \rangle^{1/2}$ of our calculations 
is probably caused by the underestimates of $P_{\rm VS}$ and $Q_{\rm VS}$ that we employ.
The analytic calculations of Tanaka and Ida (1996) give slightly lower $P_{\rm VS}$ and $Q_{\rm VS}$ 
for $e_{\rm r}/h_{ij} >4$ and $i_{\rm r}/h_{ij} < 1$ 
than those from direct three-body orbital integrations (Ida, 1990, Rafikov and Slepian, 2010). 
The influence of this difference is less important for large  $\langle e^2 \rangle^{1/2}/h_{ij}$ and $\langle i^2 \rangle^{1/2}/h_{ij}$
and that is why the hybrid simulation in Fig.~6b shows an excellent agreement with the analytic result, contrary to Fig.~6a.
 
Figure~7 shows snapshots on the $a-e$ plane for the simulation of Fig.~6a. Snapshots of the corresponding pure $N$-body
simulation are also shown for comparison. 
The hybrid code can reproduce not only  $\langle e^2 \rangle^{1/2}$ and $\langle i^2 \rangle^{1/2}$  but also 
the distributions of  $e$ and $i$. Ida and Makino (1992) carried out $N$-body simulations and found that the distributions 
of $e$ and $i$ can be well approximated by a Rayleigh distribution. 
We perform the Kolmogorov-Smirnov (KS) test to check whether the distributions of $e$ and $i$ in the hybrid simulations 
resemble a Rayleigh distribution. The test gives the significance level, $Q_{\rm KS}$ ($0 \le Q_{\rm KS} \le 1$) (Press et al., 1986). 
A large $Q_{\rm KS}$ means that the likelihood that two functions are the same is high.   
The time-averaged ($0 \le t \le 50,000$ yr) values of $Q_{\rm KS}$ for the distributions of $e$ and $i$ are 0.44 and 0.57, respectively. 
We also perform the same KS test using the outputs of the pure $N$-body simulation,  and the values of $Q_{\rm KS}$ for 
$e$ and $i$ are 0.62 and 0.19. Thus, the distributions of $e$ and $i$ from our hybrid simulations resemble a Rayleigh distribution 
at a level similar to those from pure $N$-body simulations. 
 
Figure~7 also shows that a planetesimal disk radially expands with time due to viscous diffusion at a rate similar to that for the 
pure $N$-body simulation. To compare more quantitatively, 
we plot the time evolution of the standard deviation of $1/a$ (the normalized orbital energy) of the disk in Fig.~8.
Since the radial diffusion of a low resolution hybrid simulation is found to be noisy, we take average 
over several simulations.  The off-sets between simulations at $t=0$ are caused by
random numbers that are used for choosing initial semimajor axes of tracers.
We find a good agreement in the increasing rates of the standard deviations of $1/a$
between the hybrid simulations and the pure $N$-body simulation. This comparison indicates that 
the diffusion coefficient given by Eq.~(\ref{eq:di0}) is a good approximation.

Contrary to collisional velocity evolution shown in Section~3.1, 
conservation of the angular momentum is violated by gravitational scattering as 
we do not explicitly solve pair-wise gravitational interactions in the statistical routine. 
Provided that the radial migration distance of a tracer due to random walk is about $a_i e_i$,
the error in the total angular momentum is estimated as 
$|L_{\rm tot}(t)-L_{\rm tot}(t=0)|/L_{\rm tot}(t) \sim \langle e^2 \rangle^{1/2}/N_{\rm tr}^{1/2}$,
which we found consistent with the actual errors in the test simulations.

\subsection{Planetesimal-driven migration of sub-embryos}

\begin{figure}
\begin{center}
\includegraphics[width=0.7\textwidth]{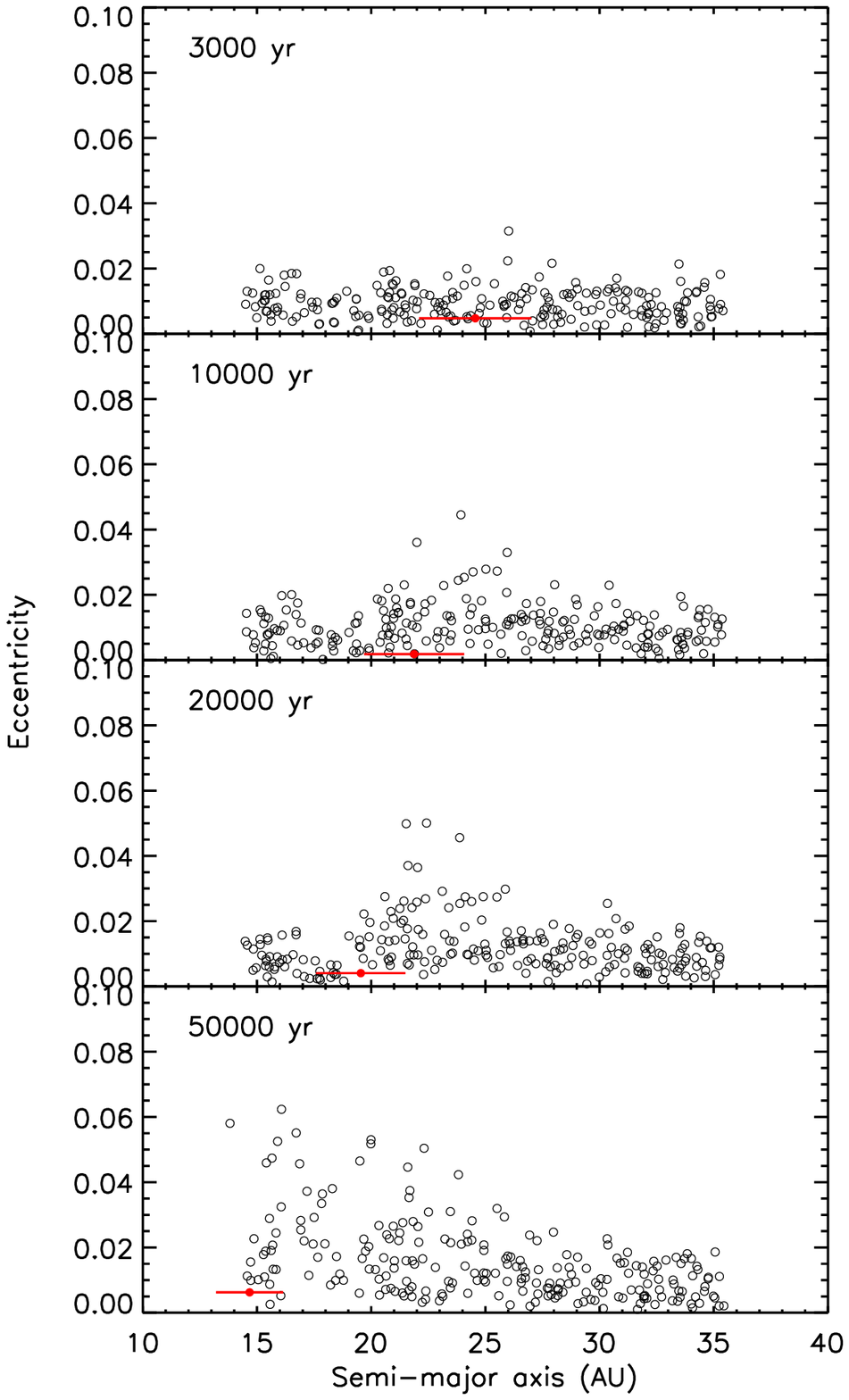}

\end{center}

Fig.~9. Self-sustained (Type III) migration of a sub-embryo.
We employ $m_{i} = m_{\rm t0} = 1 M_{\oplus}$ (where $m_i$ is the embryo mass), 
$N_{\rm tr} = 230$, and $\langle e^2\rangle^{1/2}/h_{ij} =2\langle i^2\rangle^{1/2}/h_{ij} = 1$.
The embryo is displayed as a red filled circle with 
a horizontal bar with a half length of 10 Hill radii.

\end{figure}

\begin{figure}
\begin{center}
\includegraphics[width=0.67\textwidth]{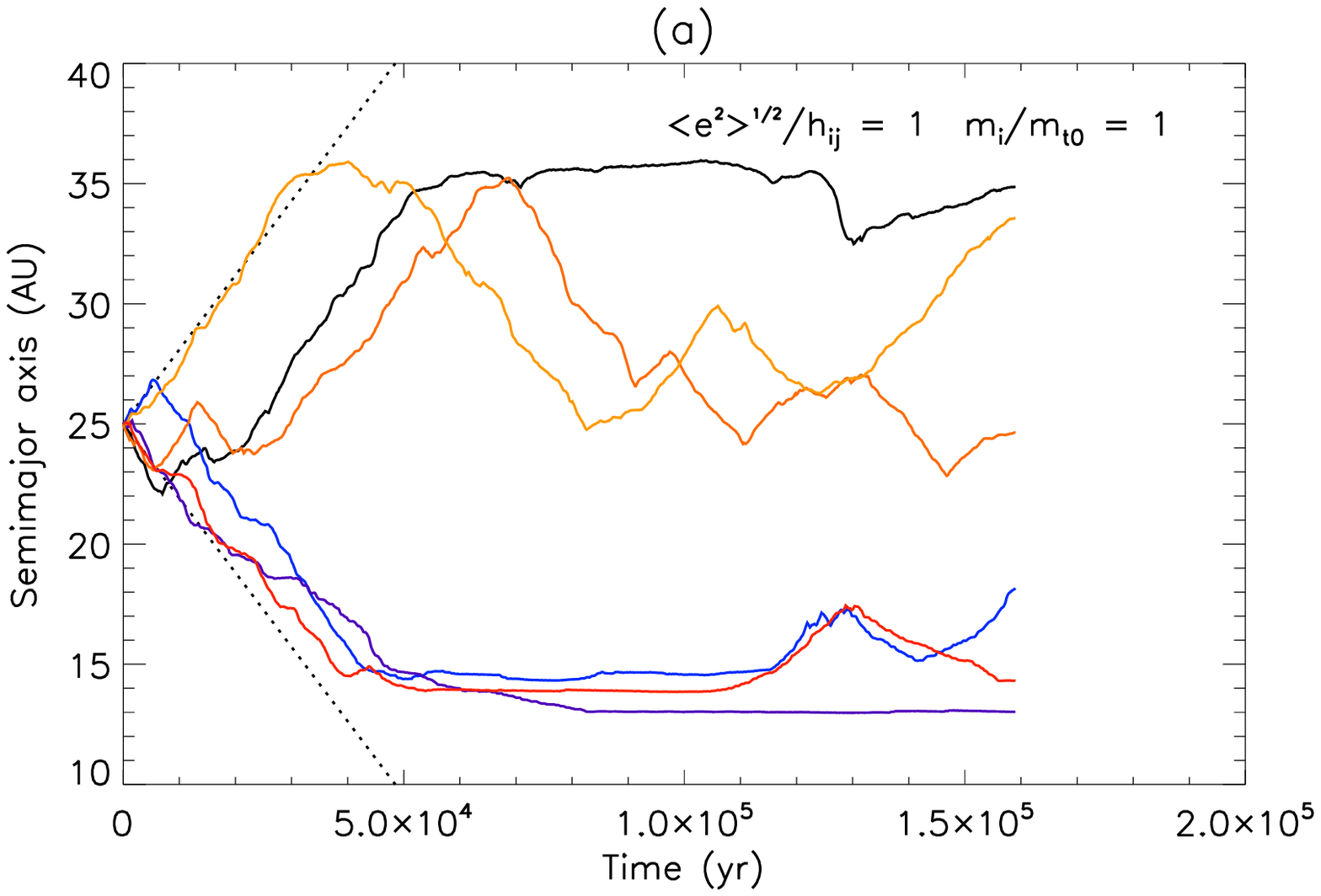}
\includegraphics[width=0.67\textwidth]{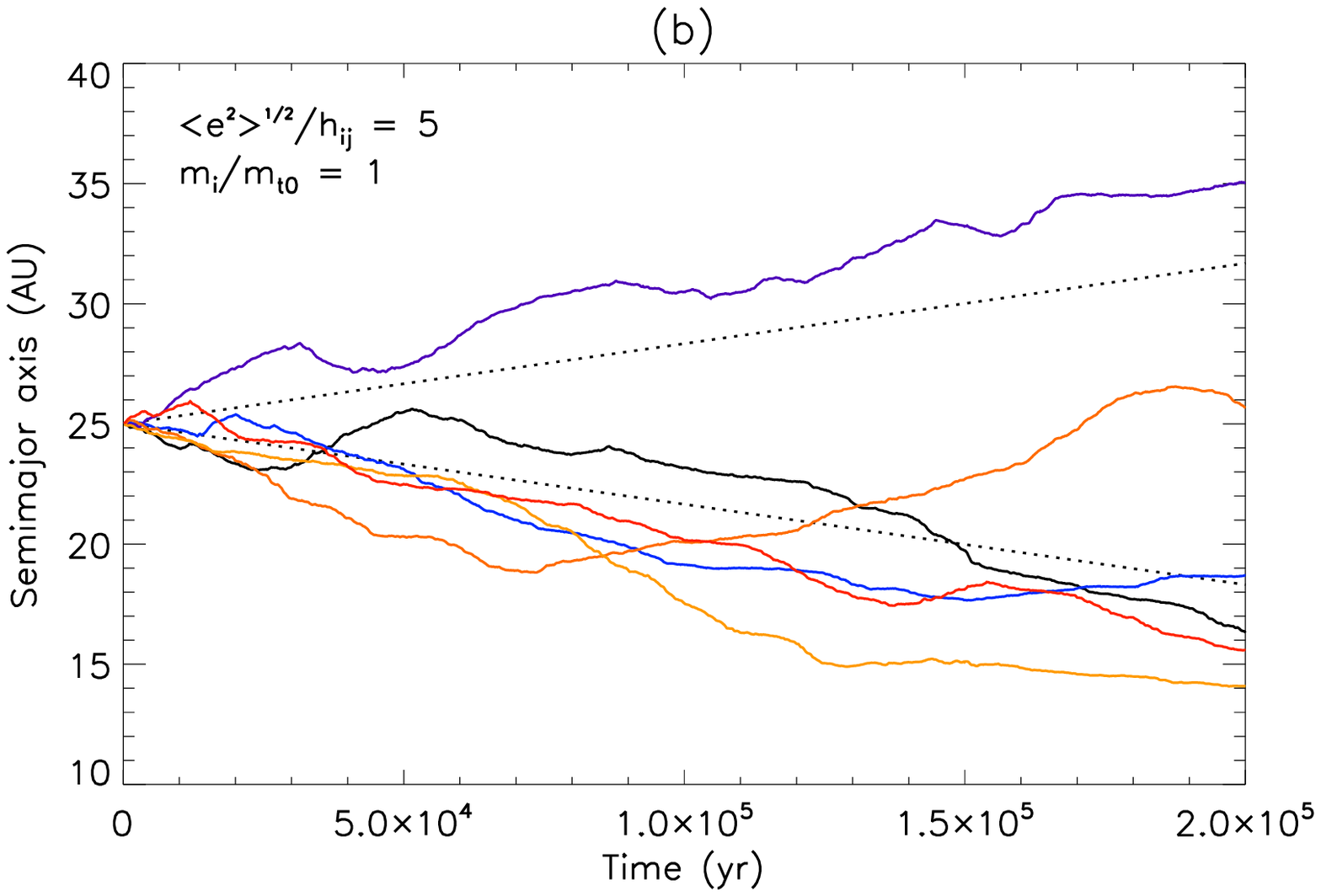}
\includegraphics[width=0.67\textwidth]{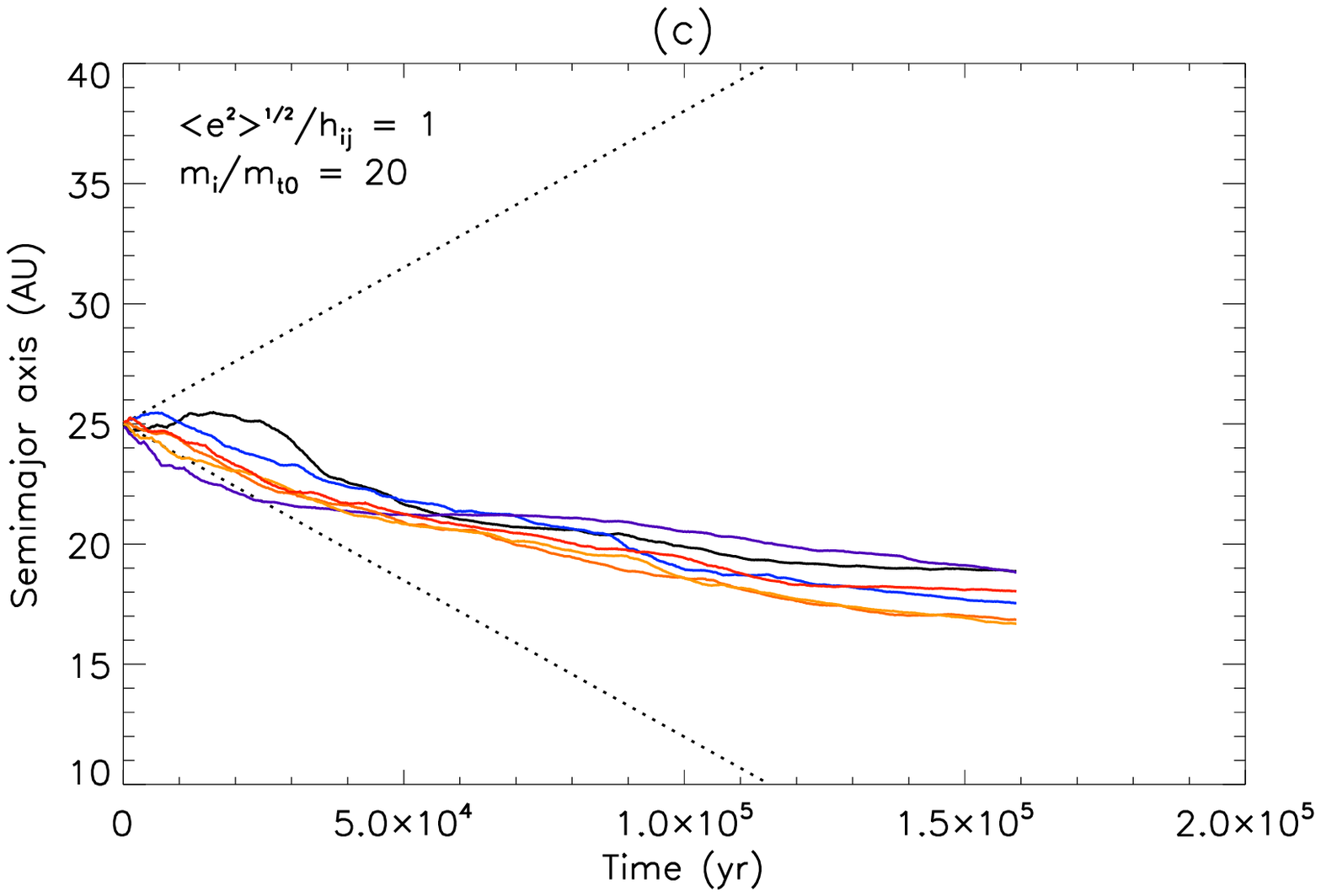}
\end{center}

Fig.~10. Time evolution of the semi-major axis of a sub-embryo due to self-sustained migration:
(a) $\langle e^2\rangle^{1/2}/h_{ij} = 1$ and $m_i = 1 M_{\oplus}$, 
(b) $\langle e^2\rangle^{1/2}/h_{ij} = 5$ and $m_i = 1 M_{\oplus}$,  and
(c) $\langle e^2\rangle^{1/2}/h_{ij} = 1$ and $m_i = 20 M_{\oplus}$.
For all cases, $N_{\rm tr} = 230$ and $m_{\rm t0} = 1 M_{\oplus}$.
Analytic migration rates (Bromley and Kenyon, 2011) are shown by dashed lines.

\end{figure}

In Section~2.6.2, we described how we handle migration of sub-embryos.
To test the routine of  sub-embryo migration, we perform a suite of 
hybrid simulations starting with initial conditions similar to 
those in Kirsh et al. (2009) and Bromley and Kenyon (2011).
The initial planetesimal disk extends from 14.5 AU to 35.5 AU and its total mass 
is 230 $M_{\oplus}$ represented by 230 tracers  (thus $m_{t0} = M_{\oplus}$).
A sub-embryo with a mass of $m_i$ is initially placed at 25 AU and
growth of the embryo is turned off. In this subsection, the index $i$ is given to a sub-embryo.
Each tracer has 10000 planetesimals and tracer-tracer interactions are quite small compared with 
gravitational effects of the embryo. 

An example of migration of an Earth-mass embryo ($m_i = M_{\oplus}$) is shown in Fig.~9 
using four snapshots on the $a$-$e$ plane. 
For this example, the migration is in the mass independent regime [$f_{\rm slow} = 1$; Eq.~(\ref{eq:slow})]. 
The embryo first has some strong encounters with tracers and a stirred region is produced. 
The embryo tends to migrate away from the stirred region due to strong torques exerted by tracers in 
the unperturbed region.  This process trigers fast self-sustained migration. 
During the migration, the embryo continuously encounters with unperturbed tracers on
the leading side while leaving the stirred tracers in the trailing side. 
The fast migration continues until the embryo reaches the disk edge.

Embryo migrations for three cases with different embryo masses and initial velocity dispersions 
are shown in Fig.~10. In each case, six simulations are performed, using different random number seeds 
to generate initial orbital elements of tracers. The migration rates suggested by Bromley and Kenyon (2011)
are shown by dotted lines. They confirmed good agreements between their formula and $N$-body simulations. 
Overall, our method can reproduce migration rates suggested by Bromley and Kenyon (2011).
However, our method gives almost the same probabilities of inward and outward migrations if
$m_i \sim m_{\rm t0}$, whereas $N$-body simulations show that inward migration is predominant 
(Kirsh et al., 2009; Bromley and Kenyon, 2011). If the embryo is small, the direction of migration is determined by 
first few strong encounters, and once migration starts it is self-sustained regardless of its direction. 
In realistic planetary accretion, even if inward migration is predominant, the embryo encounters with inner massive embryos
and then turns its migration direction outward (Minton and Levison, 2014). 
Therefore, we believe that the artificially high occurrence of outward migration is not a problem.
For a large sub-embryo, inward migration is clearly predominant even in our hybrid simulations (Fig.~10c).

To understand these trends and why our method works, we consider conditions necessary to produce 
smooth self-sustained migration. 
If planetesimals are small enough compared with the embryo, 
smooth self-sustained migration of the embryo takes place (ordered migration). 
If planetesimals are large, on the other hand, the embryo 
suffers strong kicks and the direction of its migration may change rapidly 
in a random fashion (stochastic migration).
Whether migration is ordered or stochastic is determined by the size distribution of planetesimals. 
This argument is similar to that discussed for planetary spins (Dones and Tremaine, 1993).

The relative contributions of the ordered and stochastic components are estimated as follows.
Consider that an embryo migrates over a distance $Da_i$ through $n$ encounters with surrounding planetesimals: 
\begin{equation}
 Da_i= n\langle \nu_j a_i h_{ij} \delta b \rangle,
\end{equation}
where again $\nu_{j} = m_{j}/(m_i+m_j)$ and
$\delta b$ is the change of $b$ $[= d_{ij}/(a_ih_{ij})]$ during a single encounter. 
In the above, the angle brackets mean the averaging over encountered planetesimals.
The expected value of $(Da_i)^2$ is given as 
\begin{eqnarray}
(Da_i)^2  &=& n\langle (\nu_j a_i h_{ij} \delta b)^2 \rangle + n(n-1)\langle \nu_j a_i h_{ij} \delta b\rangle^2, \nonumber \\
	      &\simeq& na_i^2 h_{ij}^2 \langle \nu_j^2  \rangle \langle (\delta b)^2 \rangle 
	      \left[1 +  \left(\frac{\langle \nu_j \rangle}{\langle \nu_j^2 \rangle}\right)   \left(\frac{\langle \delta b \rangle}{\langle (\delta b)^2 \rangle}\right) 
	      \left(\frac{\langle  Da_i \rangle}{a_i h_{ij}}\right) \right],	\label{eq:stoch}      	      	 	 
\end{eqnarray}
where we assume that $n \gg 1$ and that there is no correlation between $\nu_j$ and $\delta b$ to transform to the second row. 
If the first term in the r.h.s of Eq.~(\ref{eq:stoch}) is larger than the second term, the migration is stochastic rather than ordered. 
This means that the embryo is likely to migrate over  $Da_i$ via random walk rather than self-sustained migration. 
If we assume that the ordered torque exerted on the embryo is the one-sided (Type III) torque, 
$\langle \delta b \rangle$ is determined by the non-dimensional migration rate
$R_{\rm MG}$ (Appendices~B.3).
The value of $\langle (\delta b)^2 \rangle$ is estimated using the non-dimensional viscous stirring rates, $P_{\rm VS}$ and $Q_{\rm VS}$
(Appendix~D), and we consider contribution of planetesimals in one side assuming that large velocity asymmetry exists. 
Using these non-dimensional rates, we have 
$\langle \delta b \rangle/\langle (\delta b)^2 \rangle \simeq R_{\rm MG}/(P_{\rm VS} + Q_{\rm VS})$.
This ratio is $\sim 0.12$ for the shear dominant regime [$(e_{\rm r}^2 + i_{\rm r}^2)^{1/2} < 2h_{ij}$] and 
$\sim 0.5 h_{ij}/e_{\rm r}$ for the dispersion dominant regime (we assume $e_{\rm r} = 2i_{\rm r}$). 
Therefore, even with planetesimals as massive as the embryo ($\langle \nu_j \rangle/\langle \nu_j^2 \rangle = 2$), 
the migration is likely to be self-sustained migration rather than random walk if the embryo migrates longer than $\sim a_i(4h_{ij} + e_{\rm r})$.
This length is comparable to the feeding zone width.

In the above discussion,  the planetesimal mass $m_j$ can be replaced by the tracer mass $k_j m_j$ to discuss migration calculated 
in our $N$-body routine. 
Therefore, our $N$-body routine can reproduce self-sustained migration of the embryo even with tracers as massive as the embryo.
This is true only if the ordered torque exerted on the embryo is as strong as the one-sided torque. This is not always the case.
The torque becomes nearly one-sided if a strong velocity asymmetry of planetesimals exists around the embryo.
If the embryo jumps over its feeding zone due to a strong kick from a tracer that is as massive as the embryo,
the embryo cannot see velocity asymmetry at that radial location.
Thus, the condition to apply the one-sided torque is that radial migration distance of the embryo 
due to a single encounter is much smaller than its feeding zone.
This condition generally turns out to be $\nu_{j} \ll 1$. 
It is likely that the condition $\nu_{j} < 1/100$ or less recommended by various authors (Kirsh et al. 2009; LDT12; Minton and Levison, 2014)
comes from this criterion. 
In our method, alternatively,  instantaneous large changes of $a_i$ of the sub-embryo are suppressed by limiting  
the migration distance of the sub-embryo to the theoretical prediction
that is derived assuming that planetesimals are much less massive than the embryo.
Therefore, our method can reproduce a long distance self-sustained migration.

The condition to reproduce predominant inward migration is more severe because we need to consider migration of an embryo 
due to torques from both sides of the disk.
In an unperturbed state, the total torque from both sides 
is order of $h_{ij}$ of the one-sided torque in the shear dominant regime (Bromley and Kenyon, 2011) and 
is order of $e_{\rm r}$ of the one-sided torque in the dispersion dominant regime (Ormel et al., 2012). 
Thus, ($h_{ij}  + e_{\rm r}$) is multiplied to the second term 
in the  r.h.s. of Eq.~(\ref{eq:stoch}). 
Then, the condition that the migration is ordered when the embryo migrates over the feeding zone is roughly given by 
$(h_{ij} + e_{\rm r}) > \langle \nu_j^2 \rangle/\langle \nu_j \rangle$.
The initial condition of Fig.~10c barely satisfies this criterion provided that $m_j$ is the tracer mass instead of the planetesimal mass.
Our method indeed reproduces predominant inward migration.
Once the embryo migrates over the feeding zone, velocity asymmetry is obvious and self-sustained migration immerses.

\subsection{Runaway to oligarchic growth}

\begin{figure}
\begin{center}
\includegraphics[width=0.49\textwidth]{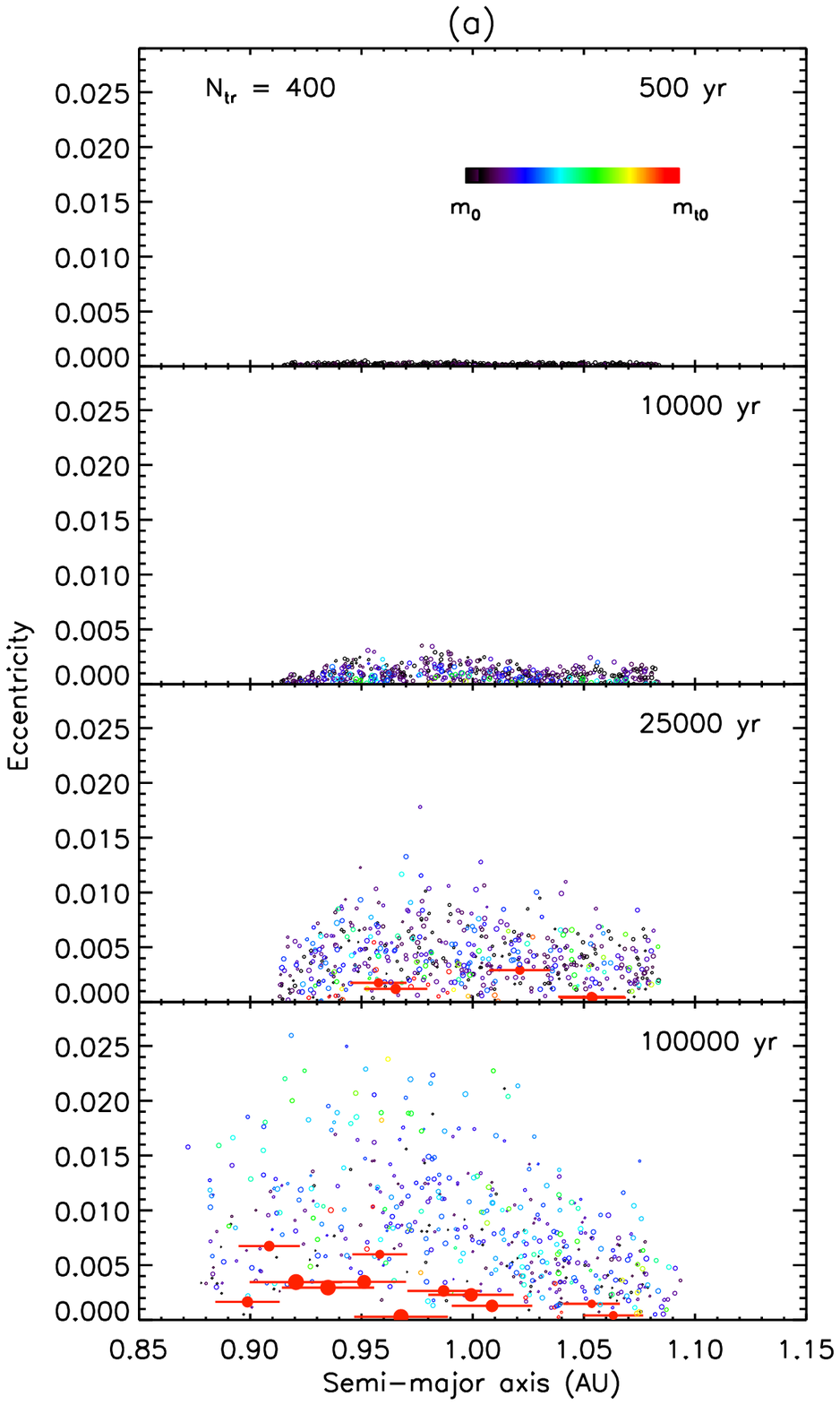}
\includegraphics[width=0.49\textwidth]{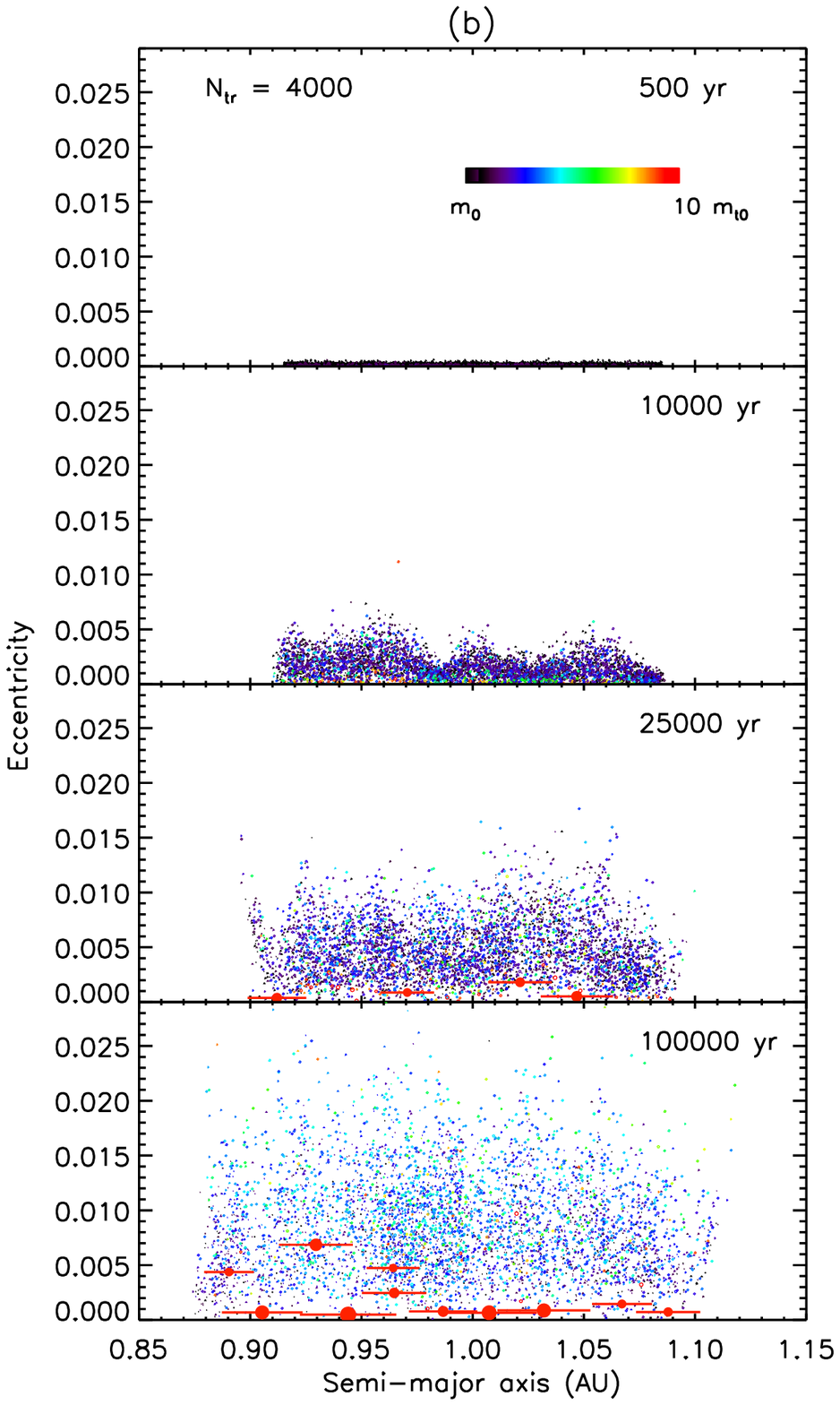}
\end{center}

Fig.~11. Snapshots on the $a-e$ plane for the hybrid simulations;  (a) $N_{\rm tr}$ =  400 and (b) $N_{\rm tr}$ = 4000.
Tracers are displayed as colored circles and their colors represent the mass of a planetesimal as shown by the color bar. 
Sub-embryos are displayed as red filled circles with a horizontal bar with a half length of 10 Hill radii
[for the panel (b), only sub-embryos more massive than $10 m_{t0}$ have a horizontal bar for consistency with the panel (a)].
A radius of a circle is proportional to $(k_im_i)^{1/3}$.

\end{figure}

\begin{figure}
\begin{center}
\includegraphics[width=0.49\textwidth]{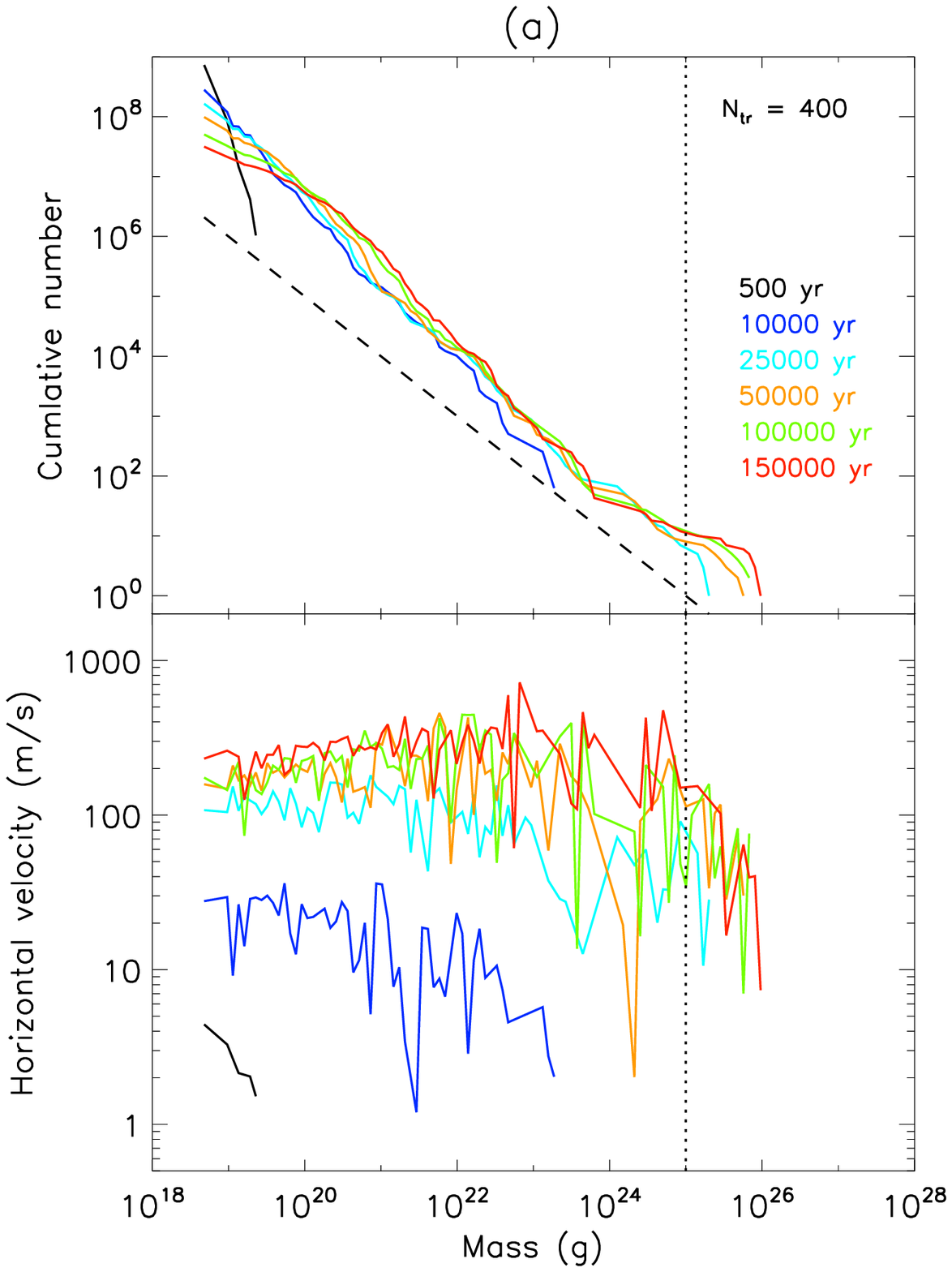}
\includegraphics[width=0.49\textwidth]{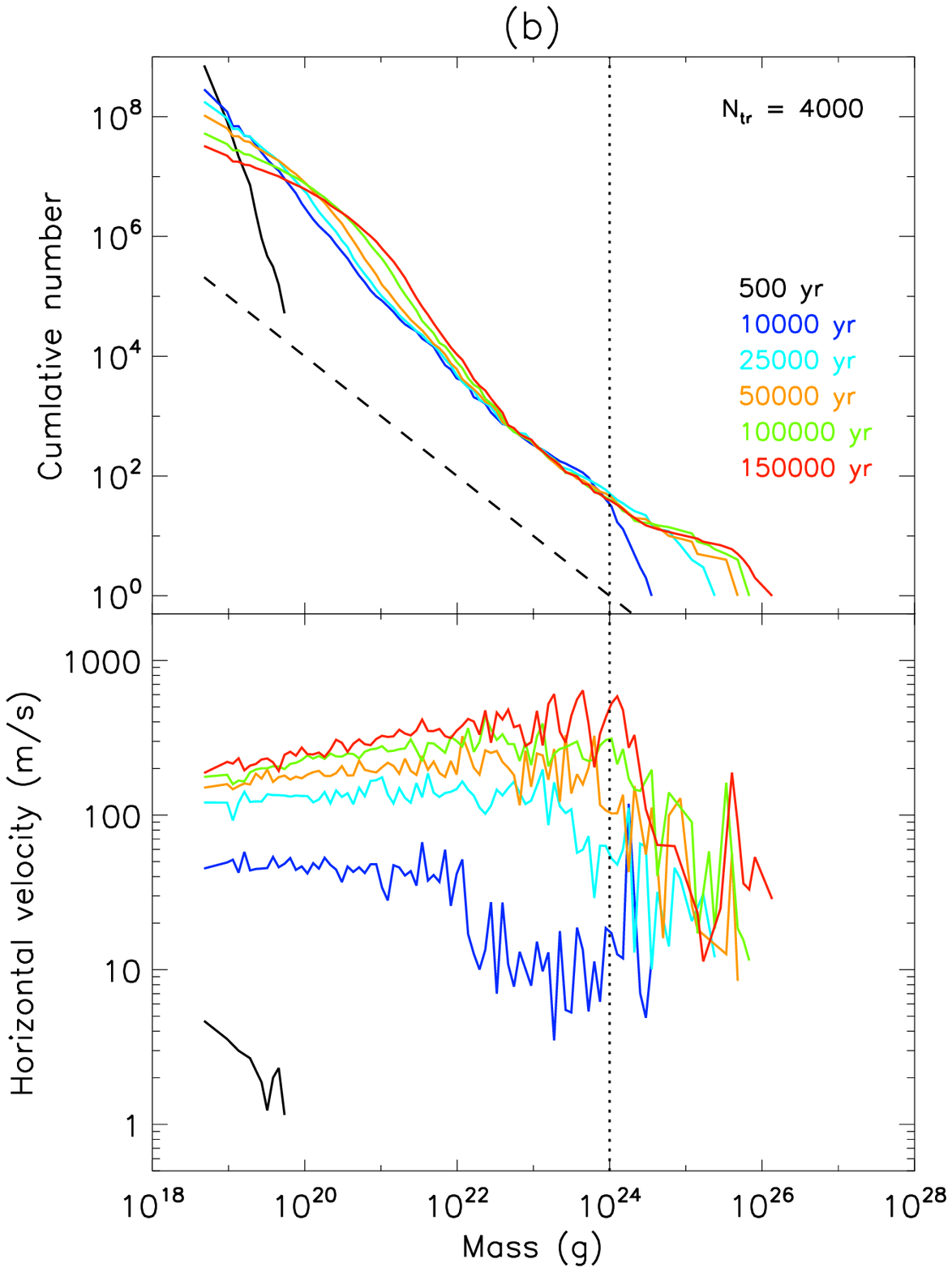}
\end{center}
Fig.~12. Snapshots of the cumulative number and the horizontal velocity ($=(5/8)^{1/2}e_iv_{\rm K}$, where $v_{\rm K}$ is the Keplerian velocity at 1AU)
for (a) $N_{\rm tr}$ =  400 and (b) $N_{\rm tr}$ = 4000.
The diagonal dashed line is the number of planetesimals represented by a single tracer with a mass of $m_{t0}$ for each run.
The vertical dotted line represents $m_{t0}$. 
For comparison, the results from (c) Wetherill and Stewart  (1993) and (d) Inaba et al. (2001) are shown; both from Fig.~10 of Inaba et al. (2001). 

\end{figure}

\begin{figure}
\begin{center}
\includegraphics[width=0.49\textwidth]{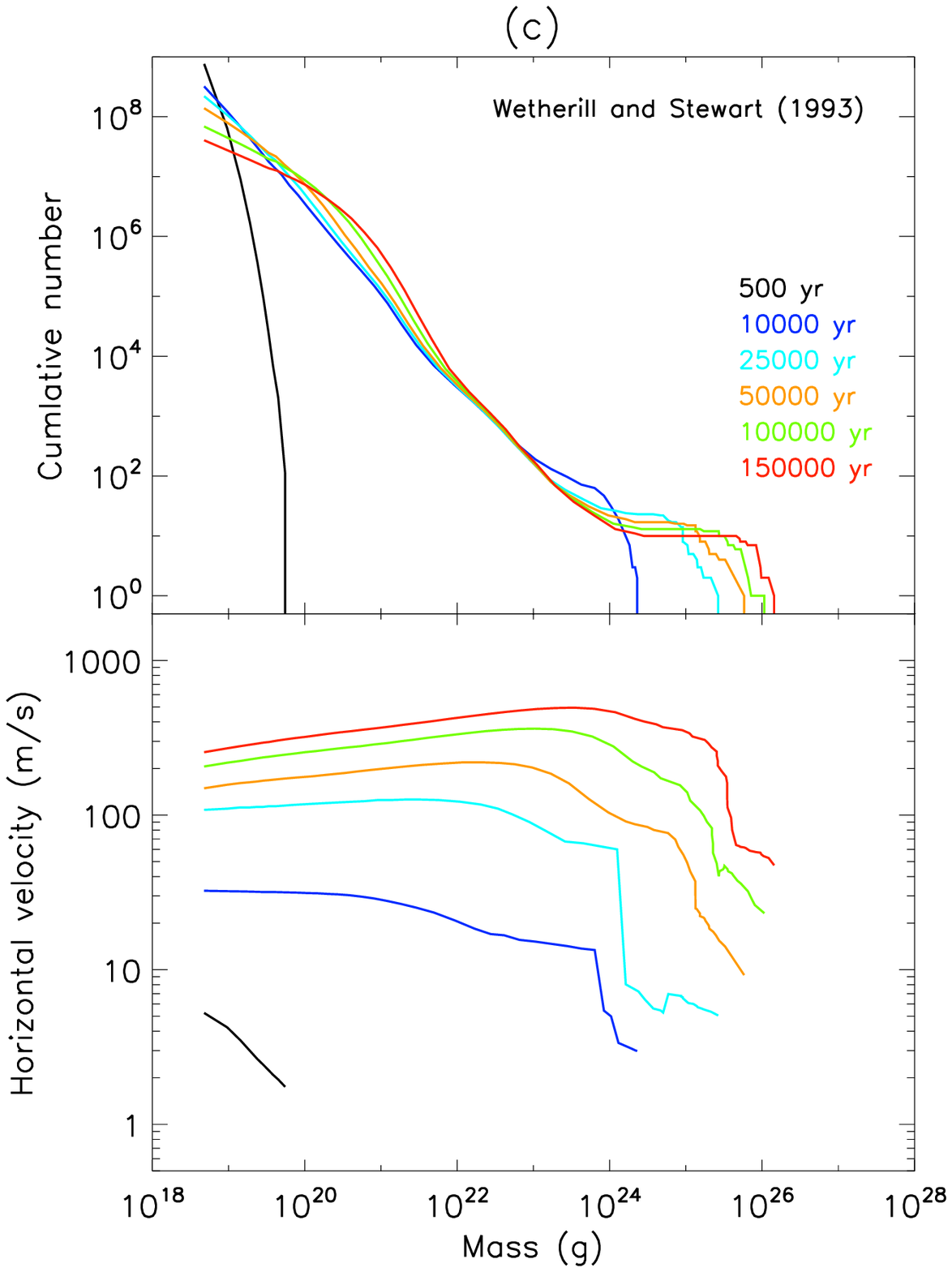}
\includegraphics[width=0.49\textwidth]{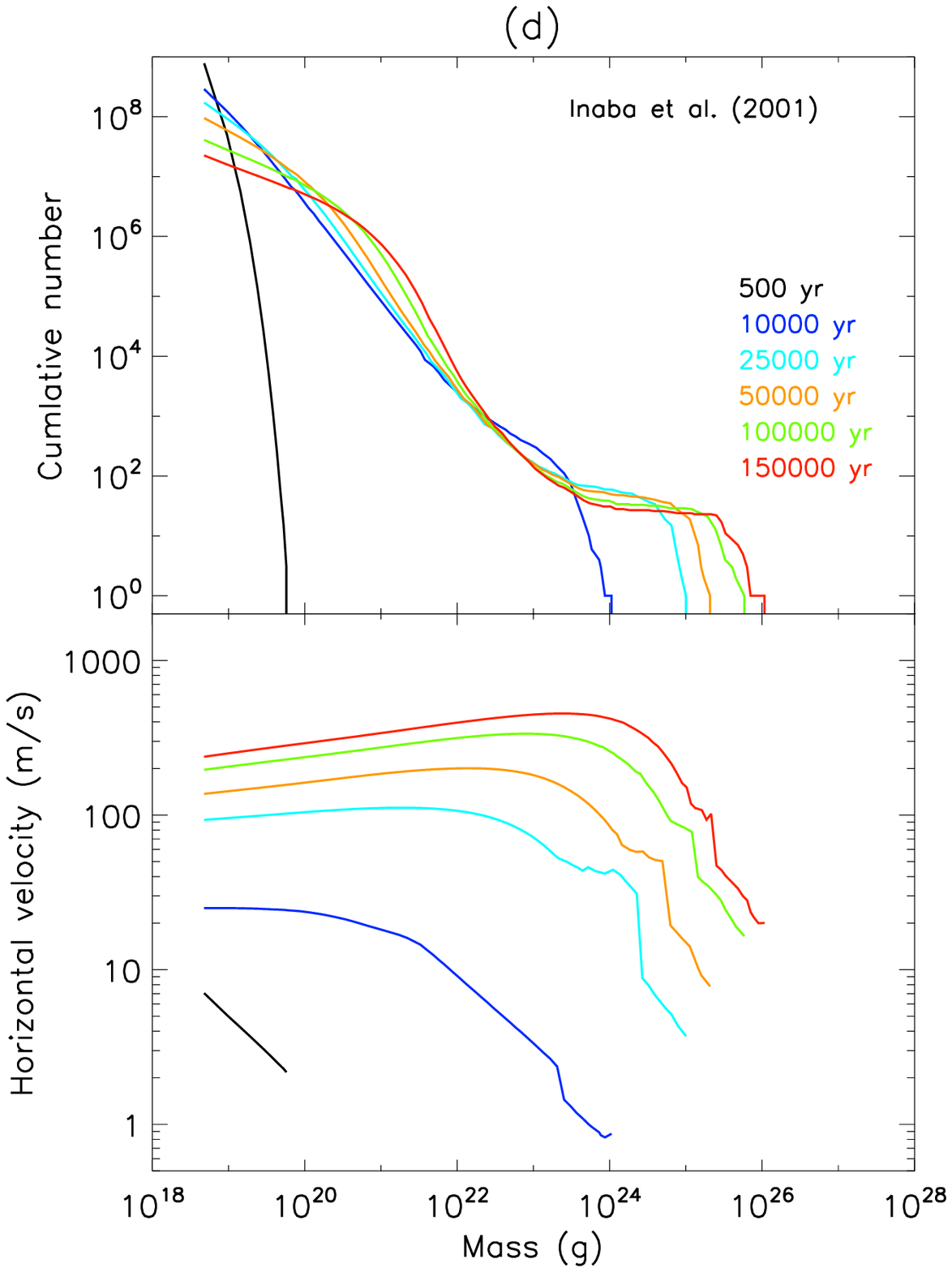}
\end{center}

 Fig.~12. Continued.
\end{figure}

Most of the authors who developed statistical planetary accretion codes (Weidenschilling et al., 1997; Kenyon and Luu, 1998; 
Inaba et al., 2001; Morbidelli et al., 2009; Ormel et al., 2010a; LDT12; Amaro-Seoane et al., 2014)  tested their codes using 
Fig.~12 of Wetherill and Stewart (1993) as a benchmark. 
We also perform simulations starting with the same condition of Wetherill and Stewart (1993).
Inaba et al. (2001) performed the same statistical simulations using their own coagulation code and the code of Wetherill and Stewart (1993)
after modifying settings of these two codes as close as possible.
Both results are shown in Fig.~10 of  Inaba et al. (2001) and we compare them with our results.
As the initial condition, $8.33 \times 10^8 $ planetesimals are placed between 0.915 AU and 1.085 AU.
Individual planetesimals have an equal-mass of $m_0 = 4.8\times 10^{18} $ g.
The nebular gas is included and the gas/solid ratio is 44.4. See Table~1 of  Wetherill and Stewart (1993) for more details. 
We take into account aerodynamic gas drag (Adachi et el., 1976)\footnote{We employ the gas drag coefficient of $C_{\rm D} = 0.5$ instead of 2.0 used in Morishima et al. (2010).} 
and tidal damping (Papaloizou and Larwood, 2000), but not Type I gas-driven migration. 
The gas disk model is described in Morishima et al. (2010) in detail.
In our simulations, the planetesimal disk radially expands with time, while the solid surface density remains constant 
in the previous statistical simulations. 
However, the effect of radial expansion is small on the timescale of the simulation ($\sim 10^5$ yr).

Figure~11 shows snapshots on the $a-e$ plane for the two hybrid simulations 
with different initial numbers of tracers, $N_{\rm tr} = 400$ and 4000.  
In the early stage of evolution ($t$ = 10,000 yr), 
large embryos grow very quickly and $e$ is enhanced locally where large embryos exist.
This stage is called the runaway growth stage.
Once  large embryos dominantly control the velocity of the entire planetesimal disk ($t = 25,000$ yr),
growth of embryos slows down (Ida and Makino, 1993). This stage is called the oligarchic growth stage.
Even though large embryos tend to open up gaps around their orbits,
the gaps are not cleared out because the embryos migrate quickly due to occasional 
short-distance planetesimal-driven migrations and embryo-embryo interactions and because other embryos tend to fill gaps.
During the oligarchic growth stage, 
the orbital separation between neighboring embryos is typically 10 Hill radii as found by Kokubo and Ida (1998, 2000).
Tanaka and Ida (1997) showed that gaps open up only if the orbital separation of neighboring embryos is larger than 20 Hill radii. 
Thus,  formation of clear gaps is unlikely to occur in this stage.

Figure~12 shows the mass and velocity distributions at different times 
for two hybrid simulations shown in Fig.~11.
The mass spectra of the hybrid simulations are smooth at the transition mass $m_{t0}$ at which a tracer is promoted to a sub-embryo
(see also Appendix~G).
For comparison, the results from Wetherill and Stewart (1993) and Inaba et al. (2001) are also shown.
The hybrid simulations agree well with these statistical simulations, although some small differences are seen.
In the hybrid simulations, the number of planetesimals represented by tracers has a lower limit shown by a dashed line in Fig.~12.  
Thus, the tail of the mass spectrum in the massive side during the runaway growth is not correctly described.
This causes a systematic delay in growth of large bodies, as already discussed by LDT12. 
The delay is particularly clear in the low-resolution hybrid simulation at 10,000 yr, although the delay is small. 
This resolution effect should be seen even in the high resolution hybrid simulation while
we see a good agreement with the statistical simulations at 10,000 yr. 
This is probably because our stirring routine slightly underestimates eccentricities and inclinations 
(Section~3.2) and this effect cancels out with the resolution effect to some degree.
Once the growth mode reaches oligarchic growth ($t > $ 25,000 yr), 
the masses of the largest bodies in the low resolution simulation almost catch up with those in the high resolution simulation and those in
the statistical simulations.



\subsection{Accretion of terrestrial planets from narrow annuli}

\begin{figure}
\begin{center}
\includegraphics[width=0.49\textwidth]{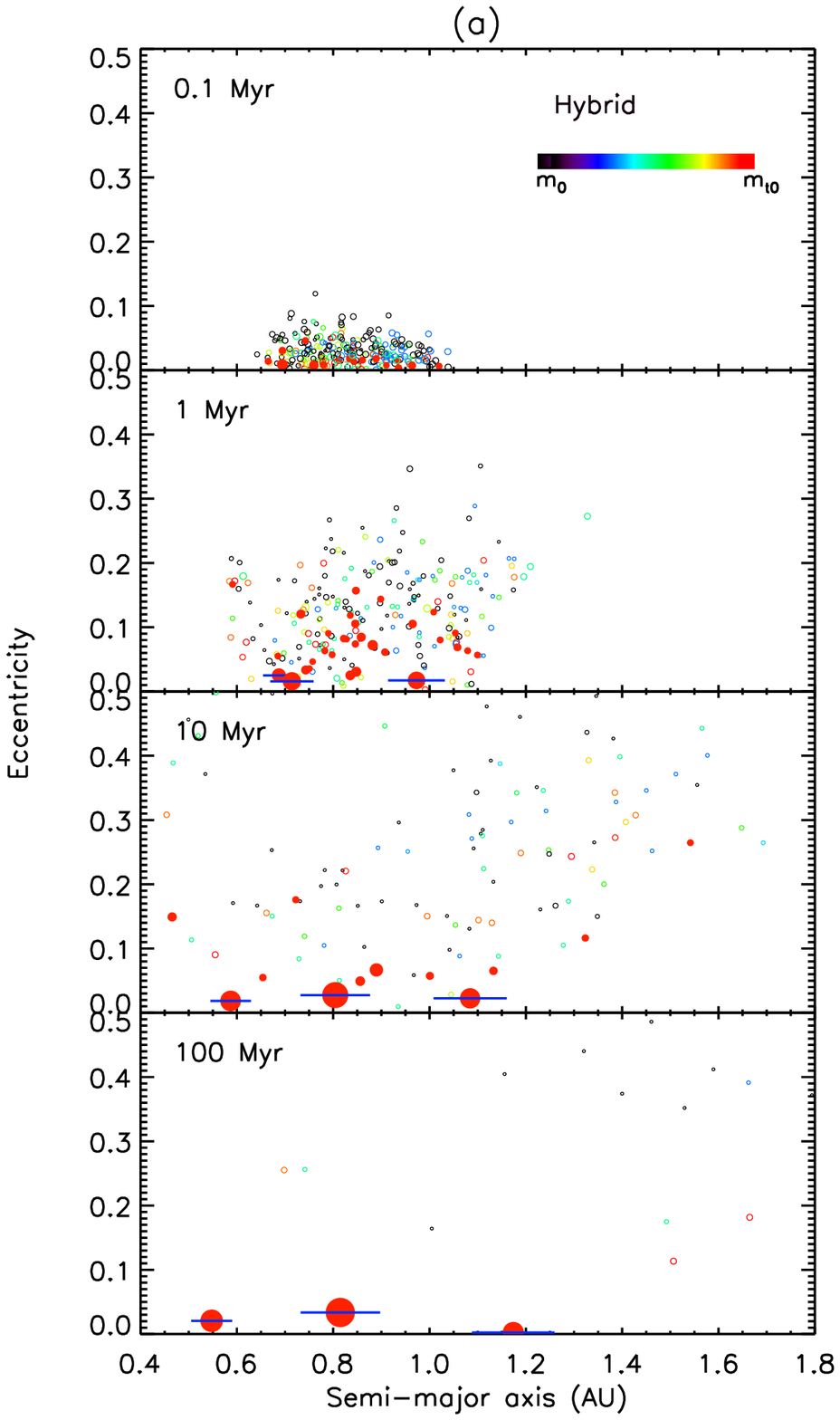}
\includegraphics[width=0.49\textwidth]{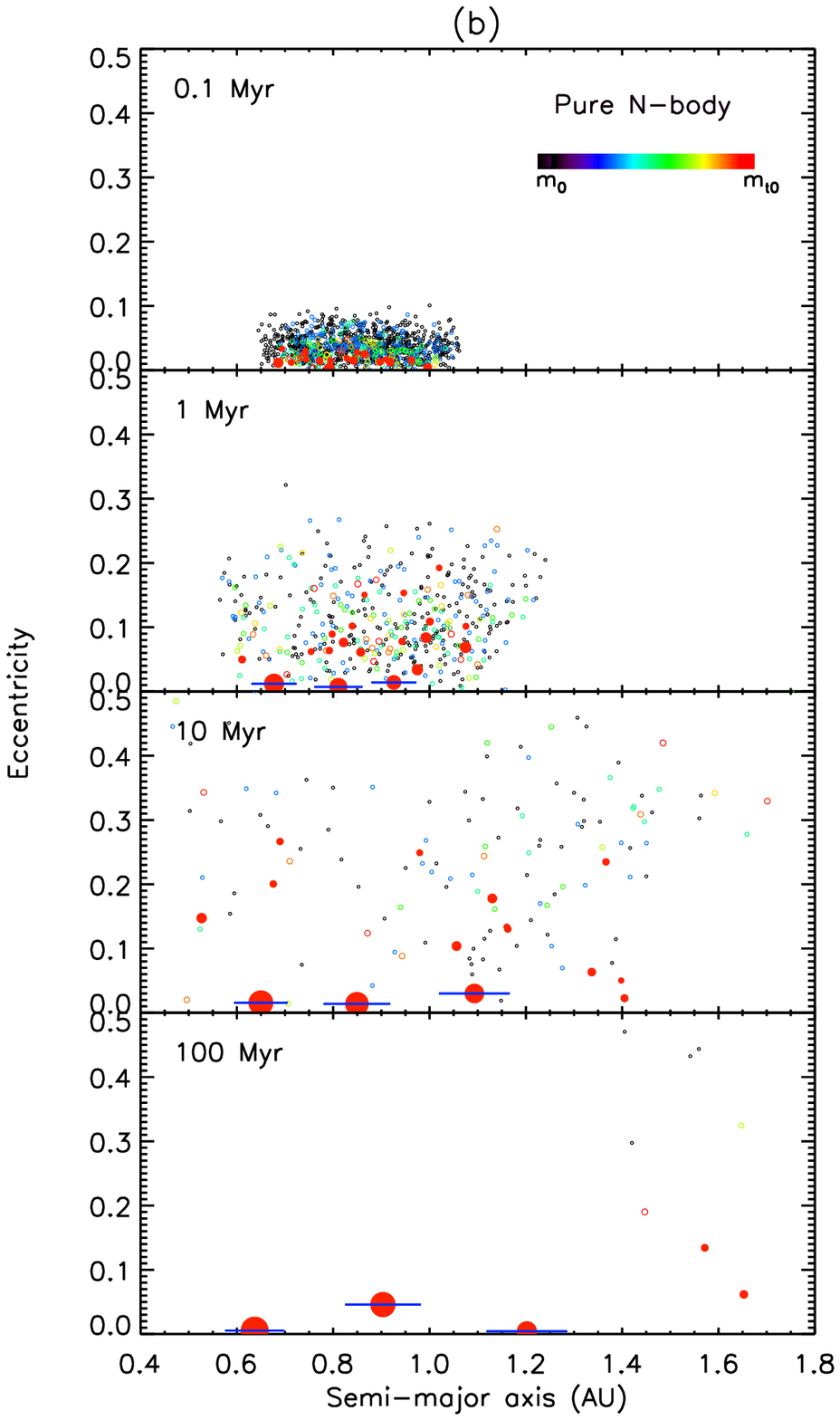}
\end{center}
Fig.~13. Snapshots on the $a-e$ plane for the simulations 
starting from equal-mass 2000 planetesimals 
in a narrow annulus between 0.7 AU and 1.0 AU:
(a) the hybrid simulation starting with 200 tracers and (b) the pure $N$-body simulation.
In the panel (a), tracers are displayed as colored circles, and the colors represent the mass of planetesimal. 
Sub-embryos are displayed as red filled circles. Sub-embryos more massive than 
$10 m_{t0}$ $(= 0.1 M_{\oplus})$ have a horizontal blue bar with a half length of 10 Hill radii.
The same color code is used in the pure $N$-body simulation.
A radius of a circle is proportional to $(k_im_i)^{1/3}$ (for the pure $N$-body simulation, $k_i = 1$ for all particles).

\end{figure}

\begin{figure}
\begin{center}
\includegraphics[width=0.8\textwidth]{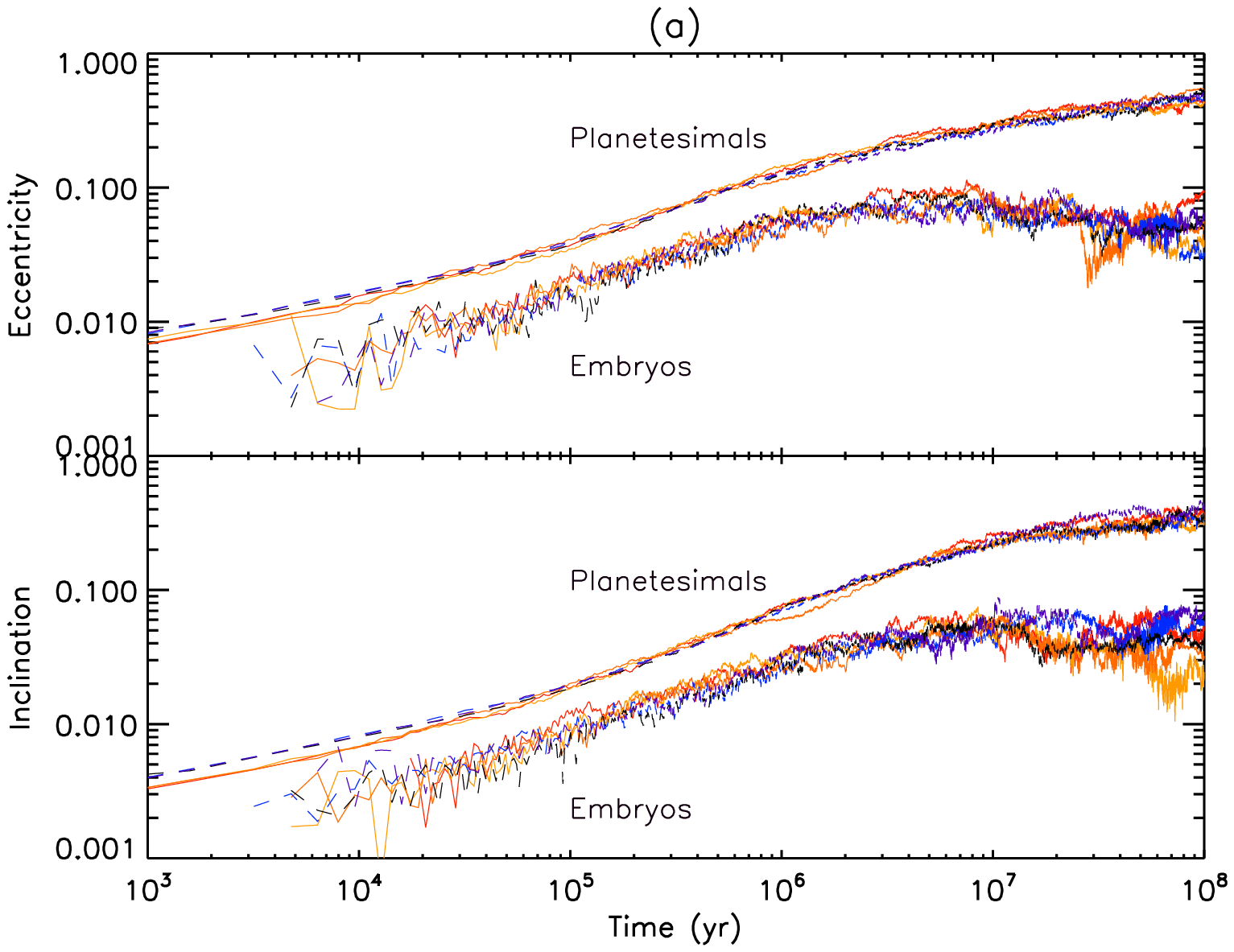}
\includegraphics[width=0.8\textwidth]{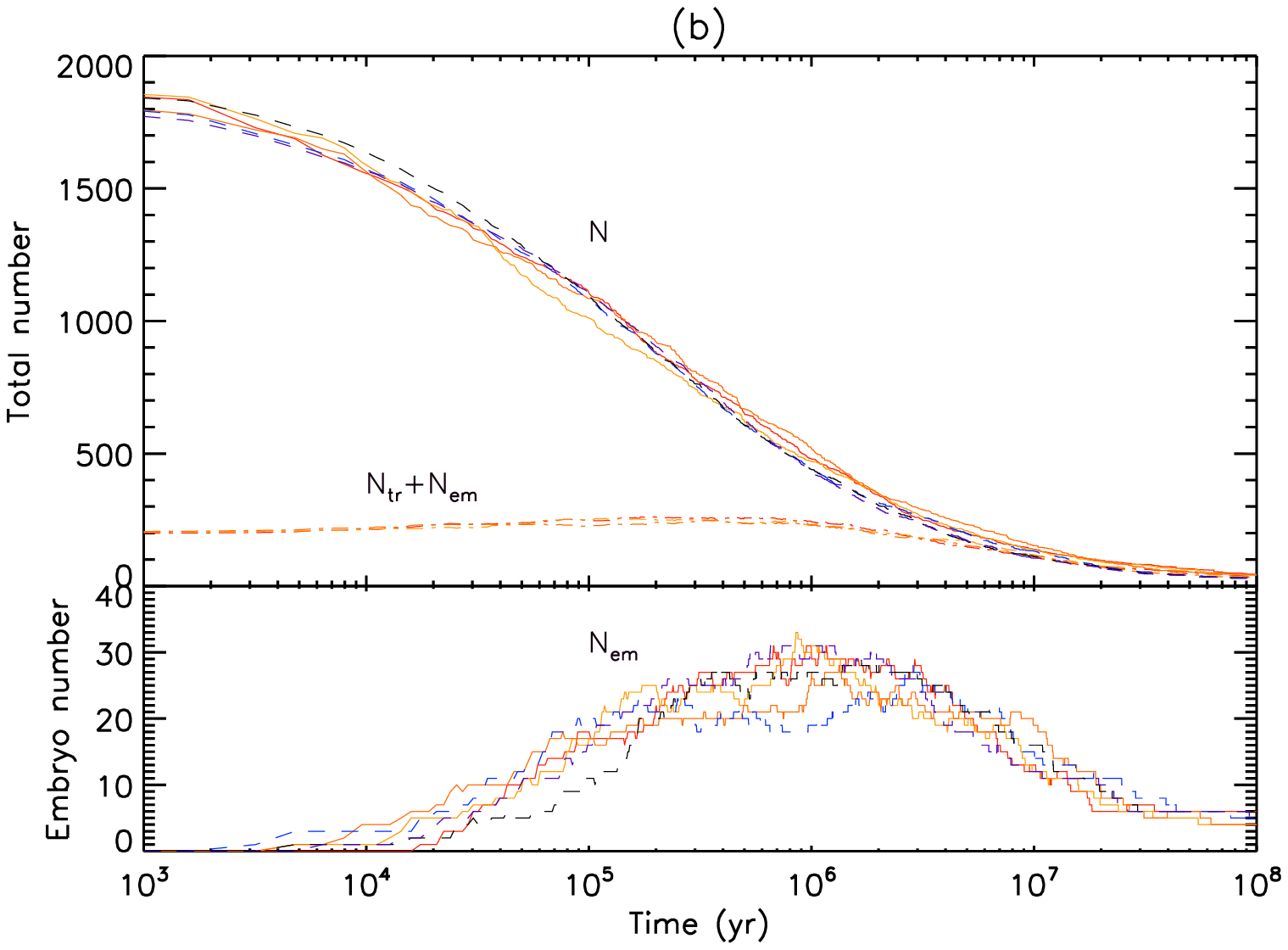}
\end{center}
Fig.~14. 
\end{figure}

\begin{figure}
\begin{center}

\includegraphics[width=0.8\textwidth]{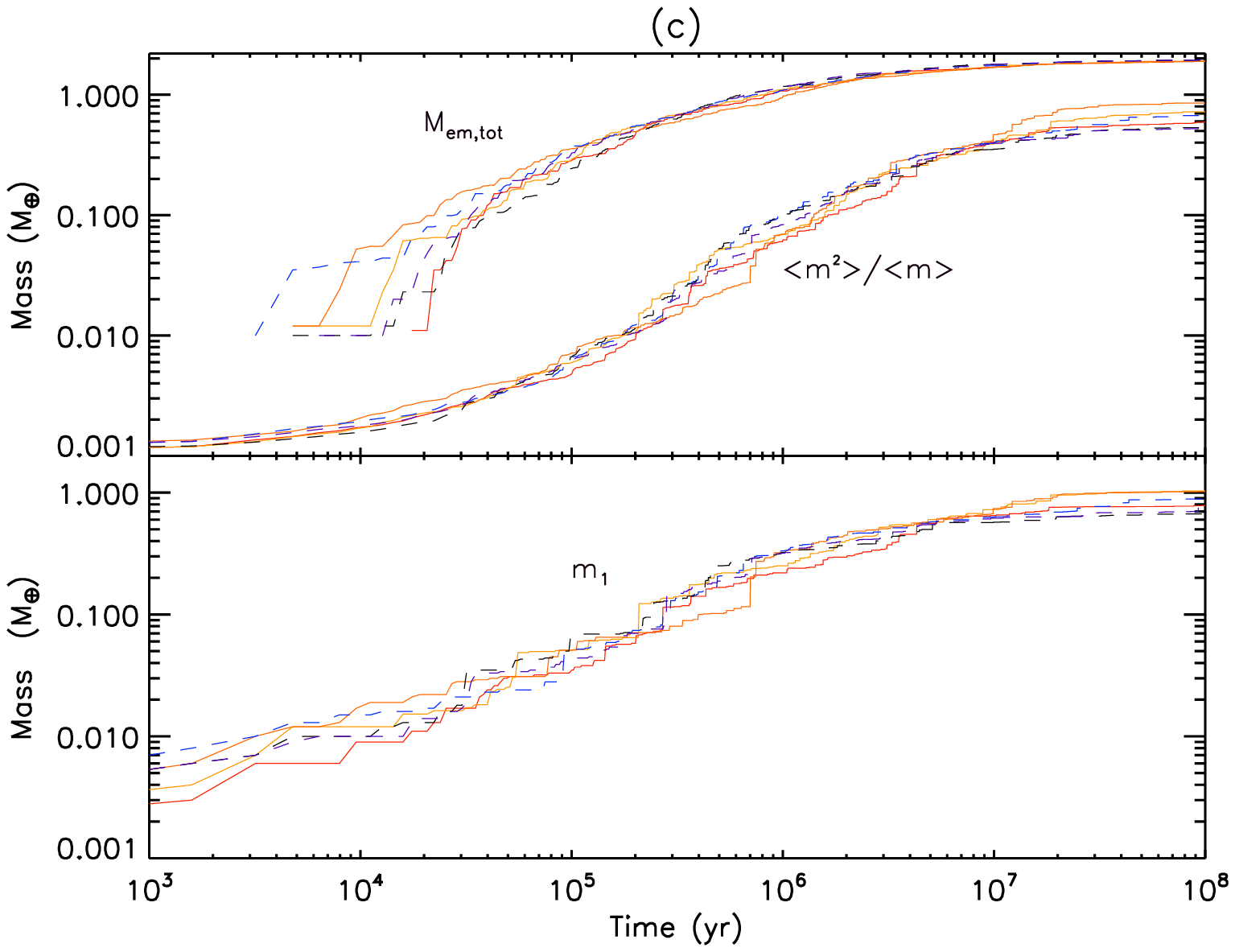}
\end{center}
Fig.~14. Time evolutions of various quantities for three hybrid simulations (orange, brown, and red solid curves) 
and three pure $N$-body simulations (blue, purple, and black dashed curves):
(a) the mass-weighted root-mean-square eccentricity and inclination,
$(\langle me^2\rangle/\langle m \rangle)^{1/2}$ and $(\langle mi^2\rangle/\langle m \rangle)^{1/2}$, for embryos and planetesimals, 
(b) the total number of all bodies $N$, the total number of embryos $N_{\rm em}$, the total number of tracers $N_{\rm tr}$, 
(c) the total mass of embryos $M_{\rm em,tot}$, the effective mass of the system $\langle m^2 \rangle/\langle m \rangle$,
and the mass of the largest body $m_1$.

\end{figure}

The last section focused only on the early stages of planetary accretion. 
Further, the damping effects due to the gas disk make it difficult to judge whether 
the code works appropriately.
Here we test our code for the entire stages of accretion without the gas disk.
We perform hybrid simulations of  accretion of terrestrial planets starting from 
a narrow annulus, similarly to those adopted in Morishima et al. (2008) and Hansen (2009), until 
completion of accretion ($t \sim 100$ Myr).
As the initial condition, 2000 equal-mass planetesimals ($m_0 = 10^{-3}M_{\oplus}$) are placed between 
0.7 AU to 1.0 AU, and the total disk mass is $2.0 M_{\oplus}$. 
The initial number of tracers is 200.
For comparison, we also perform pure $N$-body simulations with 
the same initial condition using pkdgrav2. 
A hybrid simulation takes about three days using a single core 
whereas a pure $N$-body simulation takes a few weeks using eight cores.
As found in Morishima et al. (2008), the evolution is almost deterministic due to efficient dynamical friction 
if the initial number of planetesimals is large enough ($ N > 1000$).
If the initial $N$ is small while keeping the same total mass, 
the evolution becomes rather stochastic and the final systems have relatively large diversity (Hansen, 2009). 

An example of the hybrid simulations is shown in Fig.~13 together with an example of the pure $N$-body simulations.  
Through the runaway growth stage, tens of embryos form ($t = 0.1$ Myr).  With increasing masses of embryos, 
mutual orbital crossings and impacts occur, and a few distinctively large embryos emerge ($t $= 1 Myr). In the late stage,  
the largest embryos sweep up small remnant planetesimals ($t \ge 10$ Myr).  At the end of accretion, the largest three embryos 
have orbital eccentricities and inclinations as small as those for the present-day Earth and Venus due to dynamical friction
of remnant bodies ($t = 100$ Myr). 
In the hybrid simulation, all the largest embryos are sub-embryos until the end of the simulation.
We perform three hybrid simulations and three pure $N$-body simulations in total, and 
all simulations show similar results, except one hybrid simulation produces only two distinctively large embryos. 

To see the accretion history more in detail, we plot various quantities as a function of time in Fig.~14. 
Overall agreements between the hybrid simulations and the pure $N$-body simulations are excellent.
Good agreements between the hybrid and pure $N$-body simulations suggest that our statistical routine works 
well even in the late stages of planetary accretion.
Figure~14a  shows the evolution of the mass-weighted root-mean-square eccentricity and inclination,
$e_{\rm m} = (\langle me^2\rangle/\langle m\rangle)^{1/2}$ and $i_{\rm m} = (\langle mi^2\rangle/\langle m\rangle)^{1/2}$, 
for embryos and planetesimals.
With increasing time, $e_{\rm m}$ and $i_{\rm m}$ for both embryos and planetesimals
increase primarily due to viscous stirring of largest bodies. 
After $\sim$ 10 Myr, $e_{\rm m}$ and $i_{\rm m}$ of embryos starts to decrease 
because small embryos with large $e$ and $i$ collide with large embryos with low $e$ and $i$ .
Figure~14b shows the evolution of the total number of all bodies $N$ and the total number of embryos $N_{\rm em}$.
The number $N_{\rm em}$ has a peak around 1 Myr both in the hybrid and pure $N$-body simulations.
The number of bodies which orbits are numerically integrated, $N_{\rm tr} + N_{\rm em}$, 
increases by $\sim 30 \%$ at maximum and the maximum value is taken around 1 Myr (Fig.~14b).

Figure~14c shows the total mass of embryos, $M_{\rm em,tot}$, the effective mass of the system $\langle m^2 \rangle/\langle m \rangle$,
which is sensitive to masses of largest embryos, and the mass of the largest body, $m_1$.  
It is found that 
the values of $\langle m^2 \rangle/\langle m \rangle$ and $m_1$ in the hybrid simulations are slightly lower than those 
in the pure $N$-body simulations at $t \sim 1$ Myr. 
During the giant impact stage roughly between $t$ = 0.2 Myr and a few Myr, 
the contribution of embryo-embryo impacts to mass gain of embryos is significant, comparably to embryo-planetesimal impacts.  
The embryo-embryo collision rate sensitively depends on $e$ and $i$
of embryos. The hybrid simulations give slightly higher $e$ and $i$ (Fig.~14a) and thus lower growth rates of embryos 
than those in the pure $N$-body simulations during the giant impact stage. 
This is likely to be a resolution effect. In the low resolution hybrid simulations, the number of nearby tracers around embryos fluctuates 
and that makes dynamical friction somewhat noisy. We also perform the same simulations 
but starting with a larger number of tracers ($N_{\rm tr}$ = 400) and 
find better agreements in $\langle m^2 \rangle/\langle m \rangle$ with the pure $N$-body simulations.

\subsection{Accretion of cores of jovian planets}

\begin{figure}
\begin{center}
\includegraphics[width=0.65\textwidth]{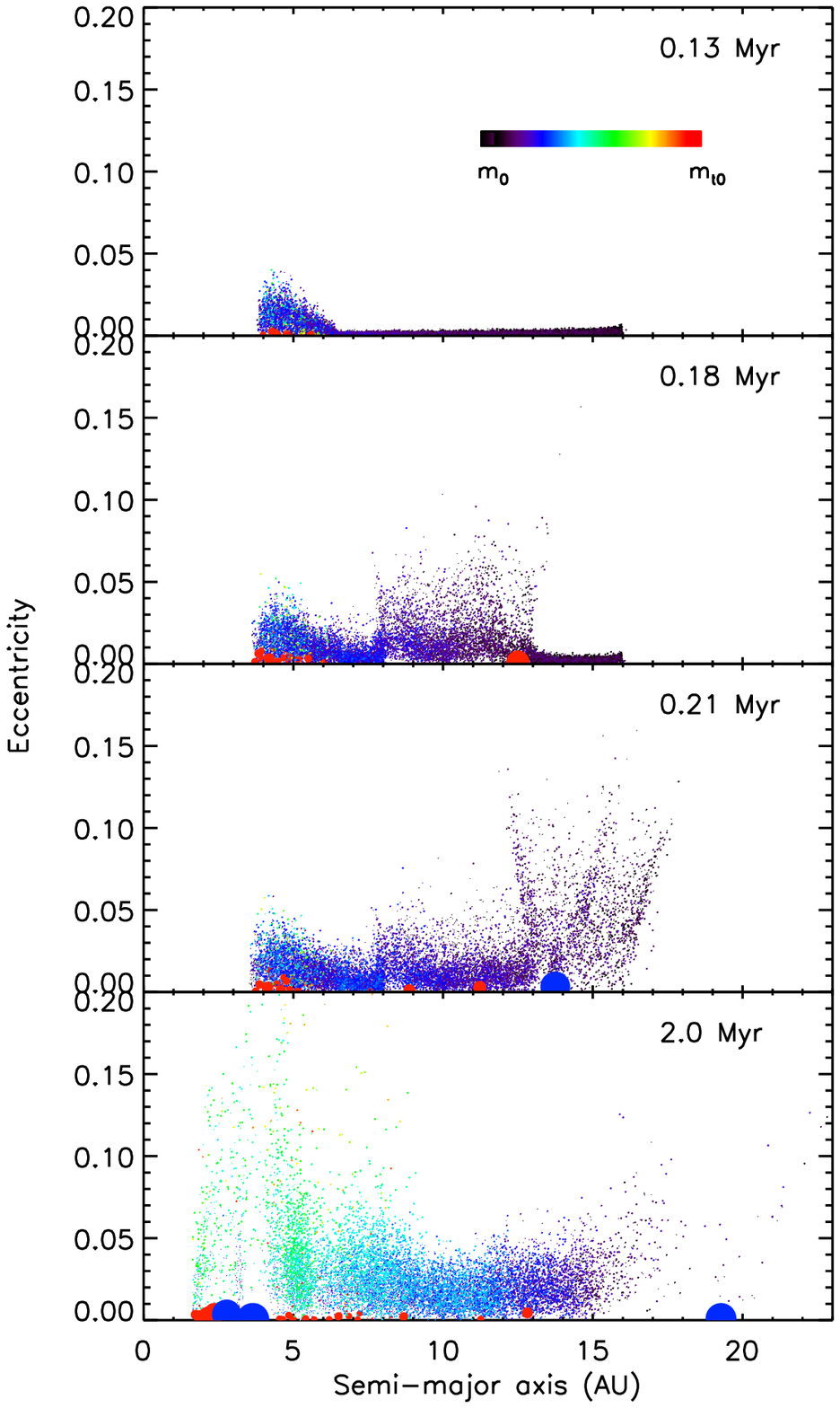}
\end{center}

Fig.~15. Snapshots on the $a-e$ plane for the hybrid simulation in the jovian planet region.
The simulation starts with equal-mass $10^{13}$ planetesimals.
Tracers are displayed as colored circles and colors represent the mass of a planetesimal.
Sub-embryos ($0.02 M_{\oplus} \le  m_i < 2 M_{\oplus}$) are displayed as red filled circles and full-embryos ($m_i \ge 2 M_{\oplus}$)
are displayed as blue filled circles. 
A radius of a circle is proportional to $(k_im_i)^{1/3}$, but enhanced by a factor of 6 for embryos for good visibility.
\end{figure}

\begin{figure}
\begin{center}
\includegraphics[width=0.69\textwidth]{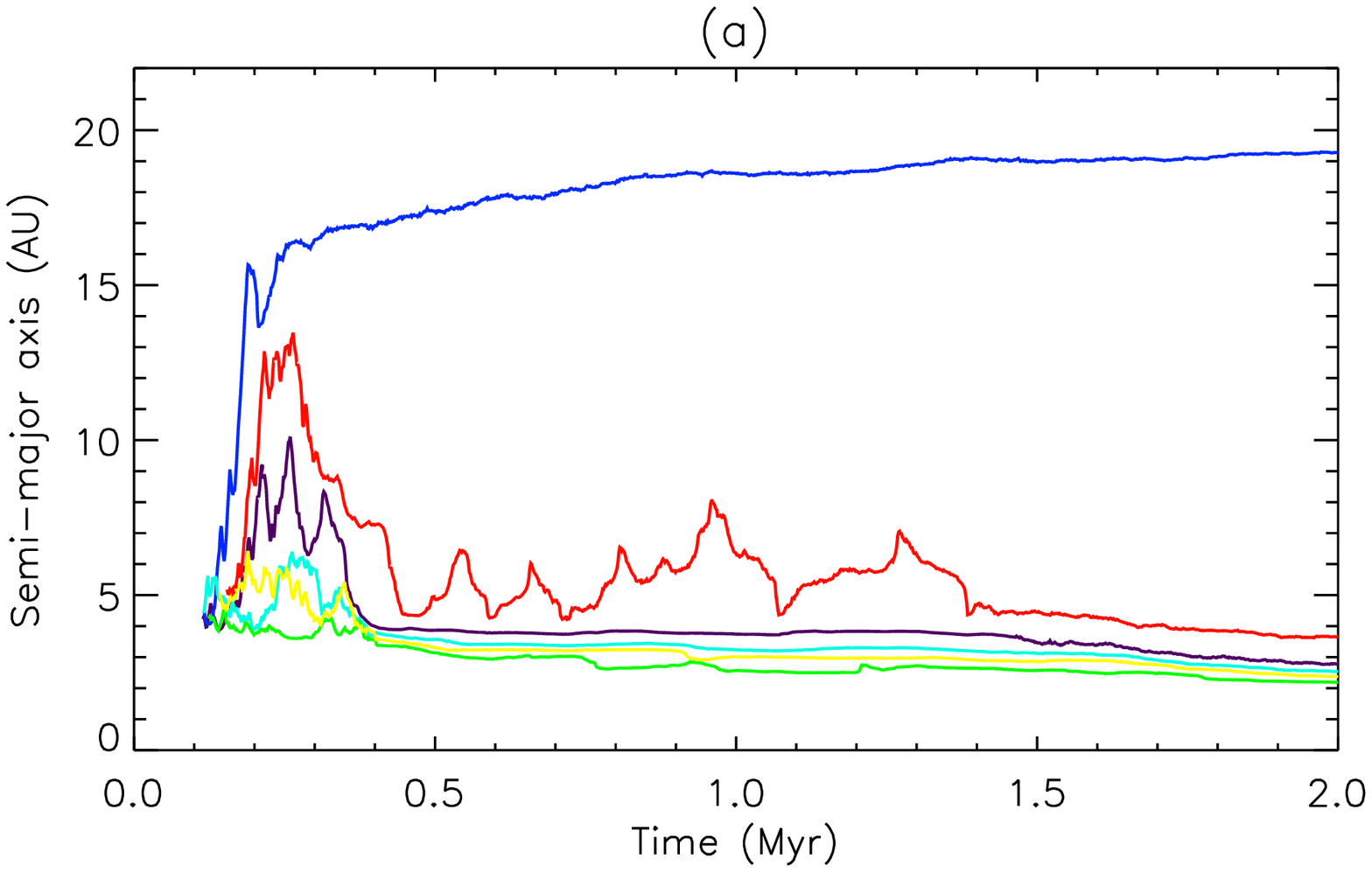}
\includegraphics[width=0.69\textwidth]{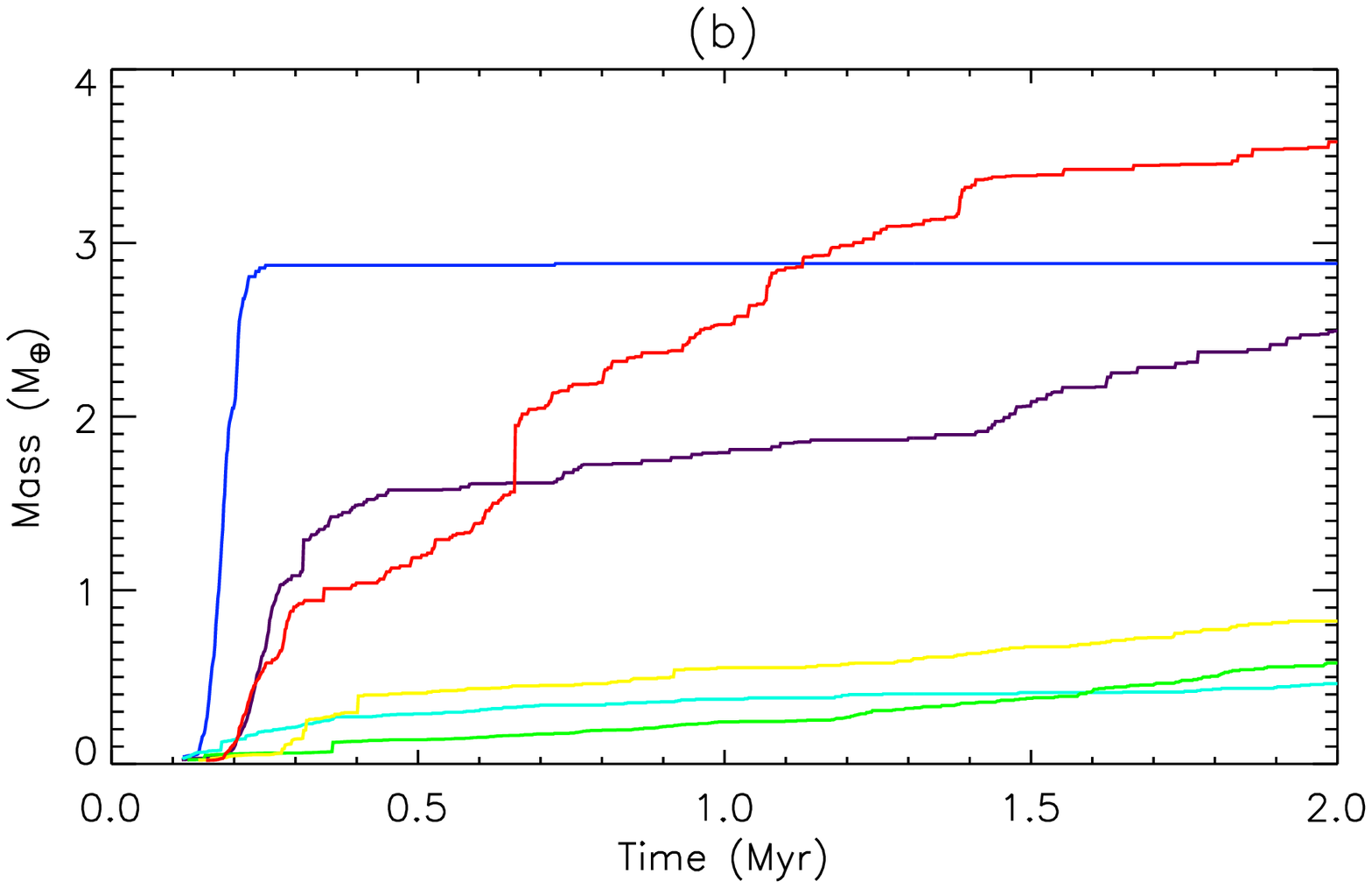}
\end{center}

Fig.~16. Time evolutions of (a) semimajor axes and (b) masses of embryos that are more massive than 0.4$M_{\oplus}$
at 2 Myr. The same color is used for the same embryo in both panels.
The embryos represented by blue, red, and purple curves are called Embryo 1, 2, and 3, respectively,
in the main text.

\end{figure}
The final test we present is accretion of cores of jovian planets starting with a wide planetesimal disk.
Levison et al. (2010) and Capobianco et al. (2011) showed importance of self-sustained outward migration of embryos, 
as they can grow rapidly via collisions with little-perturbed outer bodies.
However, these authors placed large embryos in the disk of small planetesimals in the initial conditions, and 
it is not clarified if outward migration plays a significant role in more realistic accretion scenarios. 

We set the initial condition similar to that employed by Levison et al. (2010), but without any embryos.
We place $10^{13}$ equal-mass planetesimals between 4 AU and 16 AU represented by 10000 tracers. 
The initial planetesimal mass $m_0$ is $1.19 \times 10^{17}$ g and the radius is 2.4 km.
The total mass of planetesimals is 200 $M_{\oplus}$ and this gives the solid surface density about seven times the minimum-mass solar nebula  (Hayashi, 1981).
The simulation is performed up to 2 Myr and it takes roughly five days to complete using a single core.
The nebula gas effects are included in the same manner as the simulations in Section~3.4, 
and the gas-to-solid ratio is 56.7, which is the same as the minimum-mass solar nebula 
beyond the snow line (Hayashi, 1981). We do not consider enhanced cross sections of embryos due to their atmospheres (Inaba and Ikoma, 2003).
The nebular gas is assumed to dissipate exponentially on the timescale of 2 Myr.
Capobianco et al. (2011) showed that outward migration of an embryo is usually triggered 
in the nebular gas, if the planetesimal radius is in a certain range. 
The planetesimal radius we employ is in this range in the region inside of $\sim$10 AU [see their Eq.~(21)].
Note that, however, planetesimal radii increase with time in our simulation unlike those in Levison et al. (2010) and Capobianco et al. (2011).

The snapshots on the $a-e$ plane are shown in Fig.~15 and the time evolutions of the masses and the semimajor axes of 
six largest embryos at the end of the simulation are shown in Fig.~16. 
Several embryos appear near the inner edge of the disk after 0.1 Myr. After some embryo-embryo interactions,
one of the embryos (Embryo 1) starts self-sustained outward migration. Since Embryo 1  encounters with 
unperturbed planetesimals on its way, it grows very rapidly from  0.1 $M_{\oplus}$ to 3 $M_{\oplus}$ during only 50,000 yr.
After Embryo 1 reaches the outer edge, it turns its migration direction inward. However, it encounters with another 
smaller embryo (Embryo 2) and inward migration is halted. There, Embryo 1 opens up a gap and starts slow outward migration 
due to distant encounters with inner planetesimals.  Embryo 1 is promoted to a full-embryo at $t =$ 0.2 Myr during its inward migration.

Embryo 2 and another embryo (Embryo 3) also migrate outward during the outward migration of Embryo 1, and 
they also grow more rapidly than inner embryos, but not as rapidly as Embryo 1. 
They eventually change their migration directions inward and come back to near the inner edge.   
Embryo 2 occasionally attempts to migrate outward even after that, although its outward migration is not sustained. 
Two outward migrations at $t =$ 0.55 and 0.65 Myr are halted by encounters with 
another embryo at $\sim$ 7 AU, which merges with Embryo 2 at $t =$ 0.65 Myr. 
Embryo 2 is promoted to a full-embryo at this time.
We do not clearly understand what prevents Embryo 2 from subsequent outward migrations.
The radial profile of the surface density of planetesimals is no longer smooth due to occasional gap formation by embryos,
and the outward migrations of Embryo 2 seem to be halted at local low density regions.
Another possibility is that migrations of full-embryos are halted by stochastic effects since they are not massive enough as compared with tracers. 
This possibility may be tested using different transition masses from a sub-embryo to a full embryo. 
The masses of Embryo 2 and Embryo 3 respectively reach 3.6 $M_{\oplus}$ and 2.5 $M_{\oplus}$
at 2 Myr. Their growth is slow at this time as they open up a gap in the planetesimal disk.


Obviously, we need to perform more simulations with different parameter sets 
and conduct more careful analysis.  Nevertheless, the example we showed here is a good demonstration that 
our code can produce rapid growth of embryos during their outward migrations.

Ngo (2012) \footnote{His master thesis (Ngo, 2012) together with mpeg animations is available at 
https://qspace.library.queensu.ca/handle/1974/7400.} performed simulations of accretion of cores of jovian planets using the LIPAD code.
The initial conditions of his simulations are similar to ours.
In his simulations without fragmentation (his Figs.~5.14 and 5.15),
sub-embryos are almost radially stationary even though the routine of sub-embryo migration is included. 
Once sub-embryos are promoted to full-embryos, they suddenly start rapid radial migrations.
It is unclear to us why such large differences between sub-embryos and full-embryos are seen in his simulations.
In spite of the large difference of embryo migration between his and our simulations, the final masses of the largest 
embryos are close to each other.

\section{Discussion}
\subsection{Computational load}
The cpu time $T_{\rm cpu}$ per timestep of orbital integration as a function of the number of embryos $N_{\rm em}$ and the number of 
tracers $N_{\rm tr}$ is roughly given as 
\begin{equation}
T_{\rm cpu} = (N_{\rm em} + N_{\rm tr}) \left(C_{\rm Kep}+ C_{\rm grav}N_{\rm em} + C_{\rm sta}N_{\rm int}\frac{\delta t}{\Delta t}\right), \label{eq:tcpu}
\end{equation}
where $C_{\rm kep}$, $C_{\rm grav}$, and $C_{\rm sta}$ are the cpu times for a single orbital integration (a Kepler drift),  
for a single direct gravity calculation of a pair in the $N$-body routine, and for a single set of calculations for a pair in the statistical routine,
and $N_{\rm int}$ is the typical number of interlopers in the neighboring search region (Fig.~2).
For a standard latest computer,  $C_{\rm kep} \sim 4 \times 10^{-7}$ s. 
From some of simulations,  we find that $C_{\rm grav}/C_{\rm kep} \sim 0.15$ and $C_{\rm sta}/C_{\rm kep} \sim 0.3$. 

As seen in Eq.~(\ref{eq:tcpu}), the load on the statistical routine is controllable by modifying $N_{\rm int}$ and $\Delta t/\delta t$. 
Since the radial width of the neighboring search region is 
given by Eq.~(\ref{eq:drt0}), $N_{\rm int} \propto N_{\rm tr}^{2/3} \delta \theta$ for a fixed surface density.
If we choose an azimuthal width as $\delta \theta \propto N_{\rm tr}^{-2/3}$, $N_{\rm int}$ becomes independent of $N_{\rm tr}$
and the cpu time for the statistical routine can be linearly scaled with $N_{\rm tr}$. 
Practically, one may choose a $\delta \theta$ so that 
the load on the statistical routine does not exceed the load on the $N$-body routine. 


The actual computational load on the statistical routine in some of our simulations is as follows.
Recall that in the present paper, we call the statistical routine every 30 steps of orbital integrations ($\Delta t = 30 \delta t$). 
If the time step of the statistical routine is halved, its load is doubled.  
In the hybrid simulations shown in Fig.~5 ($N_{\rm tr} = 200$ and $N_{\rm int} \sim 10$), 
the load on the statistical routine is only 10 \% relative to the total load.
The rest of 90 \% is on the $N$-body routine and more than half of it is occupied by the Kepler equation solver.
There is no direct gravity calculation in these simulations, as no embryo is included.
In the hybrid simulation shown in Fig.~15 ($N_{\rm tr} = 10000$ and $N_{\rm int} \sim 100$), 
the load on the statistical routine is about 60 \% before embryos appear.
If the number of embryos is larger than about 10, the load on the direct gravity calculation exceeds that for the Kepler equation solver,
and the relative load on the statistical routine decreases.  
At 1.6 Myr of Fig.~15, at which 21 embryos exist, the load on the statistical routine is about 45 \%, that on the direct gravity calculation is 
35 \%, and that on the orbital integration is 20 \%. 
Overall, the load of the statistical routine is comparable to or less than that on
the $N$-body routine.

We also comment on parallelization. The current version of our code is serial. However, the direct $N$-body routine is 
already parallelized and we test how much simulations are accelerated by parallelization in case without the statistical routine.
In the tests, we fix the number of embryos to be 10 and vary the number of tracers. We find that simulations
are accelerated only if the number of tracers is more than $\sim 10^5$. 
In contrast, the pure $N$-body simulations using the tree code, pkdgrav2, are accelerated by parallelization if $N > 10^3$ (Morishima et al., 2010).
The difference comes from the cpu times per step. To be benefitted by parallelization, the cpu time needs to be somewhat long
to ignore latency. 
The computational cost per step for hybrid simulations is as small as those for $N$-body simulations without particle-particle interactions.
That ironically makes parallelization inefficient. 

\subsection{Limitations of the code and future work}

We consider only random walk of tracers in tracer-tracer interactions.
Thus, self-sustained migration occurs only to embryos in our code although 
small bodies ($< m_{t0}$) may migrate in reality. To examine if migrations of 
small bodies play an important role for overall accretion history of embryos,
high resolution simulations (with smaller $m_{t0}$) are necessary. 
Minton and Levison (2014) showed that 
self-sustained migration occurs only if the embryo is much more massive
than surrounding planetesimals and sufficiently separated from other embryos. 
Such conditions are usually fulfilled near the transition from the runaway growth stage to the oligarchic growth stage.
The mass of the largest embryo at the transition is roughly given by $m_{\rm iso}^{4/7} m_0^{3/7}$ 
(Ormel et al., 2010b; Morishima et al., 2013), where $m_{\rm iso}$ is the isolation mass and $m_0$ is the initial mass of planetesimals.
Therefore, if $m_0$ is very small (e.g., Weidenschilling, 2011), 
a very large number of tracers is necessary to represent 
an embryo at the transition by a single particle in our code.
However, Minton and Levison (2014) also pointed out that the growth time scale needs 
to be longer than the migration time for an embryo to avoid encounters with other embryos during migration. 
If $m_0$ is very small,  this criterion is not generally fulfilled for the largest embryo at the transition
due to a very short growth time scale.
Thus,  masses of candidate embryos for self-sustained migration are 
probably not too small to be practically resolved by our code even if 
$m_0$ is very small.
The argument here, of course, depends on disk parameters and radial location and 
needs more careful estimates. 

Our code can probably handle eccentric/inclined ringlets that are usually produced by 
external massive bodies (see Section~2.3). 
However, we have not tested such simulations in this paper, 
and they remain for future work. A potential issue of our code in handling 
eccentric/inclined ringlets is accuracy of the global gravity forces (Section~2.7).
Only for this calculation, we assume that a ring is axisymmetric and symmetric 
with respect to the invariant plane.
These assumptions may not be appropriate if the radial/vertical thickness or the ringlet 
is much smaller than the radial/vertical excursion of individual particles. 
This is the case of some of narrow ringlets seen around solar system giant planets.
If the global self-gravity plays an important role in structure evolution of narrow ringlets, 
we need a more advanced method to calculate it.

The current code includes an option of hit-and-run bouncing (Appendix~E) 
but not fragmentation.
We plan to implement collisional fragmentation to our code in subsequent work.  
Since gas drag damps $e$ and $i$ of small fragments effectively, 
growth of embryos is accelerated by fragmentation of planetesimals, particularly in the oligarchic growth stage. 
On the other hand,  since the total solid mass available for planetary accretion decreases 
due to rapid inward migration of fragments in the gaseous disk, 
the final masses of planets also decrease, unless planetary atmospheres effectively capture fragments.  
These conclusions were derived by simulations using statistical accretion codes
(e.g., Inaba et al., 2003; Chambers, 2008; Kobayashi et al., 2010, 2011; Amaro-Seoane et al., 2014).
In these studies, the torques on embryos
exerted by small fragments were not taken into account.
Levison et al. (2010) showed that
small fragments shepherded by embryos push bash the embryos 
toward the central star, leading to rapid inward migration of the embryos. 
On the other hand, Kobayashi et al. (2010) pointed out that shepherded fragments
are quickly ground down due to mutual collisions and very small ground fragments 
collide with or migrate past the embryos before pushing back the embryos.
However, this has not been confirmed by direct Lagrangian type simulations and is of interest in future work.

\section{Summary}

We developed a new particle-based hybrid code for planetary accretion.
The code uses an $N$-body routine for interactions with embryos 
while it can handle a large number of planetesimals using 
a super-particle approximation, 
in which a large number of small planetesimals are represented by a small number of tracers. 
Tracer-tracer interactions are handled by a statistical routine.
If the embryo mass is similar to tracer masses and if
embryo-tracer interactions are handled by the direct $N$-body routine, 
 the embryo suffers artificially strong kicks from tracers. 
To avoid this issue, sub-embryos are introduced. 
Accelerations of sub-embryos due to gravitational interactions with tracers are handled by  
the statistical routine whereas accelerations of tracers due to gravitational interactions with sub-embryos are 
handled by the direct $N$-body routine. 

Our statistical routine first calculates the surface number densities and the orbital elements of interloping planetesimals 
around each target tracer (Section~2.2). 
Using the phase-averaged collision probability, whether a collision between the interloper and the target 
occurs is determined (Section~2.4).  
If a collision occurs, the velocity changes are accurately calculated by matching the positions of  the interloper and the target 
without changing their eccentricity and inclination vectors.
Using the phase-averaged stirring and dynamical friction rates, 
the change rates of the orbital elements of the tracers are calculated (Section~2.5).
These rates are converted into the accelerations of the tracers using both first and second order Gauss planetary equations. 
Planetesimal-driven migration of sub-embryos is basically handled by the $N$-body routine but 
the migration rate is limited to the theoretical prediction derived by the statistical routine (Section~2.6). 
This unique routine can reproduce smooth, long-distance migrations of sub-embryos.

We performed various tests using the new hybrid code:
velocity evolution due to collisions, and that due to gravitational stirring and dynamical friction, self-sustained migration 
of sub-embryos, formation of terrestrial planets both in the presence and absence of the gaseous disk,
and formation of cores of jovian planets.
All the test simulations showed good agreements with analytic predictions and/or pure $N$-body simulations for all cases, 
except that the last test did not have a robust benchmark.
The computational load on the portion of the statistical routine is comparable to or less than that for the $N$-body routine.
The current code includes an option of hit-and-run bouncing but not fragmentation that remains for future work.

\section*{Acknowledgments}
We are grateful to the reviewers for their many valuable comments, which greatly improved the manuscript.
We also thank Satoshi Inaba for kindly giving us his simulation data.
This research was carried out in part at the Jet Propulsion Laboratory, California Institute of Technology, 
under contract with NASA. Government sponsorship acknowledged. 
Simulations were performed using JPL supercomputers, Aurora and Halo.

\section*{Appendix~A:  Correction due to self-encounters}
In Section~2.2, we only consider encounters with interlopers. 
However, planetesimals in the target 
$i$ may encounter with other planetesimals in the same target if $k_i > 1$. 
To take into account this effect, we modify apparent planetesimal numbers in interlopers, 
if masses of interloping planetesimals are similar to those in the target tracer.
For this purpose only, we assume axisymmetric distribution of planetesimals and 
introduce the mass bin for planetesimals. 
We also use the radial bin used in the neighboring search.
We first calculate the total mass of tracers, $\sum_j^{j \ne i} k_j m_j$, in the radial and mass bins
where the target locates. The summation is done for 
tracers in any azimuthal locations, excluding  the target.
Since the actual total mass seen from an individual planetesimal in the target tracer
is $\sum_j^{j \ne i}  k_j m_j + (k_i-1)m_i$, we use a corrected $k_j'$ for the interloper,
\begin{equation}
k_j' = \left(1+\frac{(k_i-1)m_i}{\sum_j^{j \ne i}  k_j m_j}\right)k_j,
\end{equation}
if it is in the same radial and mass bins with the target.
If there is no other tracer in the same radial and mass bins, the correction is not applied.
We find that this self-encounter effect is very small at all as we do not identify any difference in simulations with and without the 
correction. 

The effect can also be checked by simulations with a same total number of planetesimals but 
using different numbers of tracers.  Although a simulation with a small number of tracers
has larger statistical fluctuations in various physical quantities, such as the mean eccentricity,
any differences in time-averaged quantities are not well identified. 
Thus, we can omit this correction, and mass bins are unnecessary in our code.

\section*{Appendix B: Non-dimensional collision, stirring, and migration rates}

With Hill's approximation, the equations of motion for the relative motion between the target 
and the interloper are reduced to Hill's equations (Hill 1878; Nakazawa et al., 1989).
The time and the distance of Hill's equations can be normalized by the inverse of orbital frequency, $\omega_{\rm k}^{-1}$,
and $a_{ij}h_{ij}$, respectively, where $a_{ij} = (a_i + a_j)/2$ is the mean semimajor axis and $h_{ij}$ is the reduced mutual Hill radius [Eq.~(\ref{eq:rhill})].
The solution of the relative motion in the absence of mutual gravity is expressed by non-dimensional relative orbital elements,
$b$, $\tilde{e}_{\rm r}$, $\tilde{i}_{\rm r}$, $\varpi_{\rm r}$ and $\Omega_{\rm r}$.
The elements $b$, $\tilde{e}_{\rm r}$ and $\tilde{i}_{\rm r}$ are given as
\begin{equation}
b = \frac{d_{ij}}{a_{ij}h_{ij}}, \hspace{0.3em} \tilde{e}_{\rm r} = \frac{e_{\rm r}}{h_{ij}}, \hspace{0.3em}\tilde{i}_{\rm r} = \frac{i_{\rm r}}{h_{ij}}, \label{eq:ne}
\end{equation}
where $d_{ij} = a_j-a_i $ is the difference of the semimajor axes,  the relative eccentricity $e_{\rm r}$ and the relative inclination $i_{\rm r}$ are 
defined in Eq.~(\ref{eq:relei}) together with the phases $\varpi_{\rm r}$ and $\Omega_{\rm r}$.
 
\subsection*{B.1. Collision rates}
The non-dimensional collision rate $P_{\rm col} (\tilde{e}_{\rm r}, \tilde{i}_{\rm r})$ 
is defined as 
\begin{equation}
P_{\rm col} (\tilde{e}_{\rm r}, \tilde{i}_{\rm r}) =  \int p_{\rm col}  \frac{3}{2} |b|db \frac{d\varpi_{\rm r} d\Omega_{\rm r}}{(2 \pi)^2},
\end{equation}
where $p_{\rm col}$ is unity for collisional orbits and zero otherwise. 


In the dispersion dominant regime [$(\tilde{e}_{\rm r}^2 + \tilde{i}_{\rm r}^2)^{1/2} > 2$],  
$P_{\rm col}$ is given as (Greenzweig and Lissauer, 1990; Inaba et al., 2001)
\begin{equation}
P_{\rm col,high}(\tilde{e}_{\rm r},\tilde{i}_{\rm r})=  \frac{\tilde{2r_{\rm p}}^2}{\pi}\frac{\sqrt{\tilde{e}_{\rm r}^2 + \tilde{i}_{\rm r}^2}}{\tilde{i}_{\rm r}}
\left(E(\zeta)  + \frac{6}{\tilde{r_{\rm p}}}\frac{K(\zeta)}{\tilde{e}_{\rm r}^2 + \tilde{i}_{\rm r}^2 } \right), \label{eq:pcol}
\end{equation}
where $\tilde{r}_{\rm p} = (s_i + s_j)/(a_{ij}h_{ij})$ with the physical radii $s_i$ and $s_j$,
 $K(\zeta)$ and $E(\zeta)$ are the complete elliptic integrals of the first and second kinds, respectively, 
and $\zeta^2 = 3\tilde{e}_{\rm r}^2/(4\tilde{e}_{\rm r}^2+4\tilde{i}_{\rm r}^2)$.

In the shear dominant regime [$(\tilde{e}_{\rm r}^2 + \tilde{i}_{\rm r}^2)^{1/2} < 2$],
if  the inclination is small enough
($\tilde{i}_{\rm r} < \tilde{r_{\rm p}}^{1/2}$; see Goldreich et al. (2004) for this criterion)
all pairs of particles entering the mutual Hill sphere collide, 
and $P_{\rm col}$ becomes independent of $\tilde{e}_{\rm r}$ and $\tilde{i}_{\rm r}$ as (Ida and Nakazawa, 1989) 
\begin{equation}
P_{\rm col,low}(\tilde{e}_{\rm r},\tilde{i}_{\rm r}) =  11.3\sqrt{\tilde{r_{\rm p}}}.
\end{equation}
In the shear dominant regime  with moderate inclinations ($\tilde{r_{\rm p}}^{1/2} < \tilde{i} < 2 $),
$P_{\rm col}$ depends only on $\tilde{i}$ as (Ida and Nakazawa, 1989) 
\begin{equation}
P_{\rm col,med}(\tilde{e}_{\rm r},\tilde{i}_{\rm r}) =  \frac{147\tilde{r_{\rm p}}}{4\pi \tilde{i}_{\rm r}}.
\end{equation}

Following Inaba et al. (2001), we connect the collision rates in different regimes as 
\begin{equation}
P_{\rm col}(\tilde{e}_{\rm r},\tilde{i}_{\rm r}) = {\rm min}[P_{\rm col, med}, (P_{\rm col, high}^{-2} + P_{\rm col, low}^{-2})^{-1/2}].
\end{equation}
The collision rate given by this expression well agrees with that derived by numerical integrations 
(Ida and Nakazawa, 1989; Greenzweig and Lissauer, 1990).  

\subsection*{B.2. Stirring rates}
The non-dimensional stirring and dynamical friction rates are defined as 
\begin{eqnarray}
P_{\rm VS} (\tilde{e}_{\rm r}, \tilde{i}_{\rm r}) &=&  \int \delta \tilde{e}_{\rm r}^2  \frac{3}{2} |b|db \frac{d\varpi_{\rm r} d\Omega_{\rm r}}{(2 \pi)^2}, \label{eq:pvs3}\\
Q_{\rm VS} (\tilde{e}_{\rm r}, \tilde{i}_{\rm r}) &=& \int \delta \tilde{i}_{\rm r}^2 \frac{3}{2} |b|db \frac{d\varpi_{\rm r} d\Omega_{\rm r}}{(2 \pi)^2},\label{eq:qvs3}\\
P_{\rm DF} (\tilde{e}_{\rm r}, \tilde{i}_{\rm r}) &=& -2 \int 
\left(\frac{\mbox{\boldmath $\tilde{e}$}_{\rm r} \cdot \delta \mbox{\boldmath $\tilde{e}$}_{\rm r}}{\tilde{e}_{\rm r}^2}\right)
\frac{3}{2} |b|db \frac{d\varpi_{\rm r} d\Omega_{\rm r}}{(2 \pi)^2}, \label{eq:pdfd}
\end{eqnarray}
where $\delta \tilde{e}_{\rm r}^2$, $\delta \tilde{i}_{\rm r}^2$, and $\delta \mbox{\boldmath $\tilde{e}$}_{\rm r}$ are 
the changes of $\tilde{e}_{\rm r}^2$, $\tilde{i}_{\rm r}^2$, and $\mbox{\boldmath $\tilde{e}$}_{\rm r}$ [Eq.~(\ref{eq:relei})] during a single encounter.
The non-dimensional dynamical friction rate for inclination is obtained by replacing 
$(\mbox{\boldmath $\tilde{e}$}_{\rm r} \cdot \delta \mbox{\boldmath $\tilde{e}$}_{\rm r}/\tilde{e}_{\rm r}^2)$
by $(\mbox{\boldmath $\tilde{i}$}_{\rm r} \cdot \delta \mbox{\boldmath $\tilde{i}$}_{\rm r}/\tilde{i}_{\rm r}^2)$ in Eq.~(\ref{eq:pdfd})
and is known to be the same with $P_{\rm DF}$ (Tanaka and Ida, 1996). 

The rates for the shear dominant regime are (Ida 1990, Ohtsuki et al., 2002)
\begin{eqnarray}
P_{\rm VS,low} &=& 75.4, \\
Q_{\rm VS,low} &=& 2\tilde{i}_{\rm r}^2 + 5\tilde{e}_{\rm r}^3\tilde{i}_{\rm r}, \\
P_{\rm DF,low} &=& 10.
\end{eqnarray}
For  $Q_{\rm VS,low}$, we adopt a functional form that is similar to the one with the Rayleigh distribution (Ohtsuki et al., 2002) 
but the factors are modified.


The rates for the dispersion dominant regime are given as (Tanaka and Ida, 1996) 
\begin{eqnarray}
P_{\rm VS,high} &=& 
\frac{36}{\pi \tilde{i}_{\rm r} \sqrt{\tilde{e}_{\rm r}^2+ \tilde{i}_{\rm r}^2}} \left[5K(\zeta)- \frac{12\tilde{e}_{\rm r}^2}{\tilde{e}_{\rm r}^2 + 4\tilde{i}_{\rm r}^2}E(\zeta)\right] \nonumber \\
&& \times \ln{(\Lambda^2+1)},\\
Q_{\rm VS,high} &=& 
\frac{36}{\pi \tilde{i}_{\rm r} \sqrt{\tilde{e}_{\rm r}^2+ \tilde{i}_{\rm r}^2}} \left[K(\zeta)- \frac{12\tilde{i}_{\rm r}^2}{\tilde{e}_{\rm r}^2 + 4\tilde{i}_{\rm r}^2}E(\zeta)\right] \nonumber \\
&& \times \ln{(\Lambda^2+1)},\\
P_{\rm DF,high} &=& 
\frac{288}{\pi \tilde{i}_{\rm r} (\tilde{e}_{\rm r}^2+ 4\tilde{i}_{\rm r}^2) \sqrt{\tilde{e}_{\rm r}^2+ \tilde{i}_{\rm r}^2}} E(\zeta) \ln{(\Lambda^2+1)},
\end{eqnarray}
where
\begin{equation}
\Lambda = \tilde{i}_{\rm r}(\tilde{e}_{\rm r}^2+ \tilde{i}_{\rm r}^2)/3.
\end{equation}
In $P_{\rm VS,high}$ and  $Q_{\rm VS,high}$,  
we omit the term $\Lambda^2/(\Lambda^2+1)$ as it is negligible 
compared with $\ln{(\Lambda^2+1)}$ in the dispersion dominant regime. 

We connect the stirring and dynamical friction rate at the shear and dispersion dominant regimes
in a similar manner of Ohtsuki et al. (2002):
\begin{eqnarray}
P_{\rm VS} &=& C_1 P_{\rm VS,low} + P_{\rm VS,high}  \\ 
Q_{\rm VS}  &=& C_2 Q_{\rm VS,low} + Q_{\rm VS,high} \\
P_{\rm DF} &=&  C_3 P_{\rm DF,low} + P_{\rm DF,high}. 
\end{eqnarray}
The coefficients are 
\begin{eqnarray}
C_1 &=& \ln(\Lambda^2 + 1)/\Lambda^2, \\
C_2 &=& \ln(10\Lambda^2\tilde{e}_{\rm r}+1)/(10\Lambda^2\tilde{e}_{\rm r}), \\
C_3 &=&  \ln(10 \Lambda^2 + 1)/(10 \Lambda^2).
\end{eqnarray}
The approximated formulae well agree with the rates derived by three-body orbital integrations [Fig.~8 of Ida (1990) and Fig.~1 of 
Ohtsuki et al. (2002)], except for the region with $\tilde{e_{\rm r}} >4$ and $\tilde{i}_{\rm r} < 1$ (see Section~3.2).

\subsection*{B.3. Migration rates}
The non-dimensional migration rate $R_{\rm MG}$ for one-sided torque is defined as 
\begin{equation}
R_{\rm MG} = -\int^{\infty}_{0} \frac{3}{2}b db \int \frac{d\varpi_{\rm r} d\Omega_{\rm r}}{(2 \pi)^2} \delta b, \label{eq:rmig}
\end{equation}
where $\delta b$ is the change of $b$ during a single encounter and the negative sign represents the back reaction on the embryo.
At $\tilde{e}_{\rm r} = \tilde{i}_{\rm r} = 0$, $R_{\rm MG} = 9.4$ [Ida et al., 2000; their Eq.~(10)]. 
It is expected that $R_{\rm MG} $ in the shear dominant regime is about the same value
as the migration rate is almost constant in this regime (Kirsh et al., 2009):
\begin{equation}
R_{\rm MG,low} = 9.4.
\end{equation}
For the dispersion dominant regime, Ida et al. (2000) derived the migration rate averaged over $\varpi_{\rm r}$  and $\Omega_{\rm r}$ as  [their Eq.~(18)]
\begin{equation}
\int \frac{d\varpi_{\rm r} d\Omega_{\rm r}}{(2 \pi)^2} \delta b
= - \frac{12}{\pi \tilde{i}_{\rm r}}\frac{b}{|b|}\frac{\ln{(\Lambda^2+1)}}{\sqrt{(\tilde{e}_{\rm r}^2-b^2)(\tilde{e}_{\rm r}^2+\tilde{i}_{\rm r}^2-3b^2/4)^3}}.
\end{equation}
Inserting this equation into Eq.~(\ref{eq:rmig}), we have 
\begin{equation}
R_{\rm MG,high} =   \frac{72}{\pi \tilde{i}_{\rm r}\tilde{e}_{\rm r}}\frac{\ln{(\Lambda^2+1)}}{ \sqrt{\tilde{e}_{\rm r}^2+\tilde{i}_{\rm r}^2}}\frac{\tilde{e}_{\rm r}^2}{\tilde{e}_{\rm r}^2+4\tilde{i}_{\rm r}^2},
\end{equation}
where the integration range of $b$ is limited between $0$ and $\tilde{e}_{\rm r}$, and 
$ \ln{(\Lambda^2+1)}$ is approximately treated as a constant in the integration.
Analogous to the stirring rates, we connect the migration rates at the shear and dispersion dominant regimes but 
the value is assumed not to exceed $R_{\rm MG,low}$
\begin{equation}
R_{\rm MG} = {\rm MIN}[(C_1 R_{\rm MG,low} + R_{\rm MG,high}), R_{\rm MG,low}].
\end{equation}
While this expression must be correct in the low and high velocity limits, its accuracy 
at the intermediate regime needs to be confirmed by comparing with the rate derived from three-body orbital integrations.

\section*{Appendix C: Splitting of dynamical friction}

When the non-dimensional collisional and stirring rates are derived, uniformities of 
$\varpi_{\rm r}$ and $\Omega_{\rm r}$ are assumed. However, we do not assume 
uniformities of the relative longitudes, $\varpi_{ij} = \varpi_{j}-\varpi_{i}$ and $\Omega_{ij} = \Omega_{j}-\Omega_{i}$.
Alignments of the longitudes occur, for example,  
if the system is perturbed by a giant planet and the forced eccentricity 
is much larger than the free eccentricities.

Ida (1990) showed that the eccentricity evolution of a particle $i$ due to dynamical friction of particles $j$ is 
\begin{eqnarray}
\left(\frac{de_i^2}{dt}\right)_{{\rm DF},j} &=& \frac{C_1}{\nu_{j}h_{ij}^2} 
\left[(\nu_{i} \mbox{\boldmath $e$}_i  +  \nu_{j} \mbox{\boldmath $e$}_j)\cdot (\mbox{\boldmath $e$}_j-\mbox{\boldmath $e$}_i) \right]P_{\rm DF}  \label{eq:dfj0}   \\
&=& \frac{C_1}{\nu_{j}h_{ij}^2}\left[\nu_{j}{e}_{j}^2 - \nu_{i}e^2_{i} +(\nu_{i}-\nu_{j})e_{i}e_{j}\cos{\varpi_{ij} } \right]P_{\rm DF}, \label{eq:dfj1} 
\end{eqnarray}
where the characters appeared in these equations are defined around Eqs.~(\ref{eq:dedfs}) and (\ref{eq:dedfd}). 
If $\varpi_{ij}$'s of interlopers are randomly distributed and the distribution of $e_{j}$ is the Rayleigh distribution, 
the term with $\cos{\varpi_{ij}}$ becomes small enough 
after averaging over interlopers (Ida, 1990). 
The equilibrium state for this case leads to the well-known energy equipartitioning 
($\nu_{j}{e}_{j}^2 = \nu_{i}e^2_{i}$) with dynamical friction only. With viscous stirring, however, 
the energy equipartitioning is achieved only if the mass distribution is very steep (Rafikov, 2003). 

If the distribution of $\varpi_{ij}$ is not random, as may be the case of a forced system, the term with  $\cos{\varpi_{ij}}$
in Eq.~(\ref{eq:dfj1}) cannot be ignored.
In a forced system,  the eccentricity vectors may be split
as  $\mbox{\boldmath $e$}_i =  \mbox{\boldmath $e$}_{\rm f} +  \mbox{\boldmath $e$}_{i0}$
and $\mbox{\boldmath $e$}_j =  \mbox{\boldmath $e$}_{\rm f} +  \mbox{\boldmath $e$}_{j0}$,
where $\mbox{\boldmath $e$}_{\rm f}$ is the forced eccentricity vector,  
$\mbox{\boldmath $e$}_{i0}$ and $\mbox{\boldmath $e$}_{j0}$ are the free eccentricity vectors, respectively (Murray and Dermott, 1999).
Inserting these forms into Eq.~(\ref{eq:dfj0}) and setting 
$\mbox{\boldmath $e$}_{\rm f} \cdot (\mbox{\boldmath $e$}_{j0} - \mbox{\boldmath $e$}_{i0}) = 0$ due to the uniformity of  $\varpi_{\rm r}$, 
we can find an equation similar to Eq.~(\ref{eq:dfj1}) but $e_i$ and $e_j$
replaced by the free eccentricities $e_{i0} (= |\mbox{\boldmath $e$}_{i0}|)$ and 
$e_{j0} (= |\mbox{\boldmath $e$}_{j0}|)$ and $\varpi_{ij}$ replaced by the relative longitude of the free eccentricities.
Since there is no correlation between the directions of $\mbox{\boldmath $e$}_{j0}$ and $\mbox{\boldmath $e$}_{i0}$,
the third term in the square bracket of the r.h.s of Eq.~(\ref{eq:dfj1}) can be neglected.
Thus, dynamical friction tends to lead the energy equipartitioning of the free eccentricities, if $\mbox{\boldmath $e$}_{\rm f}$ is shared 
for all particles. A similar conclusion can be derived for inclinations.

Now we consider splitting Eq.~(\ref{eq:dfj1}) into two terms: a stirring term that 
is preferably larger than the viscous stirring term [as required in Eq.~(\ref{eq:vrs})] and a damping term that decreases the free eccentricity.
We find the forms that can fulfill such conditions as
\begin{eqnarray}
\left(\frac{de_i^2}{dt}\right)_{{\rm DFS},j} &=& \frac{C_1}{h_{ij}^2} {e}_{\rm r}^2 P_{\rm DF}, \label{eq:dedfsap} \\
\left(\frac{de_i^2}{dt}\right)_{{\rm DFD},j} &=& - \frac{C_1}{\nu_jh_{ij}^2} \left(e_i^2 - e_i e_j\cos{\varpi_{ij}} \right)  P_{\rm DF}. \label{eq:dedfdap}  
\end{eqnarray}
The same equations are given in Eqs.~(\ref{eq:dedfs}) and (\ref{eq:dedfd}). 
In a forced system,  the terms in the parenthesis of the r.h.s. of Eq.~(\ref{eq:dedfdap}) are reduced to $e_{i0}^2$ 
using $\mbox{\boldmath $e$}_{\rm f} \cdot (\mbox{\boldmath $e$}_{j0} - \mbox{\boldmath $e$}_{i0}) = 0$ and 
$\mbox{\boldmath $e$}_{i0} \cdot \mbox{\boldmath $e$}_{j0} = 0$ 
due to the uniformity of  $\varpi_{\rm r}$.  
The l.h.s. of Eq.~(\ref{eq:dedfdap}) is also reduced to $de^2_{i0}/dt$ because the orientation 
of $\mbox{\boldmath $e$}_{i0}$ changes randomly due to gravitational interactions with other tracers  
 [$\mbox{\boldmath $e$}_{f} \cdot (\mbox{\boldmath $e$}_{i0}/dt) = 0$]. 
Thus, the damping term [Eq.~(\ref{eq:dedfdap})] decreases the free eccentricity, as required.
Usually in astrophysics, only this damping term is called dynamical friction (Binney and Tremaine, 1987).  
A similar splitting is possible for inclinations, and the split terms are given 
by Eqs.~(\ref{eq:didfs}) and (\ref{eq:didfd}).


\section*{Appendix D: Radial diffusion coefficient}
The classical theory of random walk shows that the diffusion coefficient $D_i$ for the one dimensional problem 
is $\langle (\Delta a_i)^2 \rangle /(2 \Delta t)$, where $\Delta a_i$ is the change in the semimajor axis for the tracer $i$
during the time step $\Delta t$ and the angle brackets mean averaging over multiple changes. 
The change of $a_i$ due to a single encounter with other planetesimal is defined to be $\delta a_i 
= \nu_j a_{ij}h_{ij} \delta b$, where
 $\delta b$ is again the change of $b$ during an encounter. 
Thus, 
the diffusion coefficient $(D_i)_j$ for the tracer $i$ due to encounters with planetesimals in the tracer $j$ is given as 
\begin{equation}
(D_i)_j = (\nu_j a_{ij}h_{ij})^2 \frac{\langle (\Delta b)^2 \rangle_j}{2\Delta t},  \label{eq:did}
\end{equation}
where $\Delta b$ is the change of $b$ during $\Delta t$ and the subscript for the angle brackets means averaging over planetesimals in the tracer $j$.

Petit and H\'{e}non (1987) performed three-body orbital integrations and showed that 
$\langle b\delta b \rangle/\langle (\delta b)^2 \rangle = - 3.07/17.72 \simeq -1/6$
for $\tilde{e}_{\rm r}=\tilde{i}_{\rm r}=0$. Although it is unclear if this factor is applicable for $\tilde{e}_{\rm r}, \tilde{i}_{\rm r} > 0$, we use 
the relationship for any cases. 
This gives $\langle (\delta b)^2 \rangle = 3\langle \delta b^2 \rangle$/2, using $(\delta b)^2 = -2b \delta b + \delta b^2$.
The averaged change $\langle (\Delta b)^2 \rangle_j$ is given by 
\begin{eqnarray}
\frac{\langle (\Delta b)^2 \rangle_j}{\Delta t} &=& n_{j}a_{ij}^2{h_{ij}}^2 \omega_{\rm K} \int 
\left(\frac{3}{2}\delta b^2\right) \frac{3}{2} |b|db \frac{d\varpi_{\rm r} d\Omega_{\rm r}}{(2 \pi)^2}, \nonumber \\
	&=&  2n_{j}a_{ij}^2{h_{ij}}^2 \omega_{\rm K} (P_{\rm VS} + Q_{\rm VS}), \label{eq:db2}
\end{eqnarray}
where we used $\delta b^2= (4/3) (\delta \tilde{e}_{\rm r}^2 +  \delta \tilde{i}_{\rm r}^2)$ that is derived from 
the Jacobi integral of Hill's equations [Eq.~(\ref{eq:jaco})], and $P_{\rm VS}$ and $Q_{\rm VS}$ are defined in Eqs.~(\ref{eq:pvs3}) and ~(\ref{eq:qvs3}). 

Inserting Eq.~(\ref{eq:db2}) into Eq.~(\ref{eq:did}),
the diffusion coefficient $(D_i)_j$ is given as 
\begin{equation}
(D_i)_j =  \nu_j^2 a_{ij}^4h_{ij}^4 \omega_{\rm K} (P_{\rm VS} + Q_{\rm VS}).
\end{equation}
Finally, the diffusion coefficient due to encounters with all interlopers is given by $D_i = \sum_j (D_i)_j$.  Ohtsuki and Tanaka (2003) derived the viscosity 
for the equal-mass planetesimals disk and their viscosity is $(2/9)D_i$. The diffusion coefficients derived by 
Ormel et al. (2012) and Glaschke et al. (2014) are similar to ours.

\section*{Appendix E: Hit-and-run collisions}
In Section~2.4, we show how the total number of tracers changes due to merging between planetesimals. 
On the other hand, in hit-and-run collisions, there should be no change in the number.
Before describing how to handle hit-and-run collisions between planetesimals in two tracers,
we first show how to handle a hit-and-run collision between embryos or between a full embryo and a tracer.

\subsection*{E.1. Criterion of hit-and-run collisions}
Consider an impact between the target with mass $m_i$ and the impactor with mass $m_j$ ($\le m_i$).
Let the impact velocity and the impact angle be  $v_{\rm imp}$ and $\theta_{\rm c}$ ($\theta_{\rm c} = 0$ for a head-on collision). 
If $v_{\rm imp} > v_{\rm cr}$, where $v_{\rm cr}$ is the critical velocity,  
the collision results in escape of the projectile after bouncing on the target. 
Based on thousands of SPH simulations, 
Genda et al. (2012) derived the formula of $v_{\rm cr}$ as 
\begin{equation}
\frac{v_{\rm cr}}{v_{\rm esc}} = c_1 \Gamma \Theta^{c_5} + c_2 \Gamma + c_3 \Theta^{c_5} + c_4, \label{eq:gen}
\end{equation}
where $v_{\rm esc}$ is the mutual escape velocity, $\Gamma = (m_i- m_j)/(m_i + m_j)$, and $\Theta = 1-\sin{\theta_{\rm c}}$. 
The coefficients are $c_1 = 2.43$, $c_2 = -0.0408$, $c_3 = 1.86$, $c_4 = 1.08$, and $c_5 = 5/2$.

If the impact is judged as a hit-and-run collision ($v_{\rm imp} > v_{\rm cr}$), 
the post-impact velocities of the projectile and the target are
determined as follows. We do not consider any mass exchanges 
between the target and the impactor for simplicity following Kokubo and Genda (2010).
The impact velocity vector is decomposed to the normal and tangential components ($v_n$ and $v_t$)
relative to the vector pointing toward the center of the target from the center of the impactor. 
The tangential component of the post-impact relative velocity 
$v_{t}' $ is assumed to be the same as $v_t$. 
The normal component of the post-impact relative velocity is given as 
\begin{equation}
v_n' = 
\left\{ \begin{array}{ll} 
0
& \mbox{(for $v_t'  > v_{\rm esc} $)}, \\ 
-\left(v_{\rm esc}^2-v_t'^2\right)^{1/2}
& \mbox{(otherwise)}.
\end{array}\right.
\label{eq:vnd}
\end{equation}
The velocity change described here is similar to but slightly different from that of Kokubo and Genda (2010), 
who always set $v_{n}' = 0$ and adjust $v_t'$.

The normal and restitution coefficients, $\epsilon_{\rm n}$ and $\epsilon_{\rm t}$, 
are defined as $\epsilon_{\rm n} = -v_{\rm n}'/v_{\rm n}$ and $\epsilon_{\rm t} = v_{\rm t}'/v_{\rm t}$.
While the above case basically assumes $\epsilon_{\rm n} = 0$ and $\epsilon_{\rm t} = 1$,
these coefficients can be changed to any values in simulations (in Section~3.1, we use $\epsilon_{\rm n} = 0$ and 0.8).



\subsection*{E.2. Procedure of hit-and-run collisions between tracers}

Now we consider hit-and-run collisions between planetesimals in two tracers. We will handle them as consistent as 
possible with those between embryos. 
If a collision is judged to occur in the statistical routine, 
the target and the impactor
are moved to the impact position without changing their eccentricity and inclination vectors, as described in Section~2.4.2.
The relative velocity, $v_{\rm r} =  |\mbox{\boldmath $v$}_{\rm r}|$, at infinity (without acceleration due to mutual gravity)
is calculated at this impact position.
If $k_{jb} > 0$,  only some fraction of planetesimals in the tracer $j$ are involved in the collision (Fig.~4).
The impact velocity is given as $v_{\rm imp} = (v_{\rm r}^2+v_{\rm esc}^2)^{1/2}$.
We ignore deflection of the relative orbit during acceleration due to mutual gravity so that the direction 
of the impact velocity vector is identical to that of the relative velocity vector $\mbox{\boldmath $v$}_{\rm r}$.
The problem is that we cannot directly use the relative position vector of the tracers
because their mutual separation is not exactly 
the sum of the radii of a colliding pair after matching their positions.
Instead of using the actual relative position vector, we make a fictitious relative position vector $\mbox{\boldmath $r$}_{\rm rf}$.
The fictitious impact angle $\theta_{\rm c}$ is given by 
$\cos{\theta_{\rm c}} = (\mbox{\boldmath $v$}_{\rm r} \cdot \mbox{\boldmath $r$}_{\rm rf}) / (v_{\rm r} |\mbox{\boldmath $r$}_{\rm rf}|)$.
Assuming there is no correlation between $\mbox{\boldmath $v$}_{\rm r}$ and $\mbox{\boldmath $r$}_{\rm rf}$,
the probability that the impact angle is between $\theta_{\rm c}$ and $\theta_{\rm c} + d\theta_{\rm c}$ is given as 
$df = 2 \sin{\theta_{\rm c}}\cos{\theta_{\rm c}}d\theta_{\rm c}$ (e.g., Shoemaker and Wolfe, 1982).
We choose an arbitrary $\theta_{\rm c}$ between 0 and 90 degrees following 
this distribution and also choose the tangential direction of $\mbox{\boldmath $r$}_{\rm rf}$ relative to $\mbox{\boldmath $v$}_{\rm r}$ at random.  
Then, we judge whether the collision results in merging or bouncing using 
Eq.~(\ref{eq:gen}) in which $m_j$ is replaced by  $\Delta m_i$.
If it is merging, we follow the process described in Section~2.4.2, using $\mbox{\boldmath $v$}_{\rm r}$ as the impact velocity vector.

If it turns out to be a hit-and-run collision, we calculate the post impact velocities using $v_{t}' = v_t$ and $v_{n}'$ from Eq.~(\ref{eq:vnd}).
Then their velocities are modified so that their relative velocity at infinity is given as $v_{\rm r}' = (v_t'^{2}+v_n'^{2}-v_{\rm esc}^2)^{1/2}$, again ignoring 
deflection of the orbits due to mutual gravity. If the interloper is split into two tracers [Case (a) in Eq.~(\ref{eq:delm})]
before the collision, we merge them by summing their momenta after matching their positions.
The merged tracer has  the same $m_j$ and $k_j$ as those of the original interloper before splitting. 
The semimajor axis of the interloper is adjusted so that the $z$-component of the total angular momentum of is conserved. 
Finally, the mean longitudes of the target and interloper are placed to the original values.

\section*{Appendix~F. Duration of a tracer in the neighboring search region around a sub-embryo}
Here we explain how to derive the correction factor $T_j/T_{j,{\rm in}}$ appeared in Eq.~(\ref{eq:nj2}).
Let us define the inner and outer radii of the region $i$ for neighboring search around the embryo be $r_{\rm in}$ and $r_{\rm out}$ (Fig.~2).
The interloper $j$ has the semimajor axis $a_j$, the eccentricity $e_j$, and the inclination $i_j$.
The maximum and minimum distances of the interloper $j$, $r_{\rm max}$ and $r_{\rm min}$, 
from the central star projected on the invariant plane ($z=0$) are given by
\begin{eqnarray}
r_{\rm max} = a_j(1+e_j)\cos{i_j} \\
r_{\rm max} = a_j(1-e_j)\cos{i_j}.
\end{eqnarray}

If $r_{\rm max} > r_{\rm out}$, we define the eccentric anomaly $E_{\rm out}$ ($< \pi$) at  $r =  r_{\rm out}$ as
\begin{equation}
r_{\rm out} = a_j(1-e_j\cos{E_{\rm out}})\cos{i_j} 
\end{equation}
and the mean anomaly is also defined as $M_{\rm out} = E_{\rm out} -e_j\sin{E_{\rm out}}$.
If $r_{\rm max} < r_{\rm out}$, we set $M_{\rm out} = \pi$.
In a similar manner, if $r_{\rm min} < r_{\rm in}$, we define the eccentric anomaly $E_{\rm in}$ ($< \pi$) at  $r =  r_{\rm in}$ as
\begin{equation}
r_{\rm in} = a_j(1-e_j\cos{E_{\rm in}})\cos{i_j} 
\end{equation}
and the mean anomaly $M_{\rm in} = E_{\rm in} -e_j\sin{E_{\rm in}}$. 
If $r_{\rm min} > r_{\rm in}$, we set $M_{\rm in} = 0$.
Then, the period $T_{j,{\rm in}}$ in which the interloper $j$ is radially in the region $i$ 
during the orbital period $T_j$ is given as 
\begin{equation}
\frac{T_{j,{\rm in}}}{T_{j}} = \frac{M_{\rm out} - M_{\rm in}}{\pi}.
\end{equation}

\section*{Appendix~G: Comparison with Levison et al. (2012)}
Here we summarize the differences between the LIPAD code (LDT12) and our new code. 

\subsubsection*{Spatial grid for neighboring search}
In the LIPAD code, to derive the spatial density of planetesimals, tracers are placed in mass and radial grids and
the vertical distribution is assumed to be Gaussian.
Our code has neither these fixed grids nor the assumption for the vertical distribution 
(except for the global gravity calculation described in Section~2.7). 
Thus, our code has Lagrangian natures more than the LIPAD code,
although these differences are probably not fundamental for the test simulations shown in this paper.
Our scheme is  more beneficial for systems with non-axisymmetric and/or inclined structures.  

The advantage of the assumption of the vertical distribution in  the LIPAD code is that  
changes in collision and stirring rates at different $z$ are explicitly taken into account. 
Changes in collision rates with z are taken into account in our model too, although implicitly, 
as we move both the target and the interloper to the collisional point (Section~2.4);
collisions occur more often near the mid-plane than at high $|z|$.  
In the stirring routine (Section~2.5), our method ignores this effect and this probably introduces some inaccuracies 
in orbital evolutions of individual particles. Nevertheless, the velocity evolution of the system as a whole 
is reproduced reasonably well by our method as shown in the test simulations 
(see more discussion for viscous stirring and dynamical friction given below). 

\subsubsection*{Number of planetesimals in a tracer}
The number of planetesimals in a tracer, $k_i$, is an integer in our model (Section~2.1), contrary to LDT12. 
If a tracer with a non-integer number of planetesimals is promoted to a single sub-embryo,
the extra mass, $(k_i-1)m_i$ (if $1 < k_i < 2$), needs to be incorporated into the sub-embryo or 
transferred to other tracers which may have very different planetesimal masses.  
Both are not accurate although it is not clear to us how it is handled in the LIPAD code.

\subsubsection*{Orbital isolation}
In our code, orbital isolation between planetesimals does not occur
as we choose a narrow radial half width $\delta r$ of the neighboring search region 
given by Eq.~(\ref{eq:drt0}). 
If we adopt a larger $\delta r$, orbital isolation can potentially occur. In this case we may need to 
consider how isolated bodies interact through distant perturbations. In general, however, 
the radial resolution is sacrificed with a large $\delta r$.
In our current scheme, interactions between runaway bodies are approximately represented by occasional 
close encounters rather than distant perturbations.

The LIPAD code explicitly handles orbital isolation and distant perturbations (e.g., their Fig.~10). 
This means that they use the radial grid size larger than ours. 
Even with a large $\delta r$, the exitisting Lagrangian codes
cannot handle orbital isolation of planetesimals that are much less massive than $m_{t0}$, 
because a small number of runaway bodies cannot be represented by tracers with nearly fixed masses.

\subsubsection*{Collisions between planetesimals and sub-embryos }

Collisions between tracers and sub-embryos are handled in the $N$-body routine in the LIPAD code. 
Thus,  after the promotion the first impact of a tracer doubles the mass of a sub-embryo.
This makes artificial kinks in mass spectra at the transition mass, $m_{t0}$. 
The influence of this effect on evolution of the overall system is small as discussed by 
LDT12, particularly if the masses of the largest embryos are much larger than $m_{t0}$. 

In our method, collisions of planetesimals with sub-embryos are handled by the statistical routine (Secs.~2.4 and 2.6). 
The mass spectra are smooth at $m_{t0}$ (see Fig.~12), because
only some of planetesimals in a tracer collide with a sub-embryo.
Note that we assume uniformities in phase space (Section~2.3) that seem to be valid at least in test simulations shown in this paper, 
while no assumption for phase-space is necessary for the $N$-body routine.

\subsubsection*{Collision velocity between tracers}
The impact velocity between tracers  
is calculated more accurately in our code than the LIPAD code.
As described in Section~2.4, we estimate the impact velocity between tracers 
by matching their positions without changing their eccentricity and inclination vectors.

In the LIPAD code, only the interloper's longitude is aligned to that of the target
by rotating the interloper's coordinates, not letting the interloper move along its Keplerian orbit.
This changes the longitude of pericenter of the interloper by the rotation angle. 
LDT12 managed to minimize this issue by choosing the azimuthally closest interloper for the collision with the target, 
as originally done in Levison and Morbidelli (2007). 
On the other hand, our method allows any interlopers to collide with the target.


The timescale of collisional damping shown in Fig.~2 of LDT12 is clearly too short, probably 
because they described the initial condition of a wrong simulation.

\subsubsection*{Viscous stirring and dynamical friction}
The largest difference between our code and the LIPAD code 
probably appears in the methods to handle viscous stirring and dynamical friction.
In the LIPAD code, the probability that an interloper enters within a mutual Hill radius of the target is calculated.
If an encounter occurs, the acceleration of a smaller body of the pair is given by solving a pair-wise gravitational interaction.
The acceleration of a lager body of the pair is not calculated in this routine, but the damping forces
are given by the Chandrasekhar's formula of dynamical friction.
Instead of taking into account stirring on the larger body, 
they set lower limits of $e$ and $i$ of the larger body based on energy equipartitioning.
Some inaccuracies seem to be introduced in their approach.
First, energy equipartitioning is not usually achieved except with a very steep mass spectrum (Rafikov, 2003).
Second, the Chandrasekhar's formula of dynamical friction is not applicable to the shear-dominant regime (e.g., Ida, 1990; Ohtsuki et al., 2002).

In our method,
we do not solve individual pair-wise gravitational interactions. 
Instead, we calculate the time averaged changes of the orbital elements due to multiple encounters with tracers, 
using the phase-averaged stirring rates (Section~2.5.1).
These changes of orbital elements are then converted into the accelerations of tracers (Section~2.5.2).
This approach allows us to handle viscous stirring, the stirring and damping parts of dynamical friction due to interactions between 
all pairs in the same manners regardless of their mass ratios. 
Our approach is based on uniformities of the phases (Section~2.3), and methods which 
explicitly solves pair-wise gravitational interactions (i.e., the acceleration of a smaller body in the LIPAD code)  
are probably more accurate than ours. 
We would like to point out, however, that the LIPAD code also partly assumes uniformities of the phases 
because it uses the phase-averaged collision/encounter probability of Greenzweig and Lissauer (1990); 
note that $P_{\rm col}$ we use and $F_{g}$ in the LIPAD code are the same quantity in different formats.

\subsubsection*{Planetesimal-driven migration of sub-embryos}
LDT12 included planetesimal-driven migration of sub-embryos 
by adding torques on them calculated by a series of three-body integrations 
(Sun, the target, and the interloper) during LIPAD simulations.
Each interloper is randomly chosen from one in seven Hill radii of the target. 
It has the same semimajor axis, orbital eccentricity, and inclination as one of the tracers 
but its phases are modified. 
As far as we understand, their point is that the torque on the sub-embryo is smoothed by phase averaging. 
This may be true in the dispersion dominant regime ($e_{ij}/h_{ij}, i_{ij}/h_{ij} \gg 1$). 
In the shear dominant regime, however, the phase dependence of the torque is weak 
and the averaged toque turns out to be as strong as the torque from a single interloping tracer at all 
[see Fig.~8 of Ohtsuki and Tanaka (2003), for example].  
Thus, sub-embryos should occasionally suffer strong kicks from tracers. 
Although it is unclear to us whether the point described above is indeed an issue for the LIPAD code, 
 sub-embryos in LIPAD simulations (Ngo, 2012) look radially much more stationary than those in our simulations
 (Section~3.6).


In our method, the angular momentum change of a sub-embryo is stored in the $N$-body routine but 
its release rate is limited to the theoretical prediction derived by the statistical routine (Section~2.6.2). 
Theoretical justifications of our method are discussed in Section~3.3.


\subsubsection*{Three-body integrations}
The LIPAD code performs a series of three-body integrations during a simulation for two cases.
The first one is for the routine of sub-embryo migration.
The another one is for the stirring routine in the shear dominant regime.
Our code does not need to perform three body integrations.
Migration of sub-embryos are handled as explained above.
Stirring in the shear dominant regime is handled using the approximated
formulae of the phase-averaged stirring rates (Section~2.5 and Appendix~B).



\section*{REFERENCES}

\begin{description}
\item 
Adachi, I., Hayashi, C., Nakazawa, K. 1976.
The gas drag effect on the elliptical motion of a solid body in the primordial solar nebula.	
Prog. Theoret. Phys. 56, 1756--1771.

\item
Amaro-Seonae, P., Glaschke, P., Spurzem, R., 2014.
Hybrid methods in planetesimal dynamics: formation of protoplanetary systems and the mill condition.
Mon. Not. R. Astron. Soc. 445, 3755--3769.

\item
Binney, J., Tremaine, S., 1987.
Galactic dynamics. 
Princeton Univ. Press, Princeton, NJ.

\item
Bromley, B.C., Kenyon, S.J., 2006.
A hybrid $N$-body-coagulation code for planet formation.
Astron. J. 131, 2737--2748. 

\item
Bromley, B.C., Kenyon, S.J., 2011.
Migration of planets embedded in a circumstellar disk.
Astrophys. J. 735, 29.

\item
Bromley, B.C., Kenyon, S.J., 2013.
Migration of small moons in Saturn's rings.
Astrophys. J. 764, 192.

\item
Capobianco, C.C., Duncan, M.J., Levison, H.F., 2011.
Planetesimal-driven planet migration in the presence of a gas disk.
Icarus 211, 819--831.

\item 
Chambers, J.E.,  1999.
A hybrid symplectic integrator that permits close encounters between massive bodies.
Mon. Not. R. Astron. Soc. 304, 793--799.

\item 
Chambers, J.E.,  2008.
Oligarchic growth with migration and fragmentation.
Icarus 198, 256--273.

\item
Chambers, J.E., Wetherill, G.W.,  1998.
Making the terrestrial planets: $N$-body integrations of planetary embryos in three dimensions.
Icarus 136, 304--327.

\item
Dehnen, W., 2001.
Towards optimal softening in three-dimensional $N$-body codes - I. Minimizing the force error.
Mon. Not. R. Astron. Soc. 324, 273--291.

\item
Dones, L. and Tremaine, S., 1993.
On the origin of planetary spins. 
Icarus 103, 67--92.

\item
Duncan, M.J., Levison, H.F.,  Lee, M.H., 1998.
A multiple time step symplectic algorithm for integrating close encounters.
Astron. J. 116, 2067--2077.

\item
Genda, H., Kokubo, E., Ida, S., 2012.
Merging criteria for giant impacts of protoplanets.
Astrophys. J. 744, 137.

\item
Glaschke, P., Amaro-Seoane, P., Spurzem, R., 2014.
Hybrid methods in planetesimal dynamics: description of a new composite algorithm.
Mon. Not. R. Astron. Soc. 445, 3620--3649.

\item
Goldreich, P., Tremaine, S., 1978.
The velocity dispersion in Saturn's rings.
\textit{Icarus} {\bf 34}, 227--239.

\item
Goldreich, P., Lithwick, Y., Sari, R., 2004.
Planet formation by coagulation: A focus on Uranus and Neptune.
Ann. Rev. Astron. Astrophys. 42, 549--601.

\item 
Greenberg, R., Wacker, J.F., Hartmann, W.K., Chapman, C.R., 1978.
Planetesimals to planets: Numerical simulation of collisional evolution.
Icarus 35, 1--26.

\item
Greenzweig, Y., Lissauer, J.J., 1990.
Accretion rates of protoplanets.
Icarus 87, 40--77.

\item
Grimm, S.L., Stadel, J.G., 2014.
The GENGA code: Gravitational encounters in $N$-body simulations with GPU acceleration.
Astrophys. J. 796, 23.

\item
H\"{a}meen-Anttila, K.A., 1978. 
An improved and generalized theory for the collisional evolution of 
Keplerian systems.
Astrophys. Sp. Sci. 58, 477--520.

\item 
Hansen, B., 2009.
Formation of the terrestrial planets from a narrow annulus.
Astrophys. J.,  703, 1131--1140.

\item
Hasegawa, M., Nakazawa, K., 1990.
Distant encounter between Keplerian particles.
Astron. Astrophys. 227, 619--627.

\item
Hayashi, C., 1981.
Structure of the solar nebula, growth and decay of magnetic fields and effects of magnetic and turbulent viscosities on the nebula.
Suppl. Prog. Theoret. Phys. 70,  35--53.

\item 
Hill, G.W., 1878.
Researches in the lunar theory.
Am. J. Math. 1,  5--26, 129--147, 245--261.

\item
Ida, S., 1990.
Stirring and dynamical friction rates of planetesimals in the solar gravitational field.
Icarus 88, 129--145.

\item
Ida, S., Nakazawa, K., 1989.
Collisional probability of planetesimals revolving in the solar gravitational field. III.
Astron. Astrophys. 224, 303--315.

\item
Ida, S., Makino, J., 1992.
$N$-body simulations of gravitational interaction between planetesimals and a protoplanet. 
I. Velocity distribution of planetesimals.
Icarus 96, 107--120.

\item
Ida, S., Makino, J., 1993.
Scattering of planetesimals by a protoplanet: slowing down of runaway growth.
Icarus 106, 210--227.

\item
Ida, S., Bryden,G., Lin, D.N.C., Tanaka, H., 2000.
Orbital migration of Neptune and orbital distribution of trans-Neptune objects.
Astrophys. J. 534, 428--445.

\item
Inaba, S., Ikoma, M., 2003.
Enhanced collisional growth of a protoplanet that has an atmosphere.
Astron. Astrophys. 410, 711--723.

\item
Inaba, S., Tanaka, H., Nakazawa, K., Wetherill, G.W., Kokubo, E. 2001.
High-accuracy statistical simulation of planetary accretion: II. Comparison with $N$-body simulations. 
Icarus 149, 235--250.

\item
Inaba, S., Wetherill, G.W., Ikoma, M., 2003.
Formation of gas giant planets: core accretion models with fragmentation 
and planetary envelope.
Icarus 166, 46--62.

\item
Johansen, A., Oishi., J.S., Mac Low, M.-M., Klahr, H., Henning, T., Youdin, A., 2007.
Rapid planetesimal formation in turbulent circumstellar disks.
Nature 448, 1022--1025.

\item
Kenyon, S.J., Luu, J.X., 1998.
Accretion in the early Kuiper belt. I. Coagulation and velocity evolution.
Astron. J., 115, 2136--2160.

\item
Kirsh, D.R., Duncan, M.J., Brasser, R., Levison, H.F., 2009.
Simulations of planet migration driven by planetesimal scattering.
Icarus 199, 197--209.

\item
Kobayashi, H., Tanaka, H., Krivov, A.V., Inaba, S., 2010.
Planetary growth with collisional fragmentation and gas drag.
Icarus 209, 836--847.

\item
Kobayashi, H., Tanaka, H., Krivov, A.V., 2011.
Planetary core formation with collisional fragmentation and atmosphere to form
gas giant planets.
Astrophys. J. 785, 35.

\item
Kokubo, E.,  Ida, S.,  1996.
On runaway growth of planetesimals.
Icarus 123, 180--191.

\item
Kokubo, E.,  Ida, S., 1998.
Oligarchic growth of protoplanets.
Icarus 131, 171--178.

\item 
Kokubo, E., Ida, S., 2000.  
Formation of protoplanets from planetesimals in the solar nebula.
Icarus 143, 15--27.

\item
Kokubo, E., Genda, H., 2010.
Formation of terrestrial planets from protoplanets 
under a realistic accretion condition.
Astrophys. J. 714, L21--L25.

\item
Kortenkamp, S.J., Wetherill G.W, 2000.
Terrestrial planet and asteroid formation in the presence of giant planets.
I. Relative velocities of planetesimals subject to Jupiter and Saturn perturbations. 
Icarus 143, 60--73.

\item
Kral, Q., Th\'{e}bault, P., Charnoz, S., 2013.
LIDT-DD: A new self-consistent debris disc model including 
radiation pressure and coupling dynamical and collisional evolution.
Astron. Astrophys. 558, A121.

\item
Levison, H.F., Morbidelli, A., 2007.
Models of the collisional damping scenario for ice-giant planets
and Kuiper belt formation.
Icarus 189, 196--212.

\item
Levison, H.F., Thommes, E., Duncan, M.J., 2010.
Modeling the formation of giant planet cores. I. Evaluating key processes.
Astron. J., 139, 1297--1314.

\item
Levison, H.F., Duncan, M.J., Thommes, E., 2012 (LDT12).
A Lagrangian integrator for planetary accretion and dynamics.
Astrophys. J. 144, 119.

\item
Minton, D.A., Levison, H.F., 2014.
Planetesimal-driven migration of terrestrial planet embryos.
Icarus 232, 118--132.

\item 
Morbidelli, A., Bottke, W.F., Nesvorn\'{y}, D., Levison, H.F., 2009.
Asteroids were born big.
Icarus 204, 558--573.

\item 
Morishima, R., Schmidt, M.W., Stadel, J., Moore, B. 2008.
Formation and accretion history of terrestrial planets from runaway growth through to late time: Implications for orbital eccentricity.
Astrophys. J. 685, 1247--1261.

\item 
Morishima, R., Stadel, J., Moore, B. 2010.
From planetesimals to terrestrial planets: $N$-body simulations including the effects of nebular gas and giant planets.
Icarus 207, 517--535.

\item
Morishima, R., Golabek, G.J., Samuel, H., 2013.
$N$-body simulations of oligarchic growth of Mars: Implications for Hf-W chronology.
Earth Planet. Sci. Lett. 366, 6--16.

\item
Murray, C.D., Dermott, S.F., 1999.
Solar system dynamics.
Cambridge Univ. Press, Cambridge, UK. 

\item
Nagasawa, M., Tanaka, H., Ida, S., 2000.   
Orbital evolution of asteroids during depletion of the solar nebula.
Astron. J. 119, 1480--1497.

\item
Nagasawa, M., Lin, D.N.C., Thommes, E.W., 2005.
Dynamical shake-up of planetary systems. I. Embryo trapping and induced collisions by 
the sweeping secular resonance and embryo-disk tidal interaction.
Astrophys. J. 635, 578--598.

\item
Nakazawa, K., Ida, S., Nakagawa, Y., 1989.
Collisional probability of planetesimals revolving in the solar gravitational field. I. Basic formulation.
Astron. Astrophys. 220, 293--300.

\item
Nesvold, E.R., Kuchner, M.J., Rein, H., Pan, M., 2013.
SMACK: A new algorithm for modeling collisions and dynamics of planetesimals in debris disks.
Astrophys. J. 777, 144.

\item
Ngo, H.H.K., 2012.
Numerical simulations of giant planetary core formation.
MA thesis. Queen's Univ., Kingston, Canada.

\item
Ohtsuki, K., 1992.
Equilibrium velocities in planetary rings with low optical depth.
Icarus 95, 265--282.

\item
Ohtsuki, K., 1999.
Evolution of particle velocity distribution in a circumplanetary disk
due to inelastic collisions and gravitational interactions.
Icarus 137, 152--177.

\item
Ohtsuki, K., Ida, S., 1998.
Planetary rotation by accretion of planetesimals with nonuniform
spatial distribution formed by the planet's gravitational perturbation.
Icarus 131, 393--420.

\item
Ohtsuki, K., Tanaka, H., 2003.
Radial diffusion rate of planetesimals due to gravitational encounters.
Icarus 162, 47--58.

\item
Ohtsuki, K., Stewart, G.R., Ida, S., 2002.
Evolution of planetesimal velocities based on three-body orbital integrations and growth of protoplanets.
Icarus 155, 436--453.

\item
Ormel, C.W., Dullemond, C.P., Spaans, M., 2010a.
Accretion among preplanetary bodies: The many faces of runaway growth.
Icarus 2010, 507--538. 

\item
Ormel, C.W., Dullemond, C.P., Spaans, M., 2010b.
A new condition for the transition from runaway to oligarchic growth.
Astrophys. J. Lett. 714, L103--L107.

\item
Ormel, C.W., Ida, S., Tanaka, H., 2012.
Migration rates of planets due to scattering of planetesimals.
Astrophys. J. 758, 80.

\item
Papaloizou, J.C.B., Larwood, J.D., 2000.
On the orbital evolution and growth of protoplanets embedded in a gaseous disc.
Mon. Not. R. Astron. Soc. 315, 823--833.

\item
Petit, J.M., H\'{e}non, M., 1987.
A numerical simulation of planetary rings.
II. Monte Carlo Model.
Astron. Astrophys. 188, 198--205.

\item
Press, W.H., Teukolsky, S.A., Vetterling, W.T., Flannery B.P., 1986.
Numerical Recipes. 
Cambridge Univ. Press, Cambridge, UK.

\item
Queck, M., Krivov, A.V., Srem\v{c}evi\'{c}, M., Th\'{e}bault, P., 2007.
Collisional velocities and rates in resonant planetesimal belts.
Celestial Mech. Dyn. Astr. 99, 169--196.

\item
Rafikov, R.R., 2003.
Dynamical evolution of planetesimals in protoplanetary disks.
Astron. J. 126, 2529--2548.

\item
Rafikov, R.R., Slepian, Z.S., 2010.
Dynamical evolution of thin dispersion-dominated planetesimal disks.
Astron., J., 39, 565--579.

\item
Rein, H., Lesur, G., Leinhardt, Z.M., 2010
The validity of the super-particle approximation during planetesimal formation.
Astron. Astrohgys. 511, A69.

\item
Richardson, D.C., Quinn, T., Stadel, J., Lake, G., 2000. 
Direct large-scale $N$-body simulations of planetesimal dynamics.
Icarus 143, 45--59. 

\item
Shoemaker, E.M., Wolfe, R.F., 1982.
Cratering time scales for the Galilean satellites. 
In: Morrison, D. (Ed.), Satellites of Jupiter. 
Univ. of Arizona Press, Tucson, pp. 277--339. 

\item 
Spaute, D., Weidenschilling, S.J., Davis, D.R., Marzai, F., 1991.
Accretional evolution of a planetesimal swarm: 1. A new simulation.
Icarus 92, 147--164.

\item
Stewart, G.R., Ida, S., 2000.
Velocity evolution of planetesimals: Unified analytic formulae and comparison with 
$N$-body simulations.
Icarus 143, 28--44.

\item 
Tanaka, H., Ida, S., 1996.
Distribution of planetesimals around a protoplanet in the nebula gas. 
I. Semi-analytic calculation of the gravitational scattering by a protoplanet.
Icarus 120, 371--386.

\item 
Tanaka, H., Ida, S., 1997.
Distribution of planetesimals around a protoplanet in the nebula gas. II: Numerical simulations.
Icarus 125,  302--316.

\item
Weidenschilling, S.J., 1989.
Stirring of a planetesimal swarm: The role of distant encounters.
Icarus 80, 179--188. 

\item 
Weidenschilling, S.J., 2011.
Initial sizes of planetesimals and accretion of the asteroids.
Icarus 214, 671--684.

\item
Weidenschilling, S.J., Spaute, D., Davis, D.R., Marzari, F., Ohtsuki, K., 1997.
Accretional evolution of a planetesimal swarm.
Icarus 128, 429--455.

\item 
Wetherill, G.W., Stewart, G.R., 1989.
Accumulation of a swarm of small planetesimals. 
Icarus 77, 330--357.

\item 
Wetherill, G.W., Stewart, G.R., 1993.
Formation of planetary embryos: Effects of fragmentation, low relative velocity, and independent variation of eccentricity and inclination.
Icarus 106, 190--209.

\end{description}

\end{document}